\documentclass[draft]{agujournal2019}
\usepackage{url} 
\usepackage{lineno}
\usepackage[inline]{trackchanges} 
\usepackage{soul}
\usepackage{amsmath}
\usepackage{amssymb}

\journalname{JGR: Solid Earth}

\begin{document}
\title{Mitigation and optimization of induced seismicity using physics-based forecasting}
\authors{Ryley G. Hill$^{1,2}$, Matthew Weingarten$^1$, Cornelius Langenbruch$^3$, Yuri Fialko$^2$}

\affiliation{1}{Department of Geological Sciences, San Diego State University, San Diego, California, USA}
\affiliation{2}{Scripps Institution of Oceanography, University of California San Diego, La Jolla, California, USA}
\affiliation{3}{Institute of Geological Sciences, of Freie Universität Berlin, Berlin, Germany}

\correspondingauthor{Ryley G. Hill}{ryhill@ucsd.edu}
\begin{keypoints}
\item 1) Poroelastic and statistical model suggests induced seismicity in Raton Basin is still primarily driven by wastewater injection.
\item 2) Optimization can reduce seismic hazard for a given amount of injected fluid or maximize fluid injection for a prescribed seismic hazard.
\item 3) Optimization tends to spread out higher rate injection wells and may be a useful tool to reduce basin-scale induced seismic hazard.
\end{keypoints}

\begin{abstract}
Fluid injection can induce seismicity by altering stresses on pre-existing faults. Here, we investigate minimizing induced seismic hazard by optimizing injection operations in a physics-based forecasting framework. We built a 3D finite element model of the poroelastic crust for the Raton Basin, Central US, and used it to estimate time dependent Coulomb stress changes due to $\sim$25 years of wastewater injection in the region.  Our finite element model is complemented by a statistical analysis of the seismogenic index (SI), a proxy for critically stressed faults affected by variations in the pore pressure. Forecasts of seismicity rate from our hybrid physics-based statistical model suggest that induced seismicity in the Raton Basin, from 2001 - 2022, is still driven by wastewater injection. Our model suggests that pore pressure diffusion is the dominant cause of Coulomb stress changes at seismogenic depth, with poroelastic stress changes contributing about 5\% to the driving force. Linear programming optimization for the Raton Basin reveals that it is feasible to reduce seismic hazard for a given amount of injected fluid (safety objective) or maximize fluid injection for a prescribed seismic hazard (economic objective). The optimization tends to spread out high-rate injectors and shift them to regions of lower SI. The framework has practical importance as a tool to manage injection rate per unit field area to reduce induced seismic hazard. Our optimization framework is both flexible and adaptable to mitigate induced seismic hazard in other regions and for other types of subsurface fluid injection.
\end{abstract}

\section*{Plain Language Summary}
\noindent
The Raton Basin, in the central United States, has had a remarkable increase in seismicity coincident with large wastewater injection since 2001. This seismicity primarily occurs at depths greater than several kilometers where preexisting faults in the crystalline basement are reactivated by fluid percolation. The spatial extent and rate of the induced earthquakes can inform hazard maps which display the probability of an earthquake occurrence within a specific time period. We use the physics-based and statistical models to develop an optimization framework that may help inform well operations. The proposed method allows for the maximization of injected fluid (the economic objective) and the reduction of seismic hazard (the safety objective).

\section{Introduction}
Induced seismicity is a growing problem world-wide as it accompanies a variety of industrial activities, including hydraulic fracturing \cite{rutqvist2015modeling,bao2016fault} and wastewater disposal \cite{ellsworth2013injection,keranen2014sharp,shirzaei2016surface}, extraction and storage of natural gas \cite{grasso1990ten,van2015induced,zbinden2017physics}, CO$_2$ sequestration \cite{goertz2014combining,white2016assessing}, and renewable geothermal energy exploitation \cite{fi&si00a,giardini2009geothermal,majer2007impact,mignan2015induced}. Within the last decade, a dramatic increase in seismic activity in the Central and Eastern United States (CEUS) was caused by deep injection of water that was co-produced with oil \cite{keranen2014sharp,walsh2015oklahoma,langenbruch2016will,langenbruch2018physics}. Several moderate (M5+) events were induced in historically aseismic regions \cite{ellsworth2013injection,weingarten2015high,foulger2018global}. Like natural tectonic earthquakes, induced events occur on pre-existing critically stressed faults, primarily in the crystalline basement \cite{townend2000faulting}.

The occurrence of induced seismicity is attributed to various physical mechanisms, including pore pressure diffusion, poroelastic coupling and stress changes caused by seismic or aseismic fault slip  \cite{segall2015injection,keranen2018induced,ge2022induced}. In general, all mechanisms may contribute to the triggering of seismicity, because induced earthquakes can be triggered by stress changes just above stress perturbations caused by the Earth’s tides (1-10 kPa) \cite{bachmann2012influence,cacace2021projecting,wang2022tidal,stokes2023pore}. Modelling studies at well-characterized injection locations show that the relative significance of these mechanisms varies from site to site depending on the physical rock
properties, reservoir structure, fault geometry, seismotectonic conditions, and distance
from injection among others. Pore pressure diffusion and poroelastic stress changes
are considered primary mechanisms for induced seismicity \cite{segall2015injection,keranen2018induced,zhai2019pore,ge2022induced,stokes2023pore}.

Understanding and mitigating the seismic response to fluid injection is still a major challenge, not just for wastewater disposal, but for other types of subsurface fluid injection: CO2 sequestration, enhanced geothermal systems and hydraulic fracturing. In each region where subsurface fluid injection occurs, it is paramount to future operations to find an optimal balance of efficient yet safe injection practices. The field of hydrogeology has long used coupled groundwater simulations and management models to optimize pressure changes in multiple wells for a certain benefit \cite{gorelick1983review,gorelickzhang2015}. For example, \citeA{gorelick1982optimal} sought the optimal solution that maximized pollutant disposal while meeting spatial water quality standards at the wells over time. A similar approach in the case of wastewater injection and induced seismicity could be to maximize injection while meeting spatial fault reactivation constraints. 

Here, we present a framework that seeks to optimize the amount of wastewater injected at the basin-scale with a fully-coupled poroelastic model combined with a statistical seismicity forecasting model. Optimization is performed under a spatially varying Coulomb failure stressing rate constraint dependent on faulting orientation \cite{king94,cocco_pore_2002,jin2022lithospheric}. We first demonstrate the hybrid model's effectiveness at forecasting the observed seismicity in the Raton Basin of Colorado and New Mexico -- a long-standing and well-documented case of induced seismicity. We then demonstrate the feasibility of future induced seismicity management using optimization of injection under various constraint scenarios. 

For our simulation and management models, we take advantage of the linearity in the fully coupled poroelastic equations as well as the linearity in the Coulomb stress equation. Coupled poroelastic calculations are performed using a 3D finite element hydromechanical model \cite{blabla}. Our statistical seismicity model follows the methodology of prior work performed in Oklahoma and Kansas, where spatiotemporal variations of induced seismic hazard are calculated from pore pressure changes and spatial variations of the subsurface’s susceptibility to induced earthquakes \cite{langenbruch2018physics}. The susceptibility is described
by the spatially varying seismogenic index (SI), a proxy for the number and stress state of
pre-existing basement faults affected by stress changes \cite{langenbruch2016will,shapiro2010seismogenic}. Note that the SI model applied in Oklahoma and Kansas only considered pore pressure changes, while we consider the fully coupled problem by including poroelastic stress changes in the Coulomb stress analysis. We then form a management model using a response matrix for rate dependent model constraints provided by the SI. 

The management models considered are three 5-year prospective scenarios that use the remnant pore pressure and stress conditions from prior injection in the Raton Basin. In each scenario, the optimization chooses which injection wells to operate and at which monthly rate of injection. The first scenario optimizes induced seismic hazard for an injection strategy that tapers the overall injection by 70\% from the 2022 levels (reduction objective). The second scenario minimizes the seismic hazard for the current Raton Basin injection rate, thus optimizing seismic hazard for a given injected volume (safety objective). The third scenario maximizes the total injected volume while holding constant Raton Basin's currently forecasted seismic hazard (economic objective). The total framework serves as a flexible platform by which the optimization of injection activities are drafted to reduce the seismic hazard and maximize an economic objective.

\subsection{Raton Basin}

\begin{figure}[h]
    \centering
    \includegraphics[scale=.95]{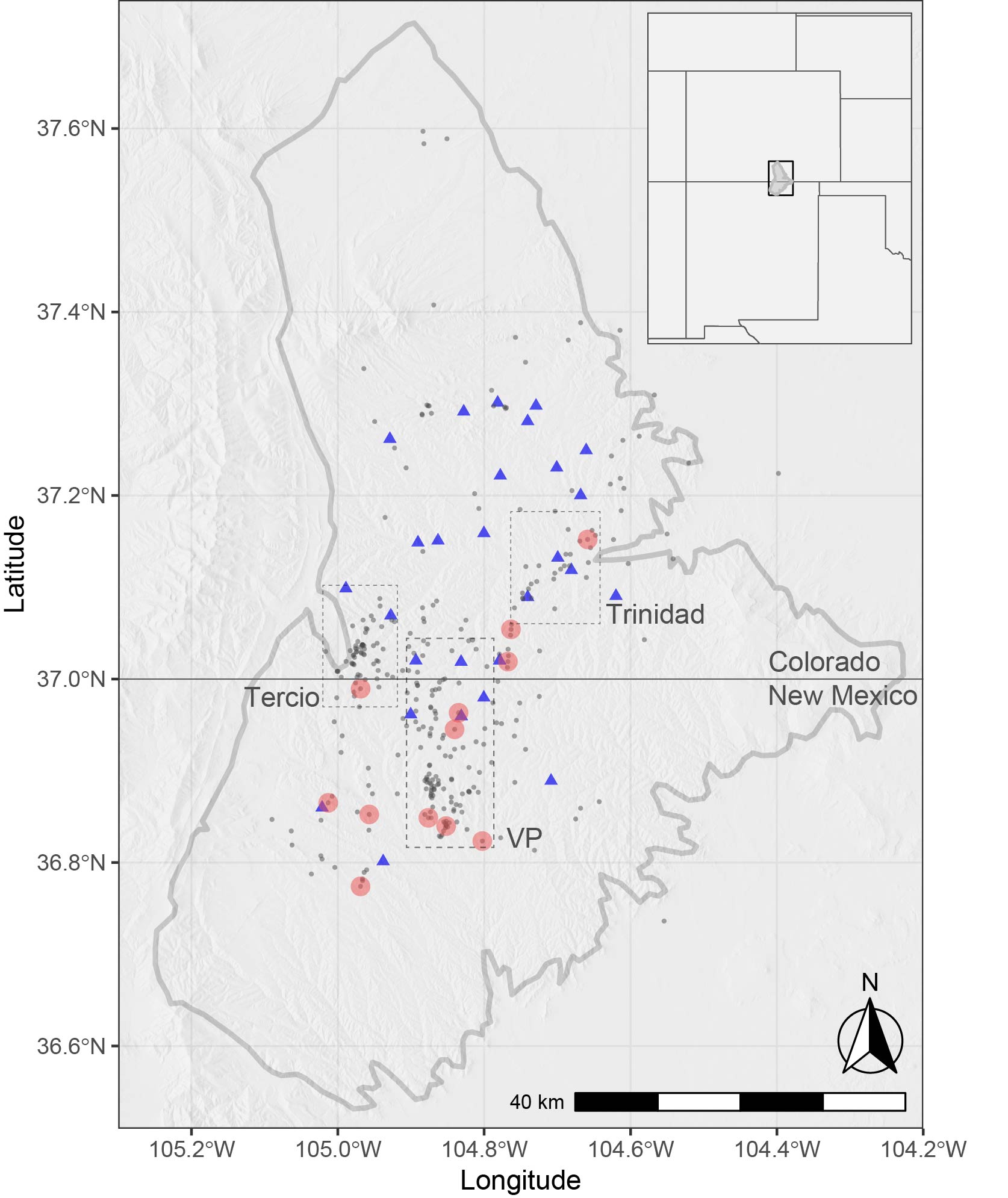}
    \caption{\textbf{Regional Context.} Light grey outline is the Raton Basin. Blue triangles are the 29 injection wells. Grey dots are earthquakes with M$\geq$2.5 and red dots are earthquakes with M$\geq$4 from Nov-2001 to July-2020. Boxed regions represent zones of seismicity: Tercio, Vermejo Park, and Trinidad.}
    \label{fig:regionalcontext}
\end{figure}

The Raton Basin, a $\sim$150 km long by $\sim$75 km wide sedimentary basin situated along the border between Colorado and New Mexico, has shown a remarkable seismic rate increase coincident with the beginning of industrial-scale wastewater injection in 2001 \cite{rubinstein20142001} (Figure \ref{fig:regionalcontext}). The rate increase was punctuated by the August 23rd, 2011 M5.3 Trinidad, Colorado earthquake, which caused structural damage in the nearby town of Trinidad, as well as 17 M4+ events, the most recent of which occurred on March 10th, 2023 [ANSS Comprehensive Catalog] (Figure \ref{fig:regionalcontext}). Previous studies have linked seismicity and wastewater injection wells operating in the basin using observational evidence and physical modeling \cite{rubinstein20142001,barnhart2014seismological,nakai2017possible}. The time-dependent seismic hazard associated with these induced events can change based on the pumping rates associated with the injection wells. Understanding both the spatial and temporal change of past seismic hazard is critical to mitigating future hazard.

\begin{figure}[h]
    \centering
    \includegraphics[scale=.55]{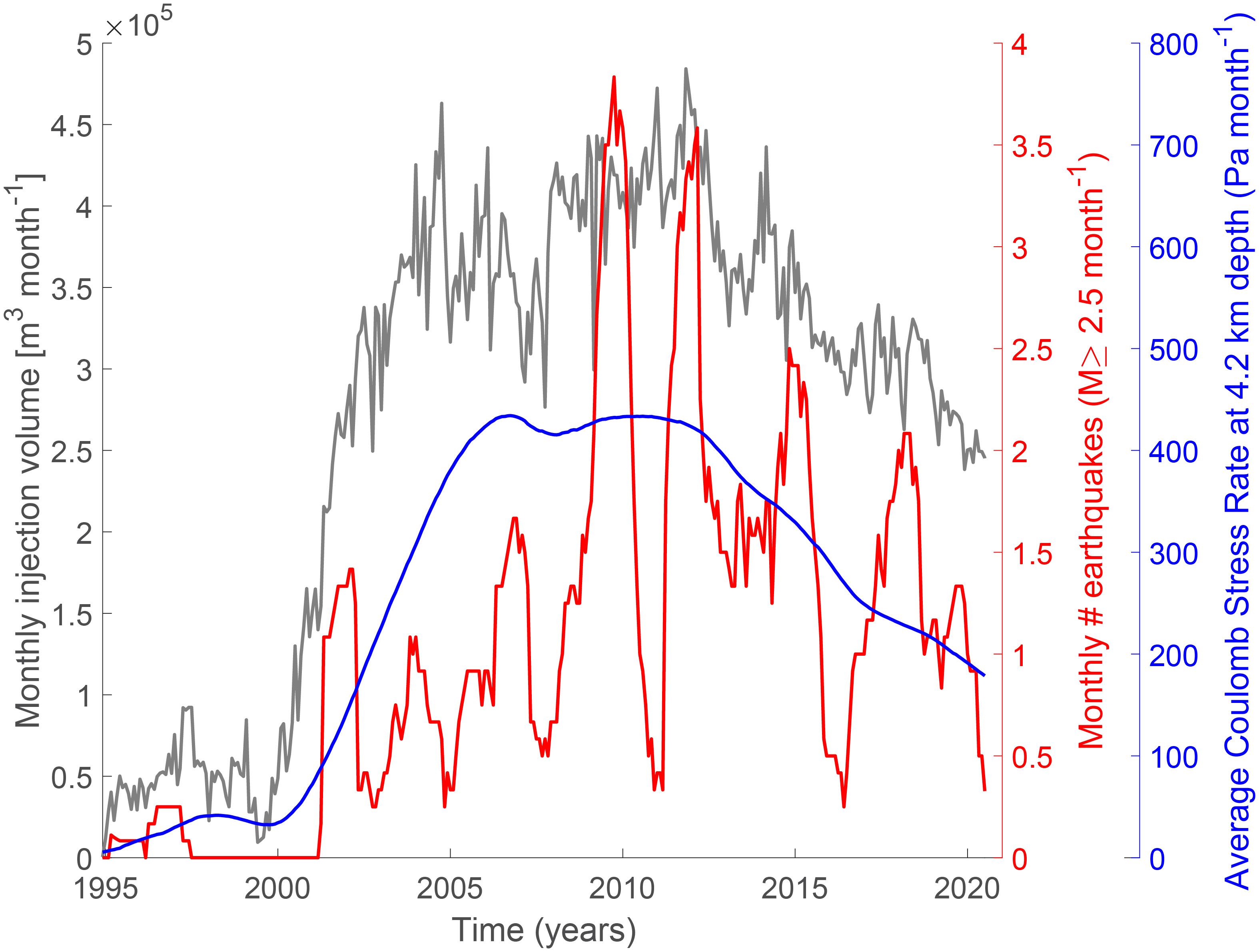}
    \caption{\textbf{Injection, induced earthquakes, and Coulomb stress rate.} Total monthly injection volume (grey), observed earthquakes M$\geq$2.5 (1 year moving mean), and the average modelled Coulomb stress rate in the study area. The Coulomb stress rate lags the injection rate due to the diffusion of pore pressure into the crystalline basement. A correlation between increased stress at depth and seismicity is observed.}
    \label{fig:injection}
\end{figure}

 Injection induced seismicity began in 2001 and peaked in late 2011 with the August 23rd, 2011 M5.3 Trinidad, Colorado earthquake (Figure \ref{fig:injection}). Since 2011, regional injection rates have declined more than $\sim$33$\%$, but the basin continues to exhibit an elevated seismicty rate with several recent M4+ events \cite{glasgow2021raton}. The regional stress field is heterogeneous, with a substantial rotation of the maximum horizontal stress from predominantly north-south to east-west directions \cite{snee2022state}. The earthquake focal mechanisms indicate a mixture of normal and strike-slip earthquakes \cite{wang2020injection,glasgow2021raton}.

\noindent

Geologic and hydrogeologic data indicate that the injection reservoir, the Dakota-Purgatoire Formation, a fractured sandstone reservoir, and underlying sedimentary units are permeable and hydraulically connected over a large lateral extent of the basin \cite{geldon1989ground,nelson2013outcrop}. The injection reservoir is also well-confined from the shallower stratigraphy within the basin by more than 700 m of poorly-permeable Pierre Shale. Additionally, the western boundary is characterized by the Sangre de Cristo Mountain thrust fault system, a complex of west-dipping, Laramide-age thrust faults that show dip-slip offsets of 0.6 to 3 km \cite{clark1966geology}. The observed seismicity in the Raton Basin is primarily found within the crystalline basement at average depths of 5 - 7 km below surface \cite{nakai2017possible,glasgow2021raton}. There is also strong evidence to suggest three prominent zones of seismicity: Tercio, Vermejo Park, and Trinidad (Figure \ref{fig:regionalcontext}) \cite{macartney2010raton,higley2007petroleum,barnhart2014seismological}. 


\section{Physics-based Forecasting Model}
\subsection{Methods}
\subsubsection{Linear poroelasticity}
To understand how injection across the Raton Basin is changing stress on pre-existing basement faults, we develop a fully coupled poroelastic model and compute the Coulomb stress changes at depth. Linear poroelasticity is essential to understanding the time-dependent coupling between the deformation of, and fluid flow in, hydrogeologic units within the Earth. The governing equations for a fully coupled linear poroelastic three-dimensional medium are defined as \cite{biot_general_1941,rice_basic_1976,wang00a}:
\begin{eqnarray}\label{eq:poro1}
G\nabla^2u_i + \frac{G}{1-2\nu}\frac{\partial^2u_k}{\partial x_i\partial x_k} &= \alpha\frac{\partial p}{\partial x_i}-F_i ,
\end{eqnarray}
\begin{eqnarray}\label{eq:poro2}
\alpha\frac{\partial \epsilon_{kk}}{\partial t} + S_{\epsilon}\frac{\partial p}{\partial t} &= \frac{k}{\mu}\nabla^2p + Q ,
\end{eqnarray}
where $G$ is the shear modulus, $u$ the displacement, $\nu$ the Poisson’s ratio, $\alpha$ the Biot-Willis coefficient, $F$ the body force, $k$ the permeability, $\mu$ the fluid viscosity, $S_\epsilon$ the constrained specific storage, $\epsilon_{kk}$ the volumetric strain, and $Q$ the fluid source \cite{wang00a}. Equations (\ref{eq:poro1}) are nearly identical to the classic equations for linear elasticity except for the coupling of pore pressure in the conservation of linear momentum equations (\ref{eq:poro1}) and the fluid flow coupled to strain by the requirement of fluid continuity (\ref{eq:poro2}). However, the system (\ref{eq:poro1})-(\ref{eq:poro2}) is more difficult to solve, with analytic solutions restricted to a few highly idealized cases. We solve the respective equations numerically using the three-dimensional finite element software Abaqus FEA \cite{blabla,labonte+09a,pearse&fi10a,hill2023major}. We validate the robustness of the numerical solution provided by Abaqus by summarizing its equivalency to that of a linear poroelastic framework \cite{jin2023saturated,jin20233d}. Additionally, we resolve the numerical pore pressure and stress outputs of a fluid mass point source compared with an analytical solution as further validation that Abaqus is a robust linear poroelastic framework \cite{RUDNICKI1986383} (see Supplementary Data).

The pore pressure diffusion is governed by an inhomogeneous diffusion equation Eq. (\ref{eq:poro2}). Because the fluid flow is coupled with the strain field pore pressure changes have direct effects on the stress and changes in the strain have direct effect on the fluid pressure. Under different assumptions, the stress field will uncouple from the pore pressure field and the diffusion equation resembles its hydrogeologic counterpart; the ground water flow equation $S\frac{\partial p}{\partial t} = \frac{k}{\mu}\nabla^2p + Q$ (where $S=S_{\epsilon}\frac{K_{v}^{(u)}}{K_v}$) \cite{detournay1993fundamentals,wang00a}.

\begin{figure}[h]
    \centering
    \includegraphics[scale=.40]{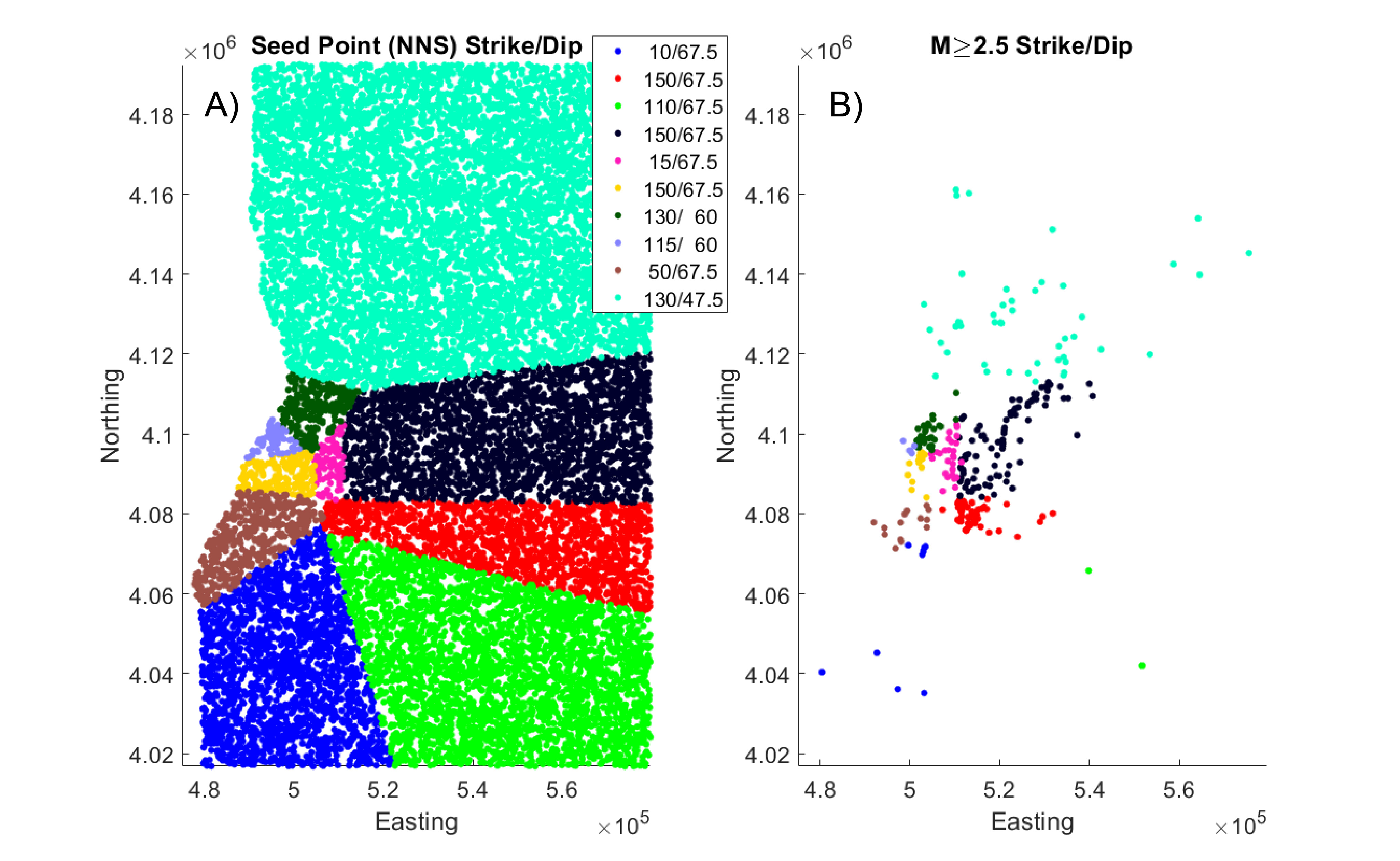}
    \caption{Map of the $\sim$25,000 seed points (A) and M$\geq$2.5+ earthquakes (B) with their associated fault geometries (strike/dip). The strikes were determined by a nearest neighbor search (NNS) across the basin by choosing several varying locations of strikes given by previous work \cite{glasgow2021raton} (their Fig. 5). Dips were determined by taking the closest large event focal mechanisms. Regardless of inaccuracies in our fault geometry assumptions, the fault geometries play a minimal role in the overall Coulomb stress rate calculations since the pore pressure rate is the largest component which is independent of fault geometry (SM Figure \ref{fig:ratioPtoCFS}).}
    \label{fig:faultGeom}
\end{figure}

Following \citeA{gorelick1982optimal} and \citeA{gorelick1993groundwater},  we use a physics-based numerical model to generate a unit source response matrix (see section \ref{resp}).  The key difference is that our simulation model incorporates the fully coupled poroelastic response (\ref{eq:poro1}-\ref{eq:poro2}), calculated using a finite element model, and generates a unit source response matrix of Coulomb stress (\ref{eq:cfs}) which is only possible due to the linearity in all the equations. The Coulomb stress is also dependent on fault geometries (SM Figure \ref{fig:faultGeom}).
\subsubsection{Stressing rate and earthquake probability}
Triggering of seismic events due to fluid injection can be adequately described by equations (\ref{eq:poro1}-\ref{eq:poro2}) and changes in Coulomb stress \cite{wang00a,cocco_pore_2002}. We define Coulomb stress $\tau$ as:
\begin{eqnarray}\label{eq:cfs}
\tau = \tau_s + \mu (\sigma_n+P),
\end{eqnarray}
where $\tau_s$ is the shear stress on a fault plane, $\sigma_n$ is the normal stress (compression is deemed negative), $P$ is the pore pressure, and $\mu$ is the coefficient of friction. An increase in pore pressure reduces the absolute value of the effective stress $(\sigma_e=\sigma_n+P)$ such that the Coulomb stress increases, corresponding to promotion of failure. In the presence of a regional stress field even modest perturbations in pore pressure may encourage slip on preexisting critically stressed faults. The diffusion of pore pressure is highly dependent on hydraulic properties. Furthermore, depending on fault geometries, the poroelastic coupling of the fluid may play a significant role in promotion or inhibition of fault failure, especially in the far field where the effects of fluid percolation are negligible \cite{segall2015injection}. 

\begin{figure}[h]
    \centering
    \includegraphics[scale=.60]{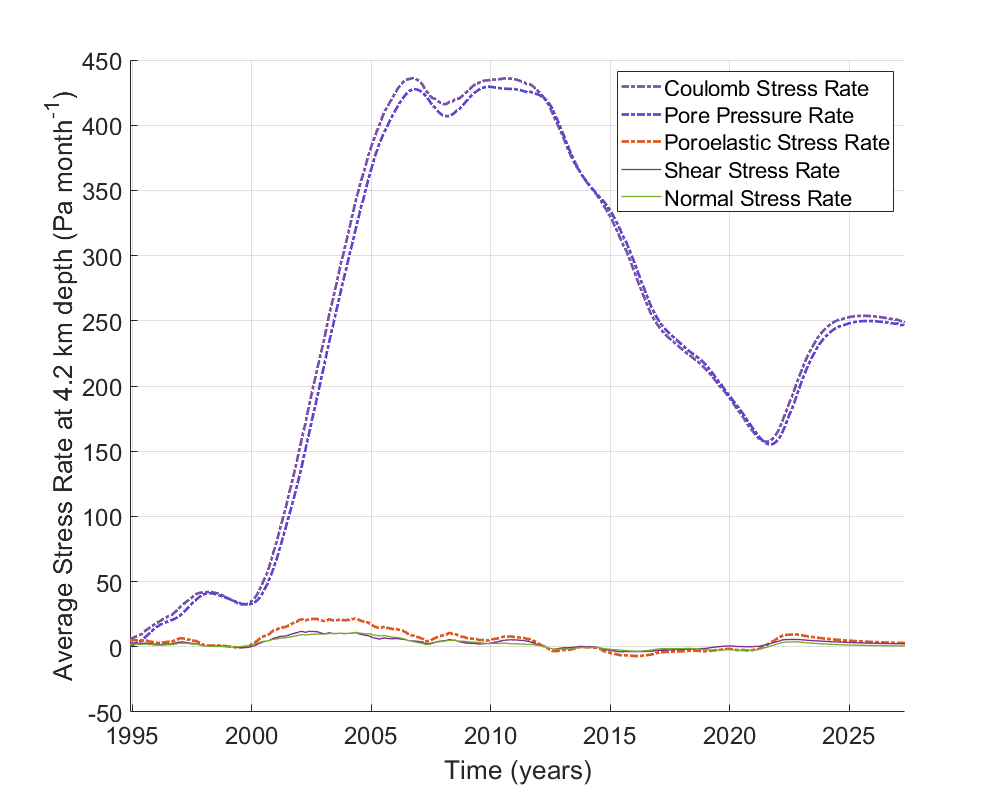}
    \caption{Average stressing rates for the central Raton basin at 4.2 km depth. The dominant signal of the Coulomb stress rate is the pore pressure rate.}
    \label{fig:ratioPtoCFS}
\end{figure}

Similar to previous work \cite{langenbruch2018physics}, which was carried out in the region of north-central Oklahoma and southernmost Kansas, seismicity data in the Raton Basin also shows the expected increase of earthquake probability with the rate of stress increase (Supplementary Methods). These observations can be used to describe the monthly earthquake rates $R_{\geq M}(\boldsymbol{r},t)$ according to a modified Gutenberg-Richter law for induced earthquakes \cite{langenbruch2018physics}:
\begin{align}\label{eq:seisrate}
    R_{\geq M}(\boldsymbol{r},t) = 10^{a(\boldsymbol{r},t)-bM} = \Bigg[ \frac{\partial}{\partial t} \tau_{\tau}(\boldsymbol{r},t)\Bigg]^2 10^{\Sigma_{\tau}(\boldsymbol{r)-bM}} ,
\end{align}
Here, we replaced the pore pressure rate, used by \cite{langenbruch2018physics} by the monthly Coulomb stressing rate $\frac{\partial}{\partial t} \tau_{\tau}(\boldsymbol{r},t)$ in space and time to add the effect of poroelastic coupling. $\Sigma_{\tau}(\boldsymbol{r})$ is the spatially varying Seismogenic Index (SI). The SI and $b$ values are evaluated through a specific calibration period (see section \ref{sec:calib}). The calibrated parameters are then used to forecast expected earthquake rates and to initialize the management model (see section \ref{sec:managementmodel}) for optimization. An important distinction from previous studies \cite{langenbruch2018physics} is the use of Coulomb stressing rate $\frac{\partial}{\partial t} \tau_{\tau}(\boldsymbol{r},t)$ as opposed to pressure rates. While pore pressure rates are still the dominant signal (SM Figure \ref{fig:ratioPtoCFS}), the fully coupled numerical model takes into account the stress field.
\subsection{Numerical Domain}
\begin{figure}[h]
    \centering
    \includegraphics[scale=.55]{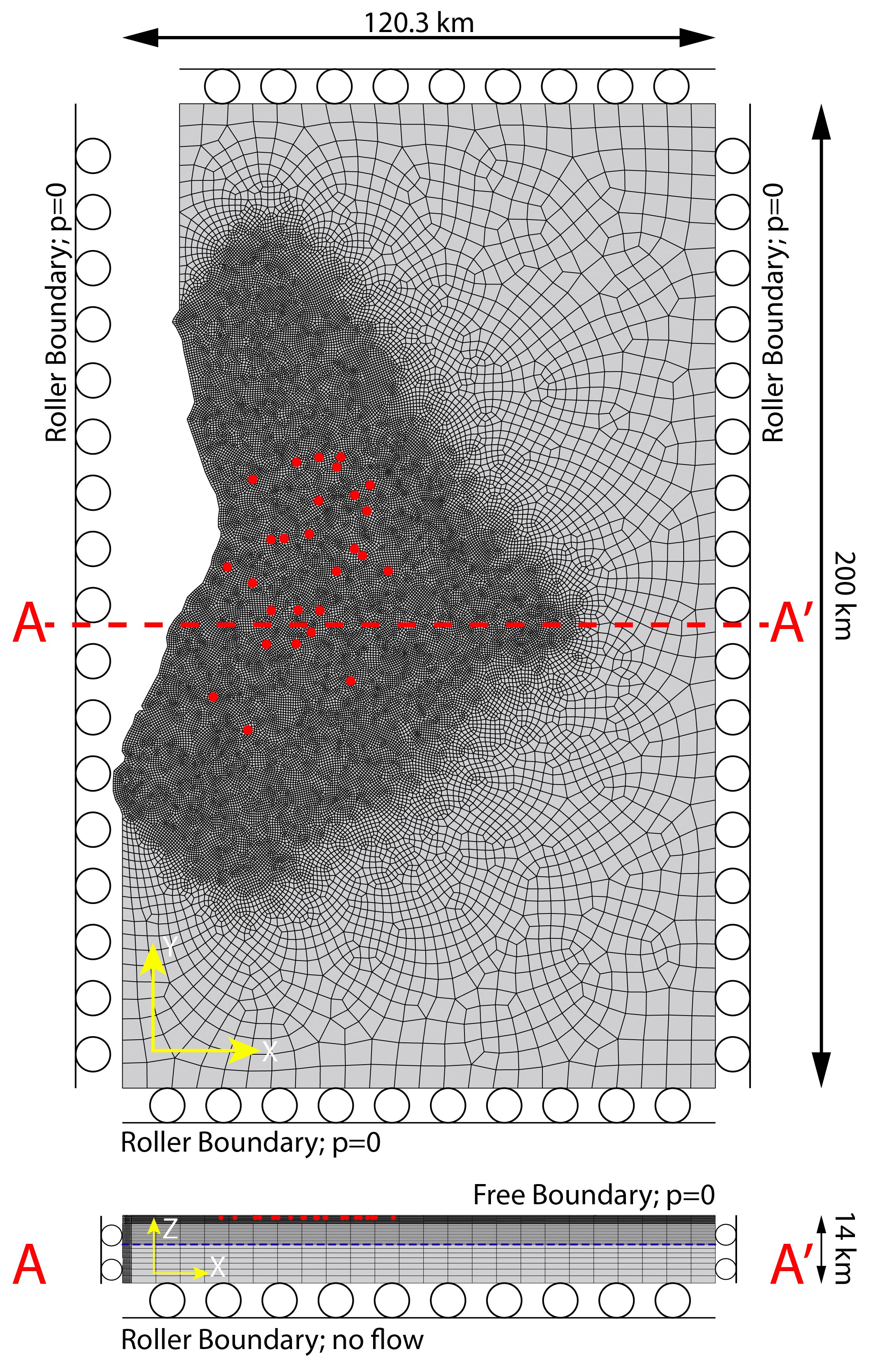}
    \caption{\textbf{Numerical Domain.} Three-dimensional finite-element model domain. The model mesh contains about 1.5 million hexahedron elements. The Red dots represent the well injection locations. The blue dotted line represents pore pressure and stress output location at the mean seismogenic depth ($\sim$7 km depth or 4240 m below the top of the crystalline basement).}
    \label{fig:numericaldomain}
\end{figure}

The numerical domain was developed and discretized in Abaqus CAE \cite<Complete Abaqus Environment, >[]{blabla}. The domain has horizontal dimensions of 120 km x 200 km and a depth dimension of 14 km, with the $y$ axis corresponding to north in the Universal Transverse Mercator coordinates (Figure \ref{fig:numericaldomain}). The finite-element mesh consists of nearly 1.5 million first-order hexahedral elements. Characteristic element sizes vary from 5,000 m in the far field to less than 500 m near the injection wells and in the vicinity of the central basin. The depth domain is partitioned into the 5 distinct hydrogeologic layers of the basin. The heterogeneous hydrogeologic properties of the model are summarized in Table 1. Permeability and storage parameters of the primary injection formations, the Dakota-Purgatoire and Morrison-Glorieta, were calibrated from analysis of injection step-rate tests (see Supplementary Materials). The permeability $k$ of the Dakota-Purgatoire formation and the Morrison-Glorietta formation is taken to be 6.4 - 6.8 $\times 10^{-14}$ and 5.8 - 8.9 $\times 10^{-14}$ m$^2$, respectively. While no wells penetrate the crystalline basement for diagnostic analysis of basement permeability, we chose a crystalline basement permeability ($k=1\times10^{-15}$ [m$^2$]) that results in the best correlation between the observed seismicity rates and modelled pressure rates (Figure \ref{fig:injection}). While this permeability is slightly higher than that inferred from small-scale field measurements of basement in other regions, it is similar to large-scale measurements made in regions of induced seismicity. In addition, it is also consistent with depth-dependent permeability models for continental crust at the mean depth of seismicity ($k\approx 3.35\times 10^{-15}$ [m$^2$]) \cite{shmonov2003permeability}, and constraints on in situ hydraulic diffusivity of the upper crust from observations of post-seismic deformation\cite<e.g.,>[]{fi04c}. The increased permeability is chosen to capture the basin-scale permeable faults that transmit fluid pressure to seismogenic depths. 

\noindent

\begin{table}[h]
\scalebox{0.65}{\begin{tabular}{l|l|l|l|l|l}
\hline
\multicolumn{1}{|l|}{Unit} & Pierre-Benton-Niobrara & Dakota-Purgatoire & Morrison-Entrada-Gloreita & Sangre De Cristo & Crystalline Basement \\ \hline \hline
Depth (km)                 & 1-1.4                  & 1.4-1.6           & 1.6 - 2           & 2 - 2.8          & 2.8 - 15                                              \\
Permeability ($m^2$)   & 1$\cdot10^{-20}$                  & 6.7$\cdot10^{-14}$ &8.9$\cdot10^{-14}$            & 8$\cdot10^{-15}$          & 1$\cdot10^{-15}$                                           \\
E (GPa)                    & 0.22                   & 38            & 32              &40.74                                & 60                                     \\
v                          & 0.3                    & 0.287              & 0.13              & 0.15            & 0.25                                                    \\
$K_s$ (GPa)                  & 0.34                   & 33.8               & 26.6               & 36.6               & 42                    \\                                      
$\phi$                  & 0.38                     & 0.25                & 0.07                & 0.06               & 0.01                   \\
\hline
\end{tabular}}
\begin{flushleft}



Table 1: \textbf{Material Properties.} Hydrogeologic material values for different units and their corresponding depths in the numerical model. Note that the model begins at 1 km depth below the surface.

\end{flushleft}
\end{table}

We assume initial conditions of equilibrium stress and pore pressure \cite[chapter 9]{segall2010earthquake}. Therefore, the model only considers the perturbing effects of the wastewater injection and does not include any tectonic loading. The bottom and sides of the model are fixed only in the surface normal direction (the roller boundary condition). The top surface of the model is stress-free. We model the Sangre de Cristo Mountain complex of thrust faults as barriers to cross-fault fluid flow and use an insulating condition at the western boundary of the model. We use the same injection depth of 1,500 m for all wells as the former is the middle depth of the modelled Dakota-Purgatoire injection reservoir. We record pore pressure and stress perturbations at the mean seismogenic depth of $\sim$7,040 m which is equivalent to $\sim$38,000 observation points for each time step. Generation of the SI map requires the full 29 well injection profile data ranging from November 1994 to December 2017, giving rise to 331 time steps, while the 5 year response matrix models require only 61 time steps.

\subsection{Seismogenic Index (SI)}\label{sec:calib}
\begin{figure}[h]
    \centering
    \includegraphics[scale=.55]{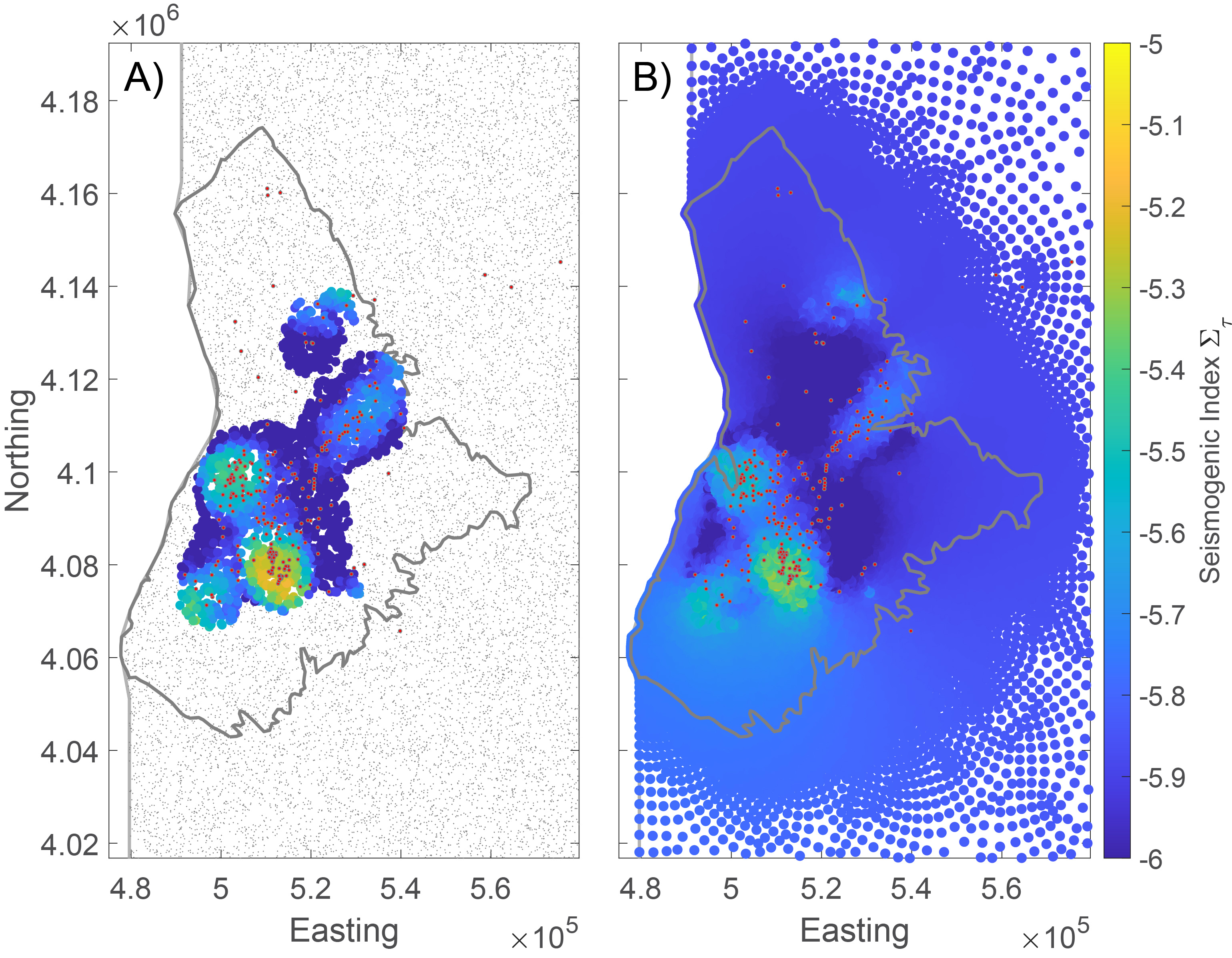}
    \caption{\textbf{Seismogenic Index $\Sigma_{\tau}$ Maps.} Mapped spatial variability of the SI in the Raton Basin. The SI is computed in local regions of 7-km radius around the 25,000 seed points (grey dots in panel A) ). The calibration time is between Nov-1994 and July-2016. See Methods for additional details. Red dots represent earthquakes M$\geq$2.5 used in calibration. Panel B) represents the inverse distance weighted interpolation of the SI to the model points used in the forward model management solutions.}
    \label{fig:SI}
\end{figure}

The SI map is a map of the seismo-tectonic state controlled by the number and stress state of pre-existing faults in the crystalline basement affected by Coulomb stress changes (Figure \ref{fig:SI})  \cite{langenbruch2018physics}. The SI ($\Sigma_\tau(r)$) is determined in local regions of 7 km radius at $\sim$25,000 seed points. The seed points represent the interpolated Coulomb stress changes produced by the model at the mean seismogenic depth within the crystalline basement. The higher the SI ($\Sigma_\tau(r)$) at each seed point, the higher the earthquake rate caused by a given Coulomb stress increase, because a higher number of (or more critically stressed) preexisting faults are affected by the Coulomb stress increase (see Eq. \ref{eq:seisrate}).

Calibration of the SI is set based on a calibration time period. In this way, future modelled Coulomb stressing rates are used to forecast expected spatiotemporal earthquake rate. We set the calibration time (Nov 1994 to July 2016) of our SI map prior to the Glasgow et al., 2020 study and find that forecasted earthquakes (July 2016 to July 2020) are well explained by basin Coulomb stressing rate, despite lowered injection rates at this time (Figure \ref{fig:seisRate}).

\begin{figure}[h]
    \centering
    \includegraphics[scale=.60]{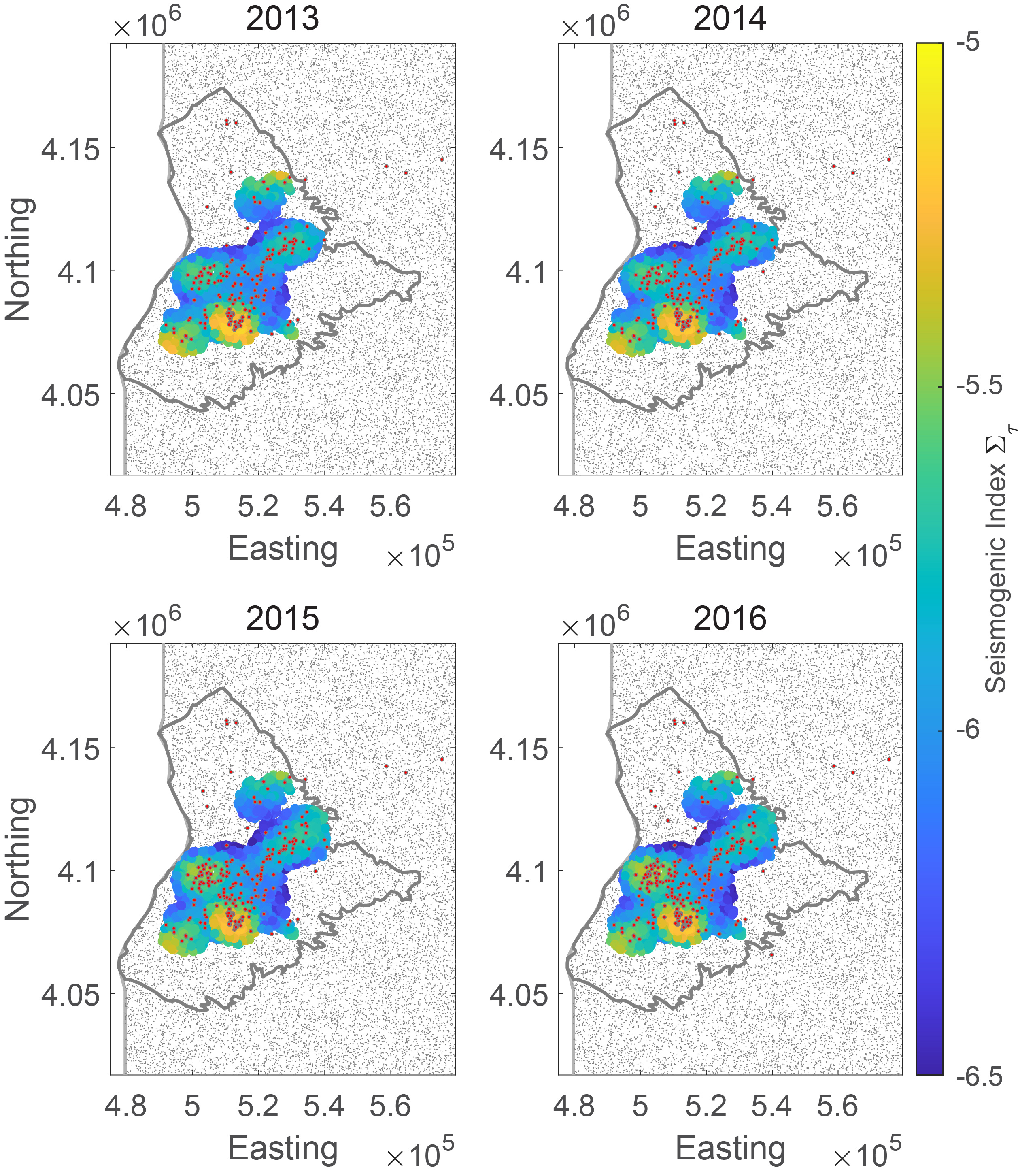}
    \caption{SI map for varying calibration period 2013-2016. With increasing earthquake count the SI improves in spatial resolution, but there is little change among the different calibration years.}
    \label{fig:simap7}
\end{figure}

\begin{figure}[h]
    \centering
    \includegraphics[scale=.60]{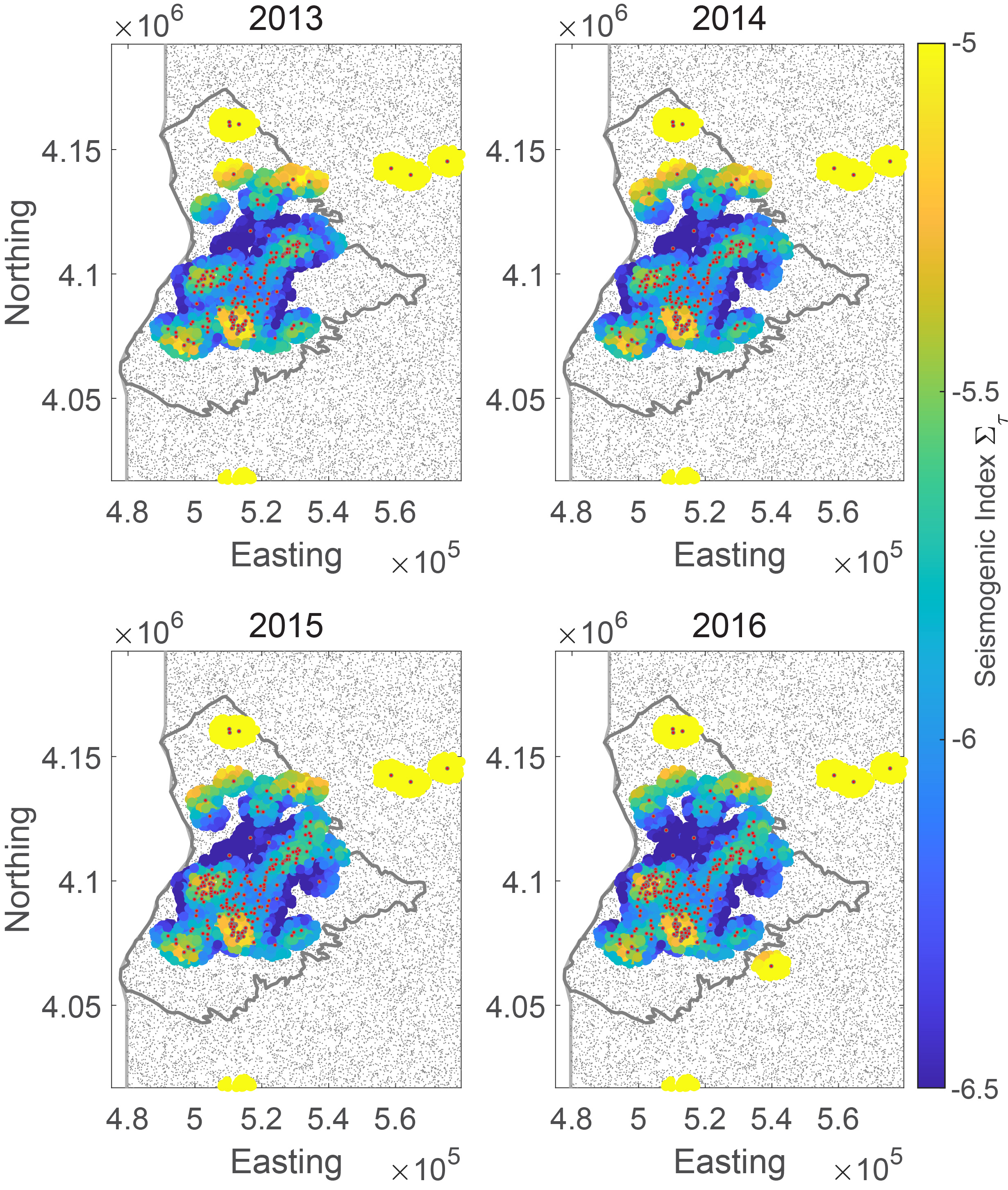}
    \caption{SI map for varying calibration period 2013-2016 for the model that uses a 5-km search radius and removes the $>$3 earthquake precondition. Outliers away from the basin provide high localized areas of enhanced SI. Notice that within the basin though, the overall structure and features of enhanced SI do not change.}
    \label{fig:simapr5}
\end{figure}

\begin{figure}[h]
    \centering
    \includegraphics[scale=.50]{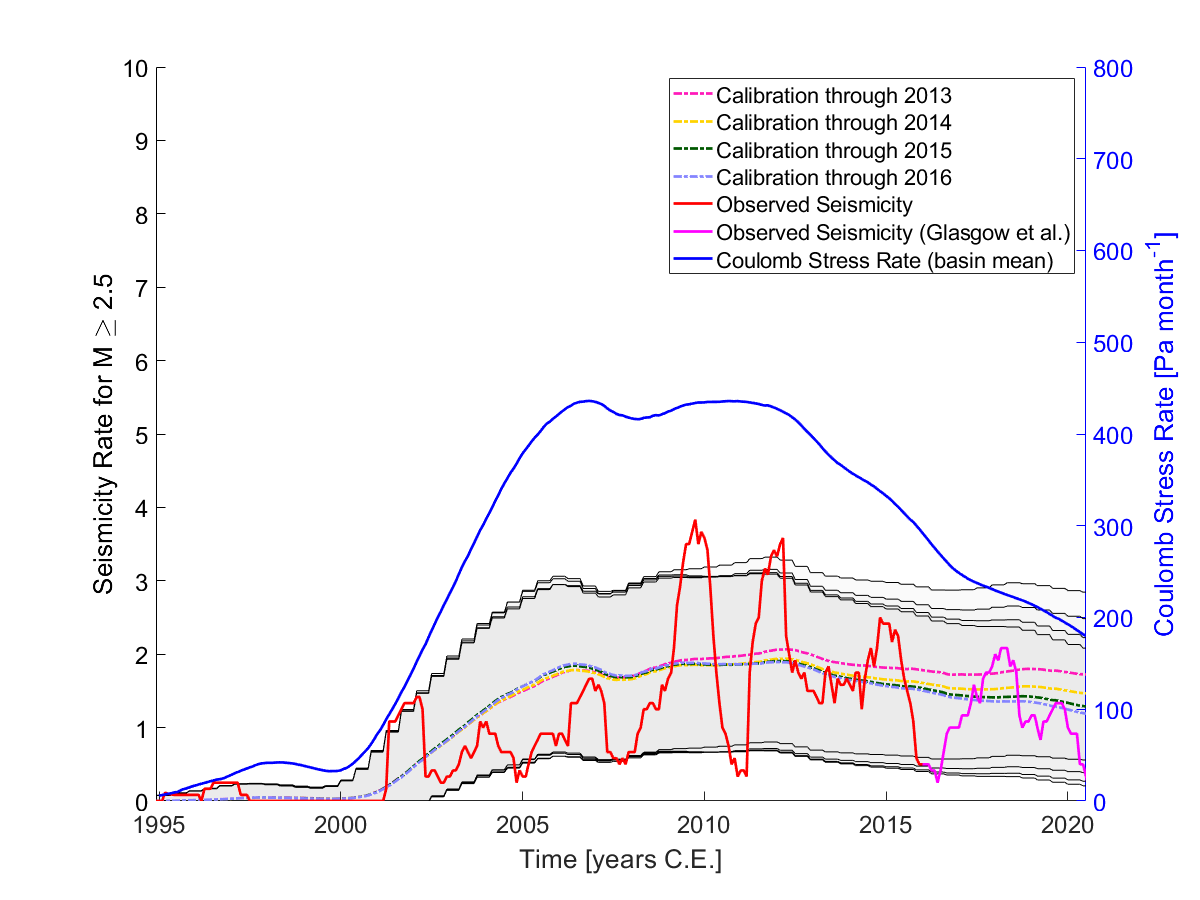}
    \caption{Seismicity rate forecasts, above our completeness magnitude M$\geq$2.5, compared to observed seismicity rate (1 year moving mean). Calibration period is from Nov 1994 to July 2016, prior to the Glagow et al., 2021 study. The earthquakes and time period used to calibrate the SI model is represented by the red line. The grey areas are the 95\% confidence bounds for the different calibration time periods for the the forecasted seismicity rate produced from the SI model that includes the inverse distance weighted interpolation (right panel of Figure \ref{fig:SI}). Magenta line represents the observed seismicity from Glasgow et al., 2021 which is well explained by the seismicity rate forecasted by our model.}
    \label{fig:forecastr5}
\end{figure}

Calibration of SI follows closely to previous methods \cite{langenbruch2018physics}. The following steps are performed to calibrate the SI maps:
\begin{enumerate}
    \item Monthly Coulomb stressing rates $\frac{\partial}{\partial t} \tau_{\tau}(\boldsymbol{r_n},t)$ at all $n$ seed points with a radius of 7-km around a selected seed point up to a given calibration time $t_c$ (we use Nov-1994 to July-2016) are extracted, squared, and summed $\sum\limits_n \big[ \frac{\partial}{\partial t} \tau_{\tau}(\boldsymbol{r_n},t\leq t_c) \big]^2$
    \item The total number $N_{M\geq M_c}(t\leq t_c)$ ($M_c=2.5$, see Supplementary Figure \ref{fig:mcplot}) of earthquakes within a 7-km radius around the current seed point observed up to the given calibration time is summed. 
    \item Estimate of the b-value is computed using all $M\geq M_c$ earthquakes recorded through the calibration time $t_c$ in the complete study area.
    \item The SI at location $\boldsymbol{r}$ is evaluated:
\begin{align}
        \Sigma_{\tau}(\boldsymbol{r}) = \log_{10}{N_{M\geq M_c}(t\leq t_c)}-\log_{10}\bigg\{ \sum\limits_n \big[ \frac{\partial}{\partial t} \tau_{\tau}(\boldsymbol{r_n},t\leq t_c) \big]^2\bigg\} + b(t_c)M
\end{align}
\end{enumerate}
Due to to the occurrence of singular earthquakes outside of the local areas of elevated seismicity one can get outlier SI values. These events are often attributed to Coulomb stressing rates that are quite low which results in significantly larger than average SI at those locations. Prior work found that as soon as two earthquakes occurred within the chosen radius of any given seed point a good estimate of the SI can be obtained \cite{langenbruch2018physics}. Our region uses a smaller radius and calibration magnitude. Therefore we precondition the SI to only be evaluated when there are more than 3 earthquakes. We evaluate the sensitivity of the SI for a smaller 5-km radius and removal of the ``more than 3 earthquakes'' precondition. These changes produce an SI map that appears different, as outliers are now included, but the overall seismicity rate remains very similar (SM Figure \ref{fig:simap7}-\ref{fig:forecastr5}).

\noindent

Within the central basin region, we find that the SI varies by about 1.5 units (Figure \ref{fig:SI}). A one unit increase in SI is the equivalent of expecting 10 times more earthquakes for the same CFS rate change at that location. A higher SI in the central basin corresponds spatially with the well known zones of seismicity: Tercio, Vermejo Park, and Trinidad.

The SI is dependent on the spatial density of the observed seismicity and the radius of inclusion. This implies that seed points without observed seismicity in a 7-km radius will not produce SI. For the purpose of forecasting seismicity and optimizing injection rates for the entire basin we use an inverse distance weighting interpolation (power=2, radius=$\infty$) (Figure \ref{fig:SI}) in areas that have no observed seismicity during the calibration period. The interpolated map helps inform the Coulomb stressing constraints in the SI dependent response matrix models.

\subsection{Results \& Discussion: Forecast Performance (2016 - 2020)}
\begin{figure}[h]
    \centering
    \includegraphics[scale=.50]{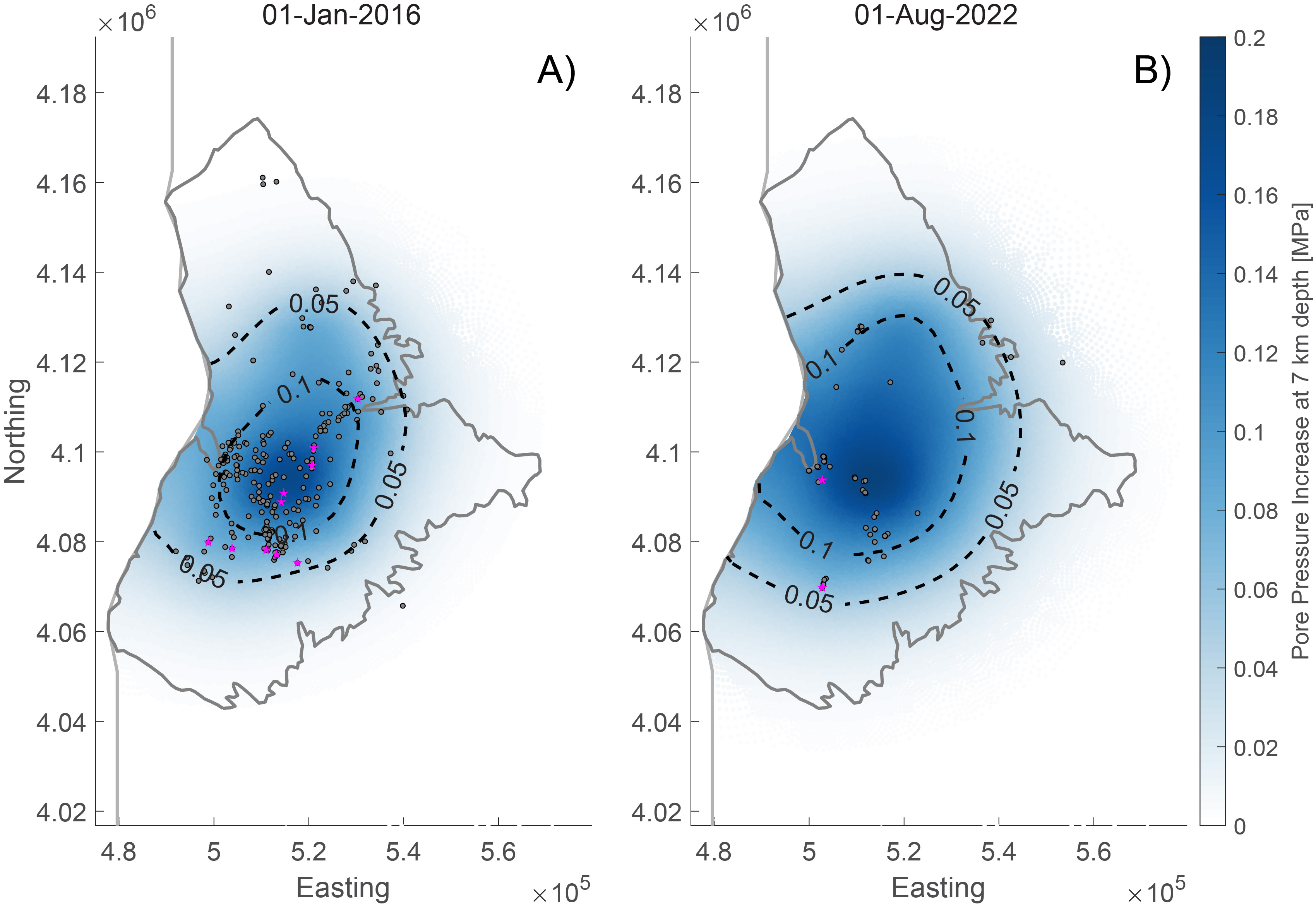}
    \caption{\textbf{Pore Pressure Increase.} A)) Pore pressure increase at mean seismogenic depth across the basin including seismicity from Dec 1994 through Jan 2016. Black dots represent earthquakes with M$\geq$2.5+ and magenta stars are earthquakes with M$\geq$4+. B) Pore pressure increase at mean seismogenic depth across the basin including seismicity between July 2016 to July 2022.}
    \label{fig:PPplot}
\end{figure}

\begin{figure}[h]
    \centering
    \includegraphics[scale=.50]{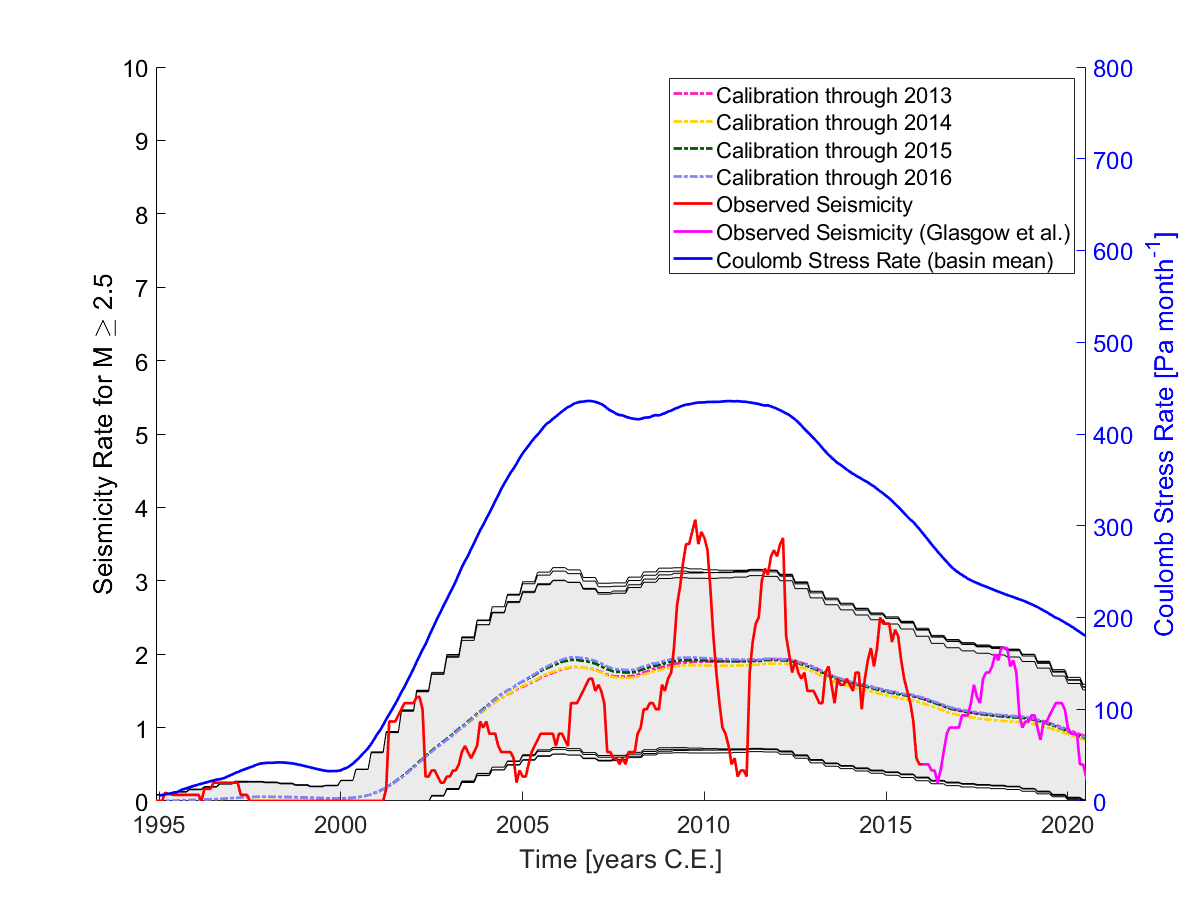}
    \caption{\textbf{Seismicity Rate Forecast.} Seismicity rate forecasts, above our completeness magnitude M$\geq$2.5, compared to observed seismicity rate (1 year moving mean). Calibration period is from Nov 1994 through 2013, 2014, 2015, and 2016 prior to the Glagow et al., 2021 study \cite{glasgow2021raton}. The earthquakes and longest calibration time period used to calibrate the SI model is represented by the red line. The varying dashed lines and grey boundaries are the 95\% confidence bounds forecasted by the seismicity rate produced from the SI model that includes the inverse distance weighted interpolation (right panel of Figure \ref{fig:SI}). Magenta line represents the observed seismicity from Glasgow et al., 2021 which is well explained by the seismicity rate forecasted by our model.}
    \label{fig:seisRate}
\end{figure}

The results of the time dependent pore pressure evolution and associated seismicity during our calibration time are shown in Figure \ref{fig:PPplot}. The pore pressure continues to increase at depth within the basin due to the diffusion of fluid pressure despite lowered injection rates during 2016-2022. The total pore pressure increases, but the rate of increase declines (Figure \ref{fig:injection}). Returning to Eq. (\ref{eq:seisrate}), we can now forecast seismicity rate beyond our calibration time using both the SI map and Coulomb stress perturbations from the numerical model. Figure \ref{fig:seisRate} depicts the seismicity rate forecasts from a variety of calibration time periods and the resulting projected seismicity rate between 2016 and 2020. There is little sensitivity of the modelled earthquake rates to the calibration time. We find that the observed seismicity rate from 2016 to 2020 is fit well by our calibrated SI model and the computed Coulomb stress changes.
\noindent

\noindent
Furthermore, assuming the occurrence of induced earthquakes follows a Poisson process  \cite{langenbruch2011inter,langenbruch2016will,shapiro2010seismogenic}, the probability of exceeding a magnitude M, that is the probability to observe one or more events of magnitude M or larger, is given by \cite{langenbruch2018physics}:
\begin{equation}\label{eq:prob}
    Pr(M) = 1 - Pr(0,M,N_{\geq M}) = 1 - \exp(-N_{\geq M})
\end{equation}
Where,  ($N_{\geq M}$) is the expected number of events of magnitude M or larger in a considered time interval (see Eq. \ref{eq:seisrate}). 

\begin{figure}[h]
    \centering
    \includegraphics[scale=.45]{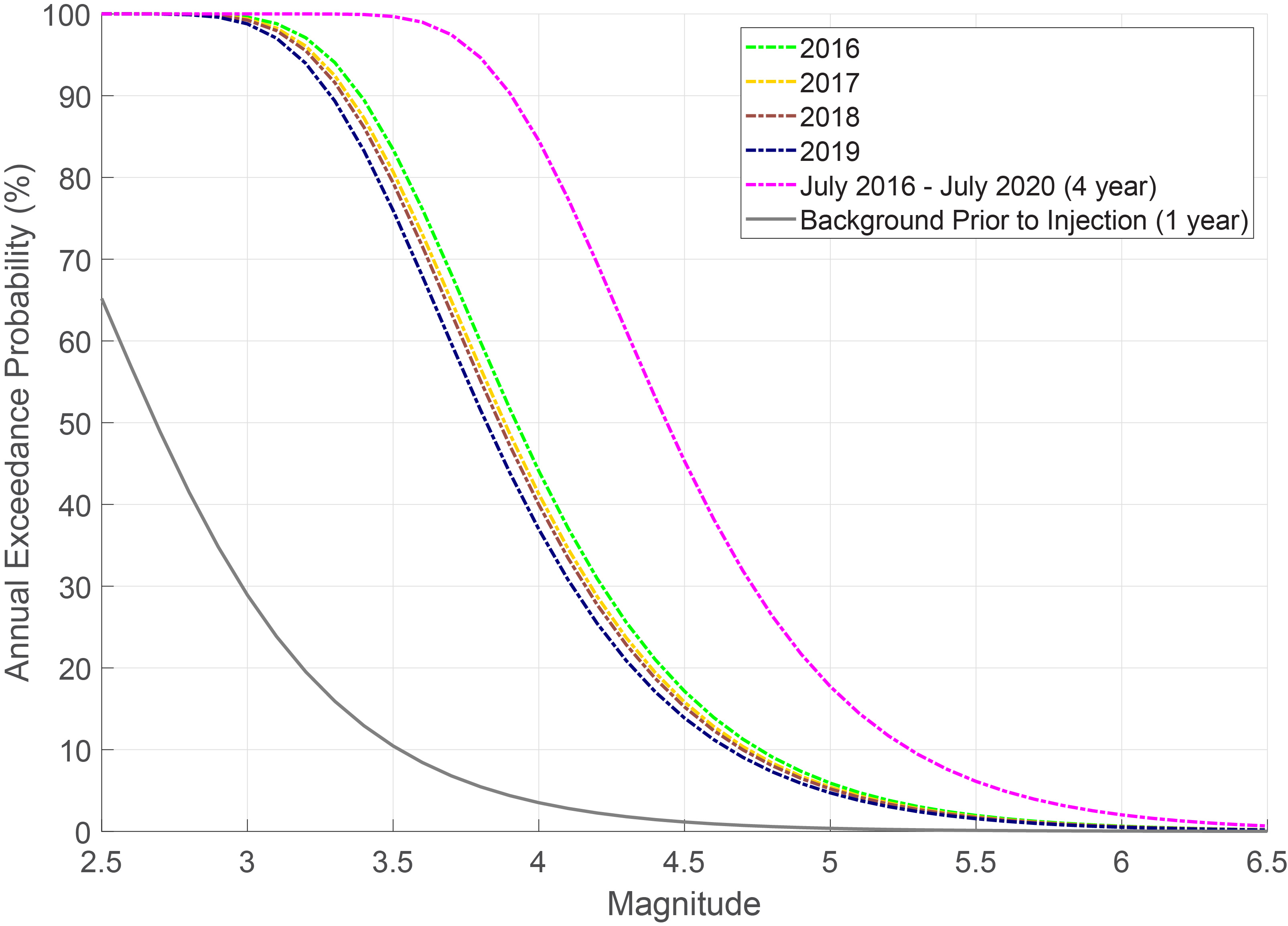}
    \caption{\textbf{Forecasted Magnitude Exceedance Probabilities.} Exceedance probabilities for magnitudes M$\geq$2.5-6.5 from our physic-based forecasting model. Each line represents the probability forecasted by our model based on the calibrated SI map and computed Coulomb stress model outputs. The forecasted probability from 2016-2020 is significantly higher than the tectonic background (grey line) and is highest in 2016. Background probabilities are derived from prior work \cite{rubinstein20142001}. Each year from 2016 to 2019 the the magnitude exceedance probabilities or decreasing, but still above the tectonic background level. From 2016 to 2020 the potential to trigger a M$\geq$5+ increases to $\sim$18\%.}
    \label{fig:annualExceed}
\end{figure}

Based on our calibrated model, we compute the annual expected number of events in the range from M 2.5-6.5 and determine magnitude exceedance probabilities using Eq. \ref{eq:prob} (Figure \ref{fig:annualExceed}). Our results suggest that between 2016-2020 there was a $\sim$85\% probability to observe one or more M$\geq$4+ earthquakes and a $\sim$18\% probability to observe one M$\geq$5+. We find that Coulomb stress rates at seismogenic depth continued to trigger seismicity between 2016-2020 although injection rates declined. Therefore, induced seismicity was still driven by wastewater injection during this time period. Declining injection rates alone are not necessarily an indicator of decreased seismic hazard as one must also consider diffusion-driven time delays in the induced seismicity process. 
\subsection{Results \& Discussion: Business As Usual Forecast (2022 - 2027)}\label{sec:bau}
\begin{figure}[h]
    \centering
    \includegraphics[scale=.35]{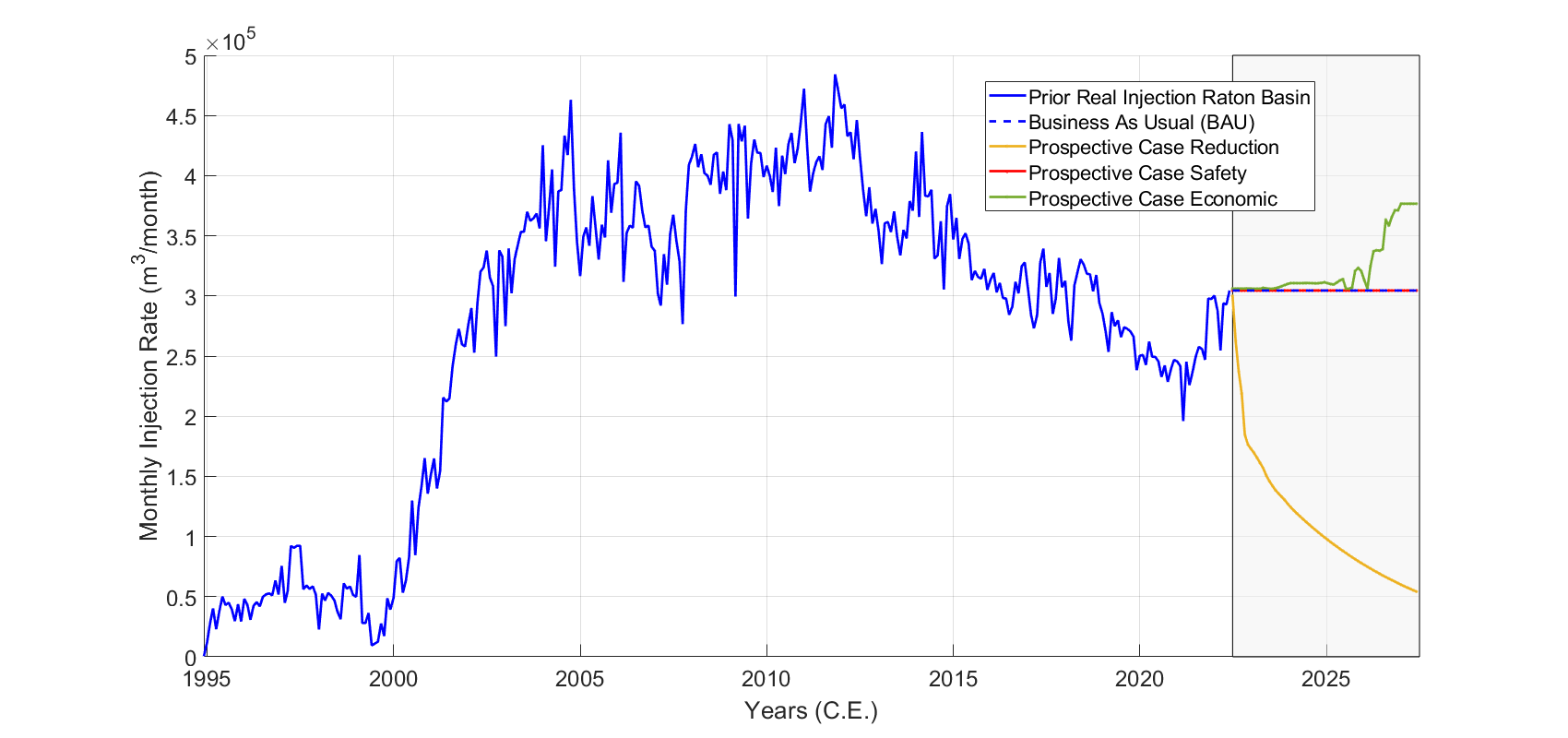}
    \caption{\textbf{Different Optimization Scenarios.} Plot shows the monthly injection rate (total of all 29 wells) for the observed data (blue). At June-01-2022, the next 5 year window (gray box) represent the forecasted injection rates. The business-as-usual rate takes the last known injection rates and holds them constant for the five years (blue-dash). The prospective case `Reduction' is the optimized injection rates subject to reducing the overall injection by 70\% in 5 years as well as a taper in individual well rates (yellow). The prospective case `Safety' is the optimized injection rates subject to the constraint that the total fluid injected must be the same as the BAU, but reduces the overall hazard (Figure \ref{fig:magExceed2}) (red). The prospective case `Economic' is the optimized injection rates subject to the constraint that the overall 5 year hazard must be the same as the BAU, but increases the overall injection (green).}
    \label{fig:comparImprove}
\end{figure}

\begin{figure}[h]
    \centering
    \includegraphics[scale=.35]{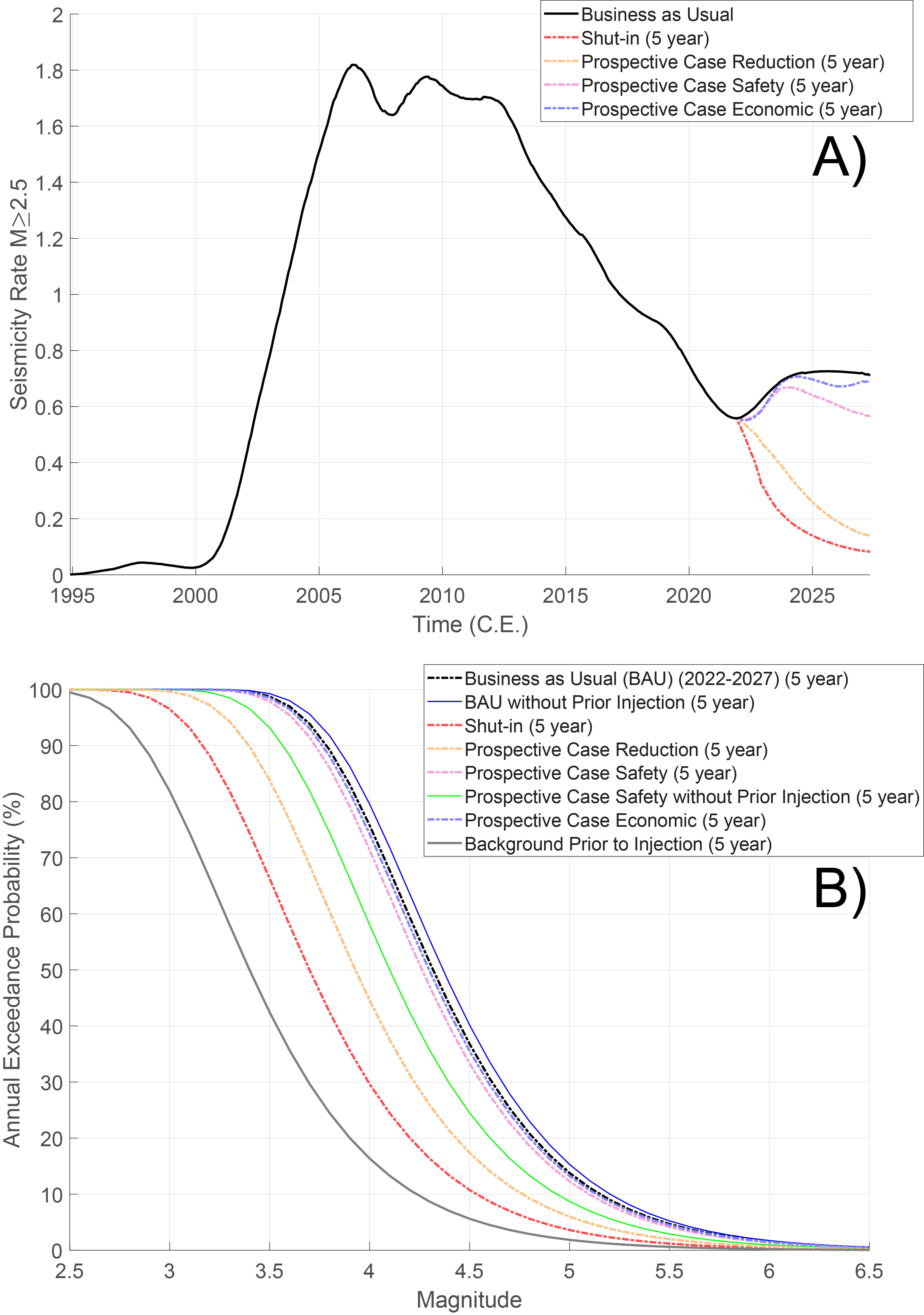}
    \caption{\textbf{Seismicity Rate Forecasts and Forecasted Magnitude Exceedance Probabilities (Optimizations).} A) Seismicity rate for M$\geq$2.5 from beginning of injection until beginning of optimization management period. Each of the 5 year optimizations have an associated exceedance probability in the next panel. B) Exceedance probabilities for scenarios projected into the future (see main text). The Business as Usual (BAU) forecast is determined by extrapolating the last observed injection well data into the next 5 years. The shut-in forecast is determined in a similar way, but for immediate shut-in of all wells in June-2022. Prospective Case `Reduction' considers reducing overall injection volume by 80\% while not allowing the probability of exceeding a M$\geq$4+ to be over 45\%. Prospective case `Safety' considers the same amount of fluid as the BAU case, but a more spatially optimized strategy based on the SI map. Prospective case `Economic' optimizes to a solution for much more fluid for the same seismic hazard as the BAU case.}
    \label{fig:magExceed2}
\end{figure}

\begin{figure}[h]
    \centering
    \includegraphics[scale=.45]{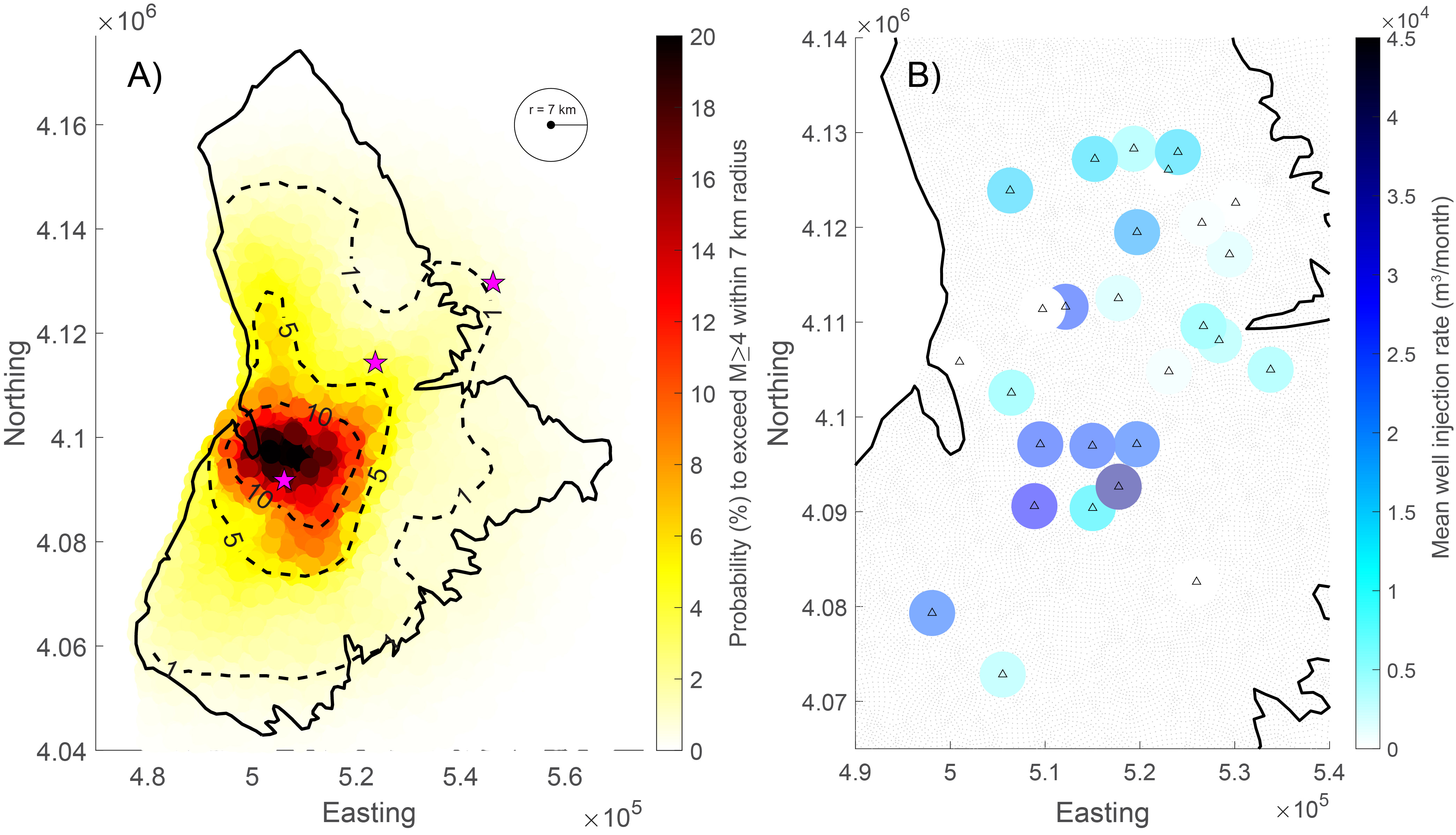}%
    \caption{\textbf{BAU Hazard and Mean Injection Rate}. A) Magnitude exceedance hazard map for M$\geq$4+ for the 5 year management window. Each location is taken as the sum in a 7 km radius. Magenta stars (3) represent the locations of actually observed M$\geq$4+ earthquakes between June-2022 and Sept-2023. B) The Mean well injection rate (m$^3$/month) for all 29 wells (triangles) in the BAU extrapolation. Grey dots represent model nodes.}
    \label{fig:bauhazard}
\end{figure}

In this section we explore the seismicity forecasted by our calibrated model from 2022 through 2027 under a 'business as usual' (BAU) injection scenario. The BAU scenario uses the last observed monthly injection rate for each well from May 2022 and holds them constant until May 2027 (Figure \ref{fig:comparImprove}). This scenario serves as the baseline comparison for the optimization scenarios presented in Section 3. We list the following important results of the BAU forecast:
\begin{itemize}
    \item The BAU forecast from 2022-2027 shows that the probability to exceed a M$\geq$5+ event is $\sim$15\% and a M$\geq$4+ event is $\sim$75\% (Figure \ref{fig:magExceed2}).
    \item Spatially, higher rate injection wells are clustered in the central portion of the basin near the Vermejo Park cluster. Injection wells in this area, just south of the CO-NM border, on average inject at rates higher than 20,000 m$^3$ per month (Figure \ref{fig:bauhazard} (B)).
    \item Seismic hazard is also mostly elevated in this same region for the BAU forecast (Figure \ref{fig:bauhazard} (A)). Within this region of clustered injection, the spatial probability to exceed a M$\geq$4+ is $\sim$20\% over the 5-year BAU forecast.
    \item Seismic hazard in the North of the basin is proportionally smaller. We interpret this as a result of lower injection rates, largely below $\sim$10,000 m$^3$ per day, and lower SI in this region.  
    \item  The two observed M4+ events that have occurred from May 2022 to September 2023 occur within the zone of elevated seismic hazard forecasted by our model (Figure \ref{fig:bauhazard}). 
    \item  In comparison to a complete shut-in of injection in May 2022, BAU injection increases the likelihood of an M$\geq$4+ event by 150\% (from 30\% to 75\%) and a M$\geq$5+ by more than 200\% (from 5\% to 15\%) (Figure \ref{fig:magExceed2}).
\end{itemize}

\begin{figure}[h]
    \centering
    \includegraphics[scale=.45]{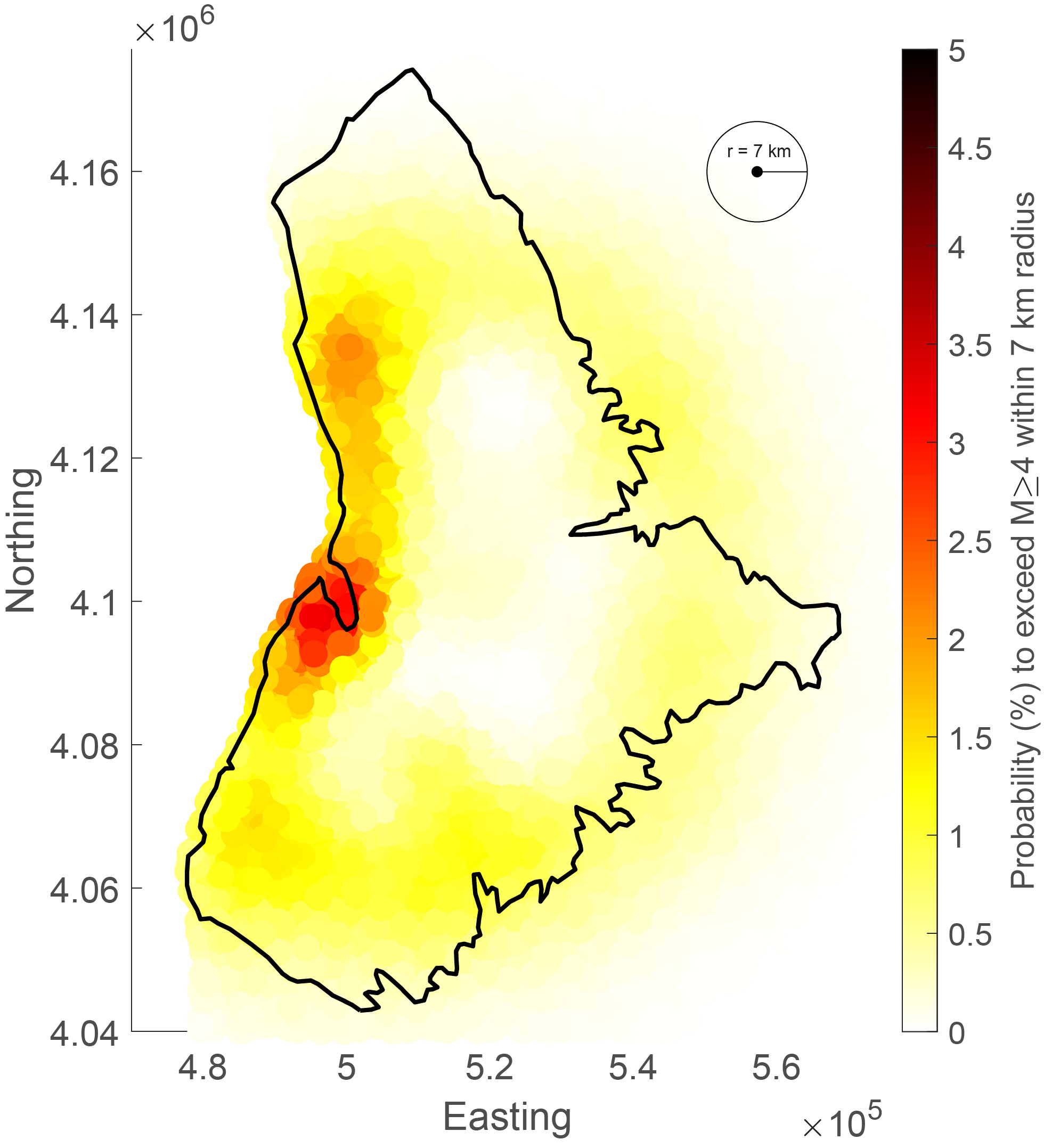}%
    \caption{\textbf{Shutin Hazard}. The 5 year hazard for the shut-in scenario (all wells cease injection in May 2022 and stay off for 5 years) is also characterized spatially for a probability of exceeding a M$\geq$4+. Shut-in represents the post-diffusion pore pressure and stress effects from the full injection history that continue to linger through the model and contribute to perturbations. Note that the colorbar axis is lower (5\%) compared to all other maps which use 20\% to clearly show the spatial distribution of the hazard.}
    \label{fig:shutinhazard}
\end{figure}

\begin{figure}[h]
    \centering
    \includegraphics[scale=.60]{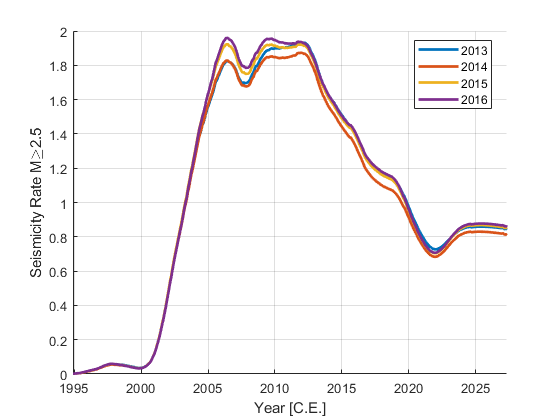}
    \caption{Seismicity rate for four different calibration periods including the BAU forecast after May 2022. This is the forecast based on our SI model shown in Figure \ref{fig:simap7}.}
    \label{fig:sr7}
\end{figure}

\begin{figure}[h]
    \centering
    \includegraphics[scale=.60]{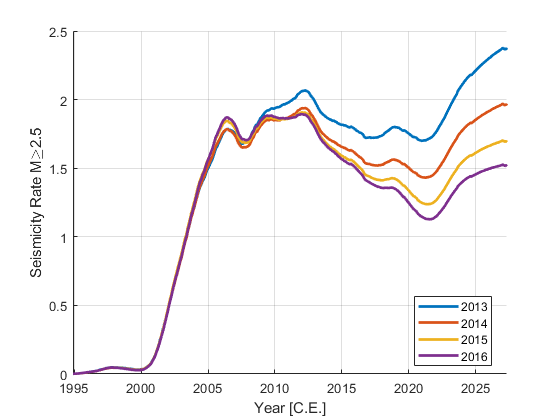}
    \caption{Seismicity rate for four different calibration periods including the BAU forecast after May 2022. This is the forecast based on our SI model shown in SM Figure \ref{fig:simapr5}. Notice that the seismicity rate increases much more than the prior model in SM Figure \ref{fig:sr7}. The reason is that the large outliers of SI now experience elevated rates of Coulomb stress rate which contribute to the overall seismicity rate considerably more.}
    \label{fig:sr5}
\end{figure}

SM Figures \ref{fig:sr7}-\ref{fig:sr5} show the seismicity rate forecasts resulting from the BAU projected injection rates. The forecasted seismicity rates are used to produce magnitude exceedance probabilities from our calibrated SI model (Figure \ref{fig:magExceed2}). Figure \ref{fig:magExceed2} also includes the lower bound on any optimization we can achieve, the shut-in scenario, which represents the post-diffusion pore pressure and stress effects from the full injection history (ie. blue line in Figure \ref{fig:comparImprove}). The 5 year hazard for the shut-in scenario is also characterized spatially for a probability of exceeding a M$\geq$4+ (Figure \ref{fig:shutinhazard}). Given enough prior seismicity to produce a SI map and a physical model to produce Coulomb stress rate any future injection scenarios can be considered in our model. We elaborate on three management models in the following sections.

\section{Physics-Based Forecasting with Optimization}\label{sec:managementmodel}
\subsection{Methods}
\begin{figure}[h]
    \centering
    \includegraphics[scale=.65]{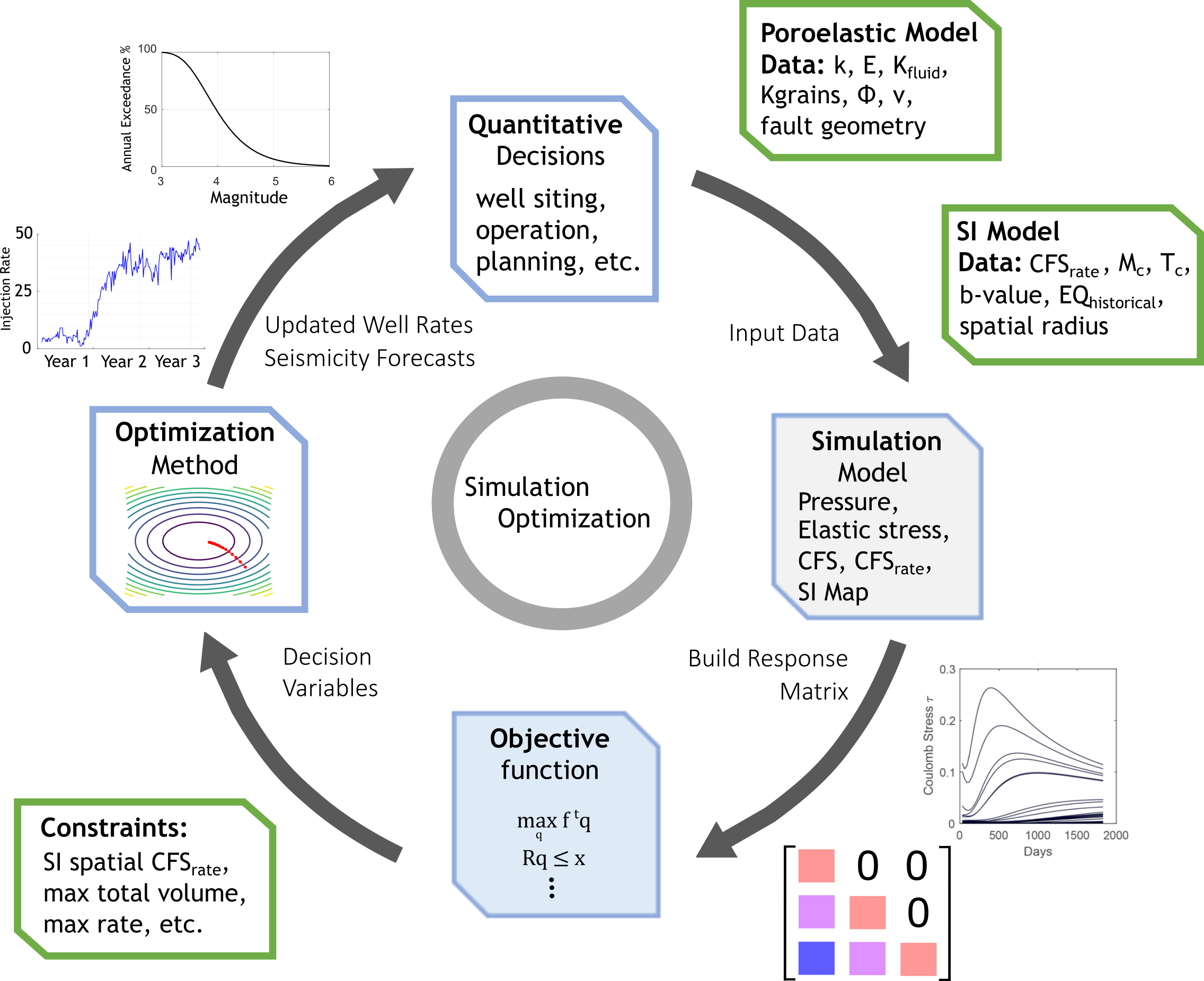}%
    \caption{\textbf{Simulation Optimization Schematic}. Beginning at the top, operations consider quantitative decisions in well placing and operation prior to injection. By developing a numerical model and SI map from current injection a simulation model is built. The simulation model is used to build a response matrix which through linear programming solves a desired objective function (maximize the fluid injected). Additional constraints further inform the optimization which arrives at informed injection rates and spatial hazard maps to then advise future operation practices.}
    \label{fig:schematic}
\end{figure}

The previous sections describe the methods to construct the simulation model built from two data sets: (1) the physics-based poroelastic model and (2) the statistical seismicity model or SI map (Figure \ref{fig:schematic}). In this section we describe the additional methods required to frame our problem as a management model that allows for varied optimizations. In our optimization model, the objective function allows for the maximization of a desired objective, i.e. total injection rate, using decision variables (monthly injection rates) subject to constraints, such as CFS rate at a particular location. In order to solve this optimization problem, we must build a response matrix of the system and use mixed-integer and linear programming to resolve our objective. An overview of the simulation-optimization procedure, including the construction of the simulation model, is provided in Figure \ref{fig:schematic}.

\subsubsection{Objective Function}
In our study of the Raton Basin, the objective function is framed to maximize a desired objective over the 5-year management period. This objective function is maximized subjected to specific constraints, i.e. Coulomb stress or Coulomb stress rate $\dot{\tau}$, below a threshold at chosen locations. Linear programming employs the unit-source solutions of the response matrix by linear superposition to acquire the optimal injection rates at each of the 29 wells in our model. The general framework of the linear program is represented as:
\begin{eqnarray}
    \min_{q} f^T q
\end{eqnarray}
subject to
\begin{eqnarray}\label{eq:linprog}
R q \leq x \\
0 \leq q \leq ub
\end{eqnarray}
where $q$ is the injection rate at each of the wells for each time step (i.e. monthly), $f^T$ is a row vector of negative ones [-1,$\cdots$,-1] so that the objective function seeks to maximize the cumulative injection, $R$ is the response matrix (see section \ref{resp}), $x$ is the constraint vector ($\dot{\tau}$) at each of the model output locations, and $ub$ is the upper bound on the monthly injection rate for each well. For all optimization scenarios presented, the upper bound for a single well injection rate is 1500 m$^3$/day, which represents the threshold of high-rate well injection nationwide \cite{weingarten2015high}. We solve the linear program using the linprog() function in MATLAB which generates optimal values of $q$, i.e. the injection rates, for each well that do not exceed the constraints at the model output points. This objective function subject to various constraints is flexible and adaptable to a wide variety of adjustments within linear programming optimization. In section \ref{sec:mixed}, we elaborate on different ways to alter the management model constraints and provide a selection of controls that may be of interest to real-world injection practices.
\subsubsection{Response Matrix}\label{resp}

Given any linear system used to describe a given simulation model, a management model can be built with a response matrix. Construction of the response matrix requires individual unit-source solutions for each well operating within the management model. A unit-source solution is generated by producing an impulse from an individual well (i.e. unit flow rate) and measuring its response at all model output locations for the duration of the management period. The impulse has a fixed value for a specified period and a value of zero thereafter. The response of the system are changes in pore pressure and stress. Due to the linearity of the Coulomb stress equation (Eq. \ref{eq:cfs}), Coulomb stress and Coulomb stress rate are derived from this response (see Appendix for rate response matrix construction). 

In our model, the Raton Basin contains 29 wells. Therefore, we must generate 29 independent, unit-source impulses (one for each well) and record the unit response at all model output locations. We must record each response for the entire 5-year management period (ie. June-2022 to June-2027). Each time step in the model is 30 days. Hence, the unit-source response is a single flow rate equivalent to 100 m$^3$/day for the first time step and then zero for the 60 months after. The result of this procedure is the unit-source response matrix of CFS rate produced by each well at every model output location (SM Fgure \ref{fig:response}). An example of this procedure is provided in the supplement (SM Methods \ref{sec:simple}; SM Figure \ref{fig:respGen}). 

\subsubsection{Considering Injection Prior to Management Time Period}
Our optimization management model optimizes injection rates under a set of given constraints for a prescribed management time period. It does not, inherently, consider injection prior to the management time period. We solve this issue by taking the difference of Coulomb stress between two simulations: (1) an ABAQUS simulation which considers all injection from Nov 1994 - 2027 (BAU rates) and (2) a response matrix simulation which considers only injection from 2022 - 2027 (BAU rates). The resulting Coulomb stressing rates represent the contribution of all prior injection during the management time period. This could be considered a `complete shut-in' scenario from 2022 - 2027.

We calculated seismicity rates and a probability of exceedance curve expected from this shut-in scenario (Figure \ref{fig:magExceed2}). SM Figure \ref{fig:shutinhazard} depicts the spatial distribution of hazard for yearly time steps. If wells were to have suddenly shut-off in May 2022 our model predicts that there would still be a $\sim$35\% probability of exceeding a M$>$4+ earthquake in the next 5 years. The shut-in Coulomb stress rate perturbations are added to the Coulomb stress rate constraints of the optimization results prior to the seismicity rate and seismic hazard calculations, thus serving as the initial conditions or starting point in the optimizations. This step is essential, otherwise the seismic hazard is underestimated by the optimizations alone.

\subsubsection{Mixed Integer Programming}\label{sec:mixed}
Mixed-integer programming (MIP) allows the optimization manager to impose constraints that simulate real-world injection practices \cite{gorelick1982optimal,hsu1989optimum}. Without MIP, the optimization solution is free to produce large swings in injection rate at individual wells. In reality, large injection wells have tolerances for injection rate changes over time. MIP allows the optimization manager to place controls what wells are operating and how the wells operate (independent or dependent on one another) through time. Injection rates can be constrained within a running average of past injection at a particular well, or monotonically increase or decrease injection through time, or exclude certain wells during certain periods.

The process of applying different types of MIP constraints is similar for most scenarios. First, a mixed-integer matrix is constructed $R^*$ such that $R^*q\leq x^*$, where $q$ is the corresponding injection well location for each management period and $x^*$ is a vector of additional constraints. Both $R^*$ and $x^*$ are concatenated with original response matrix equation, Eq. (\ref{eq:linprog}), and the objective function is maximized subject to these combined constraints ($R$ and $R^*$). A simplified example is provided in SM Section \ref{sec:simple}, and further description of applying each type of MIP constraint in the management model is provided in SM Section \ref{sec:mixedSM}.

\subsubsection{Setting a Desired Seismic Hazard}

\begin{figure}[h]
    \centering
    \includegraphics[scale=.60]{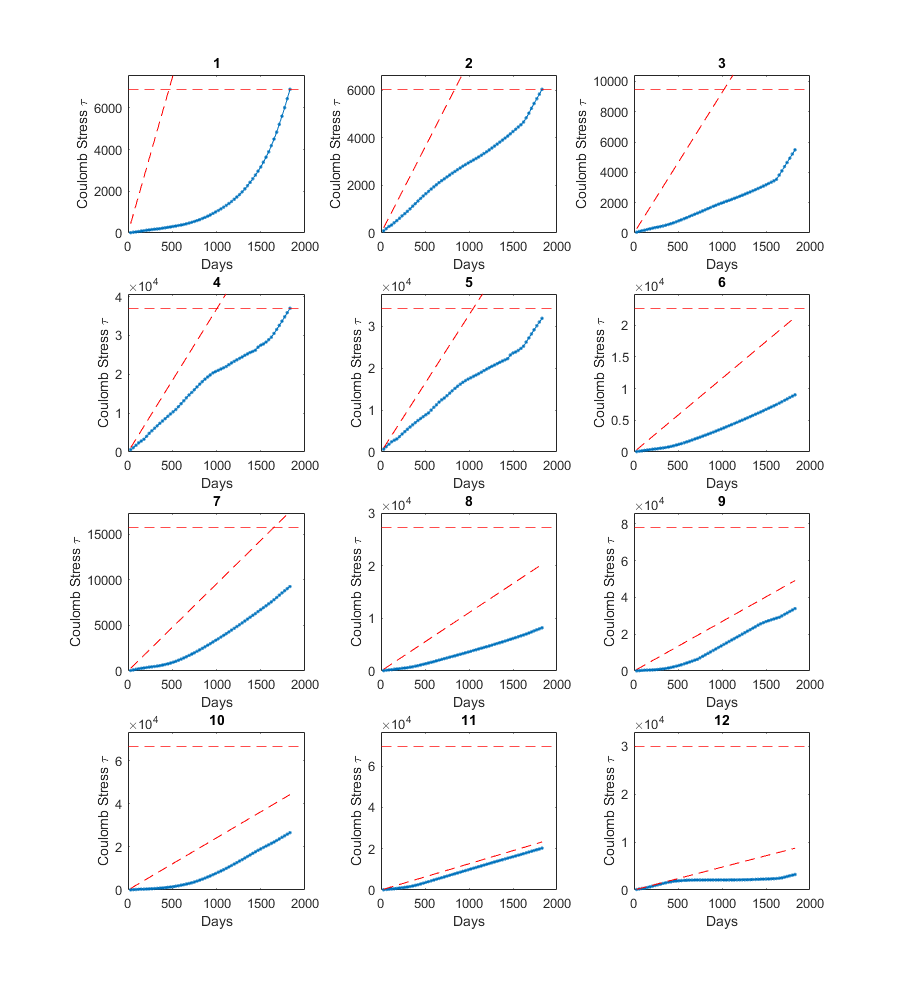}
    \caption{Simplified management model example results that does not require an SI map. The (blue line) is the optimized Coulomb stress results for the 12 model output locations compared to the maximum Coulomb stress allowed (horizontal red dash line) and compared to the maximum Coulomb stress Rate (angled red dash line) allowed at each of the model output locations.}
    \label{fig:sm12}
\end{figure}

The optimization problem described above is setup to constrain only CFS rate at specified locations through time. However, the optimization manager may still use our methodology to achieve a desired seismic hazard. This is performed by combining the calculated CFS rates with the SI model to produce seismicity rate forecasts. Optimization is still possible without coupling to a SI map if desired (See Supplementary Methods \ref{sec:simple}; SM Figure \ref{fig:12p1234}-\ref{fig:sm12}). 

For a desired magnitude exceedance probability $Pr(M)$ (Eq. \ref{eq:prob}), a user can solve for the total number of earthquakes expected during the management period ($N_{\geq M}$). This $N_{\geq M}$, in combination with spatially varying SI map $\Sigma_\tau(r)$, can be used to calculate desired Coulomb stress rate constraints $x_{\dot{\tau}}$ for the management model:
\begin{align}\label{eq:xtau}
        x_{\dot{\tau}} = \frac{\partial}{\partial t} \tau_{\tau}(\boldsymbol{r},t) = \sqrt{\frac{N_{\geq M}}{P\cdot T}10^{-\Sigma_{\tau}(\boldsymbol{r})+bM} }
\end{align}
where $P$ now refers to the total number of constraint points in the SI model and $T$ refers to the to total time chosen for the management period. This initialization assumes that each point in the model will carry a scaled portion of the total earthquake probabilistic hazard- ie. $\frac{N_{\geq M}}{P\cdot T}10^{bM}$ which is scaled by the SI (ie. $10^{-\Sigma_{\tau}(r)}$). In our case, the total number of model points exceeds the computational limitation of the linear program and a subset of the total model points must be chosen. For example, the output of our model contains $>$30,000 points across the basin, but we reduce this total to 500 constraint locations for the management model. The chosen points are based on a uniform random distribution of points within a circle that contains all of the seismicity (SM Figure \ref{fig:urandpoints}).

In practice, we have found that the CFS rate constraints provided by equation \ref{eq:xtau} always produce a basin-wide $Pr(M)$ lower than the desired threshold $Pr(M)$. The desired threshold $Pr(M)$ would only be met if the CFS rate constraint threshold is met at all points $P$ for all time $T$. To resolve this issue, we iteratively solve the optimization model while increasing the CFS rate constraints at locations within the model that reached that threshold at any time during the management period. In this way, the constraints slowly increase based on which locations require a higher CFS rate in order to produce the desired $Pr(M)$ in the basin. For our study, we set a goal of achieving the desired $Pr(M)$ in the basin to within $\pm$0.2\% (See Methods \ref{sec:iteration}).

The following steps describe the methodology, generalized for application to other studies:
\begin{enumerate}
    \item Choose a desired exceedance probability for an arbitrary magnitude threshold and solve for $N_{\geq M}$ (Eq. \ref{eq:prob}).
    \item Calculate CFS rate constraints for the management model (Eq. \ref{eq:xtau}).
    \item Find optimal injection rates for calculated CFS rate constraints.
    \item Calculate exceedance probabilities $Pr(M)$ across the basin for the optimized solution.
    \item Check if exceedance probabilities $Pr(M)$ are within $\pm$0.2\% of desired $Pr(M)$.
    \item If yes, skip steps 7 and 8.
    \item If no, adjust CFS rate constraints dependent on too high or too low of threshold.
    \item Return to step 3.
\end{enumerate}

\subsection{Prospective Case `Reduction' - Reduce the Seismic Hazard}
The first prospective case we consider is called `Reduction' (Figure \ref{fig:comparImprove} - Prospective Case Reduction). Prospective case `Reduction' is the management solution for a hypothetical well operation that seeks to reduce the overall injection and maintain the hazard within a chosen threshold. We include a constraint that the overall injection must be reduced by at least 80\% from May 2022 levels by the end of the 5 year management window. Additionally, we constrained seismic hazard such that the probability of exceeding a M$\geq4+$ event is 40\% lower than the BAU forecast (Figure \ref{fig:magExceed2}). The optimization and iterative method arrive at a solution to these constraints while maximizing the amount of fluid injected.

In order to achieve a smooth tapering of injection from the BAU initial injection rate of $\sim$10,000 m$^3$ per day we incorporate a MIP constraint to the management model. The constraint is a monotonic decrease of at least 2\% each month for all injection wells (see \ref{sec:mixedSM}) (Figure \ref{fig:comparImprove} - yellow line). This constraint smoothly reduces the overall injection rate and therefore the Coulomb stress rate by the end of the five year management period.

\begin{figure}[h]
    \centering
    \includegraphics[scale=.40]{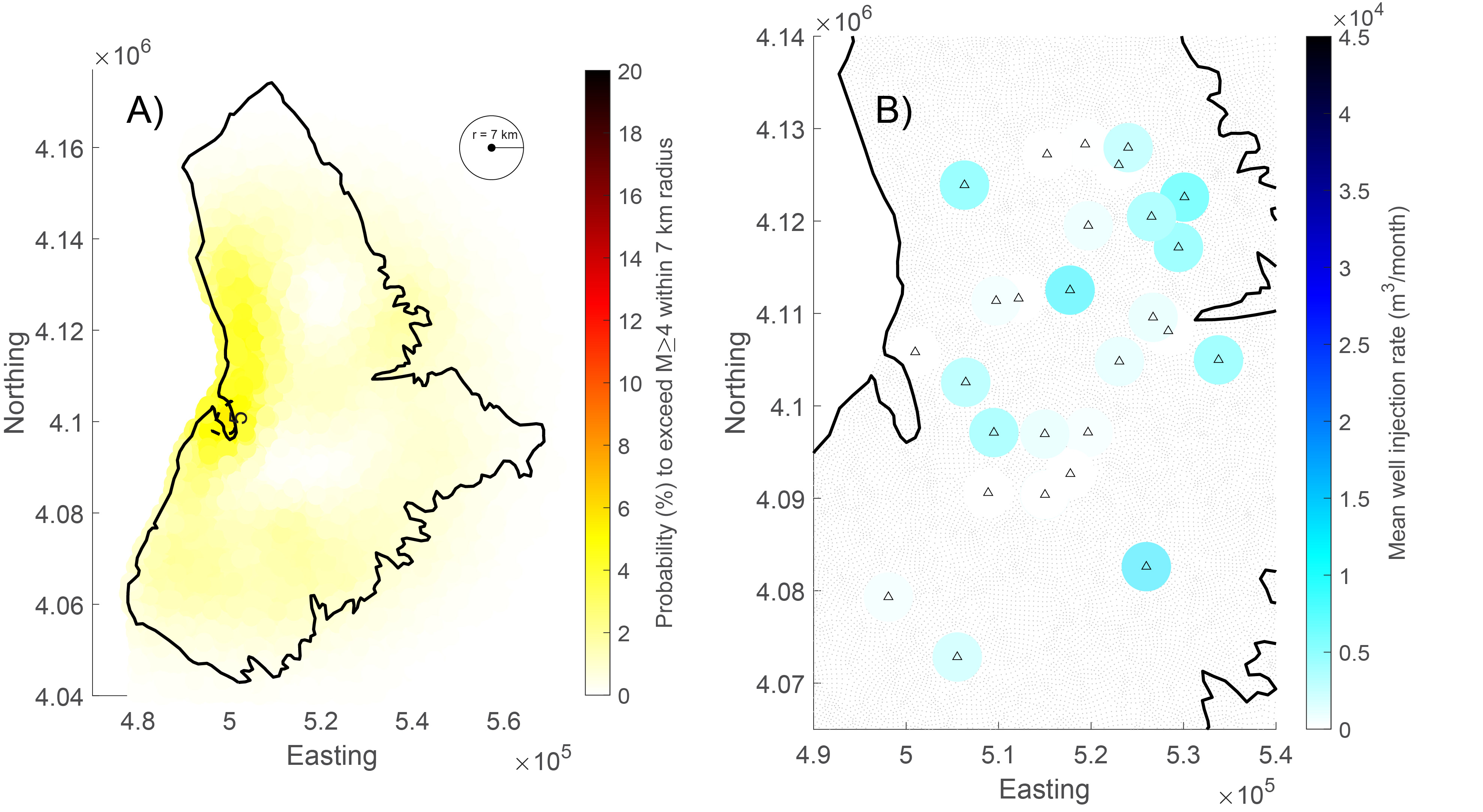}
    \caption{\textbf{Prospective Case `Reduction' Results.} A) Magnitude exceedance hazard map for M$\geq$4+ for the 5 year management window. Each location is taken as the sum in a 7 km radius. See SM Figure \ref{fig:5yearCase1} for yearly plots. B) Mean injection rate (m$^3$/month) at each well location (triangles). There are several locations where the optimization chooses not to inject. The grey dots represent the model nodes.}
    \label{fig:mm2}
\end{figure}

We find that there are several wells in the optimization that are never injecting, and that the algorithm preferentially chooses injectors towards the northeast more than other locations (Figure \ref{fig:mm2}b). The northeast portion of the basin is a relatively low SI area (Figure \ref{fig:SI}). The west-central portion of the basin, which contains the highest SI hazard, does not have large amounts of injection during the management period. The optimization preferentially chooses to spread out large injectors from one another and to regions of lower SI (Figure \ref{fig:mm2}b).

Another important observation is that prior injection still drives significant hazard due to the time delay of pressure diffusion continuing to elevate the Coulomb stress rate in the periphery of the basin (Figure \ref{fig:mm2}a). Hazard is elevated in the west-central and western portion of the basin by prior injection, despite the optimization lowering injection in these areas. Our iterative technique still slowly reduces injection at wells and areas associated with high prior hazard if hazard thresholds are not initially met. In this way, our method takes into account prior injection through iterative forward solutions without direct inclusion in the optimization constraint vector (see Section \ref{sec:iteration}). 

The enhanced hazard to the west in all of our models does not consider previously mapped faults unless they were captured by the SI map. This hazard is primarily driven by continued Coulomb stress rate increase from prior injection. The inclusion of known faults is currently a limitation to our method. However, additional spatial constraints from known faults could be implemented as additional rows/elements in the response matrix/constraint vector prior to optimization. Constraint thresholds of Coulomb stress or Coulomb stress rate could be applied to these known faults.

Visualizing the optimization at each time step is informative to the evolution of hazard and how each individual well injects over time (SM Video \ref{fig:5yearCase1}). For the prospective case `Reduction', wells inject continuously in the northeast - a low SI area - for the entire management period. Higher SI areas still receive injection but the optimization tends to spread the overall hazard across the basin.

\subsection{Prospective Case `Safety'}\label{sec:safetyAndEconomic}
Our second prospective case consider how the optimization algorithm might disperse BAU injection rates in order to minimize seismic hazard (i.e. 'Safety') (see Section \ref{sec:bau} and Figure \ref{fig:comparImprove}).

The second optimization solution, which we call prospective case `Safety', seeks an optimized solution that lowers the overall seismic hazard while the basin-wide injection rate is constrained at May 2022 levels for the 5 year management period. The optimization will preferentially increase volume in wells where SI is lower, because the Coulomb stress rate constraints will be relaxed in these areas (see Equation 10). By moving injection volume to wells and areas with lower SI, the forecasted seismic hazard is reduced. The solution therefore produces an overall annual exceedance curve that is lower for the same total injection volume (Figure \ref{fig:comparImprove} - pink line). 

\begin{figure}[h]
    \centering
    \includegraphics[scale=.40]{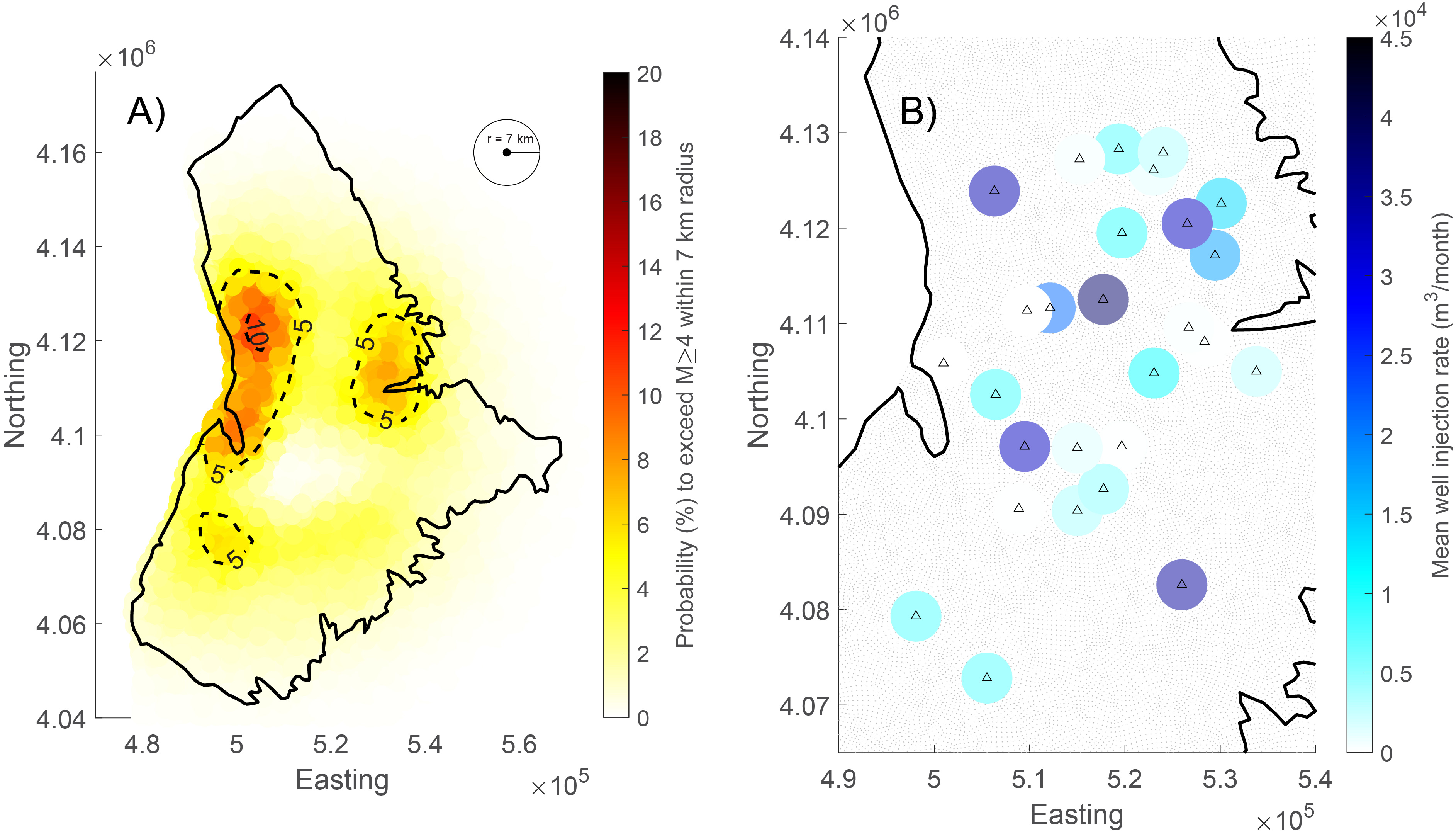}
    \caption{\textbf{Prospective Case `Safety' Results.} A) Total probability of exceeding a M$\geq$4+ earthquake across the entire basin during the total 5 year management window. Hazard is spread more evenly throughout the model and in less than the BAU case in areas that contribute to high hazard. See SM Figure \ref{fig:5yearCase2a} for yearly plots. B) Mean injection rate in m$^3$/month at each well location (triangles). There are several locations where the optimization chooses not to inject. The grey dots represent the model nodes.}
    \label{fig:prosp2a}
\end{figure}

Figure \ref{fig:prosp2a} describes the optimization results across the basin for prospective case `Safety'. When the spatial distribution of injection is compared to the Business As Usual case, we find that the optimization spread injection volume out more evenly throughout the basin, instead of clustering injection in the central region. At the same, seismic hazard increases on the peripheries of basin away from the higher SI zones in the central basin. In the central basin, forecasted haard is reduced greatly, with less than 2\% probability to exceed an M$\geq$4+ within 7 km. This is compared to nearly 20\% probability to exceed an M$\geq$4+ within 7 km in the Business As Usual case in the central basin. Forecasted hazard is highest in the northeast portion of the basin, with 10\% probability to exceed an M$\geq$4+ within 7 km. 

Our solution, during the 5 year management window, reduces the basin-wide annual exceedance probability M$\geq$4+ from 75\% to 71\%. This optimized result is a relatively small reduction in the annual exceedance probabilities. However, we found that injection prior to the management period contributes to a large portion of the overall hazard observed during the 5 year window. If the prospective case 'Safety' is run without prior injection, the optimization can reduce the annual exceedance probability M$\geq$4+ from 75\% to 58\% (Figure \ref{fig:magExceed2} - green line). This reduction in seismic hazard is due to the optimization shifting injection to areas of lower SI.

Simply excluding prior injection does not, in and of itself, reduce the overall exceedance probabilities. We ran a seismic hazard forecast for the Business As Usual case excluding prior injection and found the annual exceedance probability for a M$\geq$4+ earthquake increased from 75\% to 80\% (Figure \ref{fig:magExceed2} - BAU without prior injection line). The reason for this increase in overall seismic hazard when excluding prior injection is that prior injection was on a long-term decline, especially in areas with high SI. These declining injection rates prior to the management time period actually reduce the Coulomb stress rate in areas where the BAU injection is high. Therefore, counter intuitively, excluding prior injection increases the seismic hazard in the BAU case and decreases in the 'Safety' case.

The results from the 'Safety' case reveal that prior injection can have a large influence on how much the optimization method reduces overall seismic hazard. Furthermore, it highlights the importance of optimizing injection as early as possible in the course of an induced seismic sequence. In the case of Raton Basin, injection and induced seismicity have been ongoing for multiple decades, which reduce the positive safety effects of minimizing seismic hazard during the management period.

\subsection{Prospective Case `Economic'}\label{Economic}

The third optimization solution, which we call prospective case `Economic', seeks to increase the overall injection rate but maintain the same basin-wide seismic hazard as the BAU case (see Section \ref{sec:bau} and Figure \ref{fig:comparImprove}). In this case, we allow the optimization freedom to increase the overall volume that can be injected in any month of the 5 year management window. An optimal solution is found when the basin-wide annual exceedance probabilities are within $\leq 2$\% of the BAU probability of exceedance for M$\geq$4+ ($\sim$75\%). We include two constraints on individual wells in this solution: (1) no individual well injection rate can exceed 1,500 $m^3/day$, and (2) an MIP constraint that limits individual well injection rates to within a 6-month running average so that the optimization cannot drastically front-load or back-load the management period with injection volume. Again, the Coulomb stress rate constraints derived from the SI map force the optimization to preferentially increase volume in areas away from the largest seismic hazard (i.e. lower SI). 

An optimal solution was found for the 'Economic' case, which increased the overall injection rate basin-wide compared to the BAU case (Figure \ref{fig:comparImprove} - green line). The solution shows a gradual increase in basin-wide injection rate from $\sim$300,000 m$^3$/month in 2022 to $\sim$375,000 m$^3$/month in 2027. The increase in cumulative volume injected in the 'Economic' case is more than 1,080,000 $m^3$ ($\sim$6,750,000 barrels) when compared to the BAU case.

\begin{figure}[h]
    \centering
    \includegraphics[scale=.40]{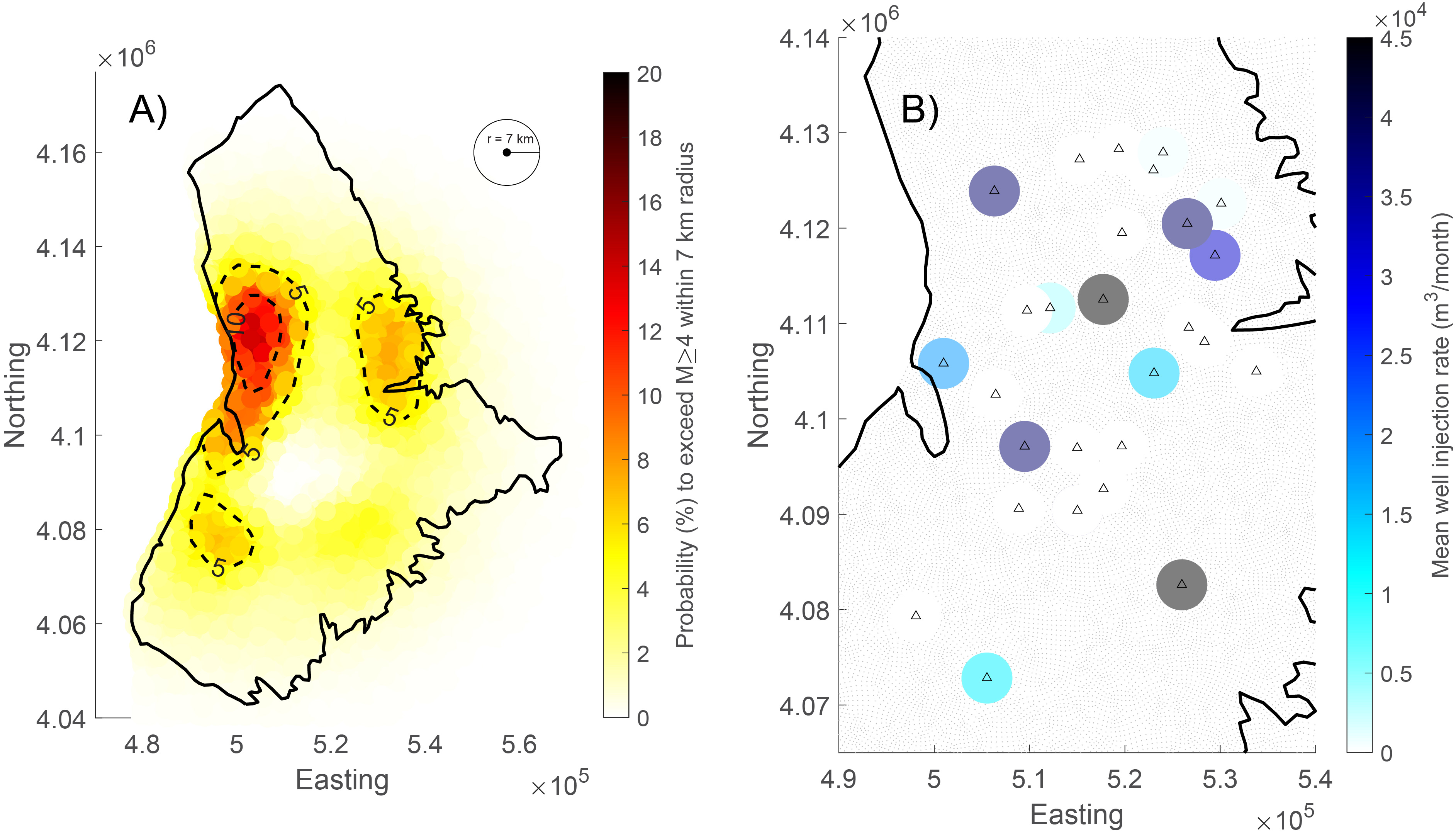}
    \caption{\textbf{Prospective Case `Economic' Results.} A) Total probability of exceeding a M$\geq$4+ earthquake across the entire basin during the total 5 year management window. The highest probability western part of the basin is associated with the large fluid injection. See SM Figure \ref{fig:5yearCase2b} for yearly plots. B) Mean injection rate in m$^3$/month at each well location (triangles). There are several locations where the optimization chooses not to inject. The grey dots represent the model nodes.}
    \label{fig:prosp2b}
\end{figure}

The spatial distribution of injection in the 'Economic' case shows a substantial change in the how the field would be operated during the 5 year management period (Figure \ref{fig:prosp2b}b). Of the 29 potential injection wells, the optimization chooses to inject at only 12 wells, while the remaining 17 are completely shut-in. Of the 12 wells which operate during the 5 year window, only 6 inject at rates higher than 20,000 m$^3$/month. These 6 injectors, where the vast majority of fluid is injected, are spread out across the entirety of the well field and to regions of lower SI. These 6 wells inject at a more or less a constant rate for the entire management time (SM Video \ref{fig:5yearCase2b}). Clustering of injection is held to a minimum when compared to the 'Reduction' or 'Safety' case. 

This case highlights what the optimization method ultimately attempts achieve: spatially distributed injection across regions of lower SI. By spreading out injectors, the basin-wide Coulomb stress rate is reduced by minimizing superposition of clustered injectors. By concentrating injection in regions of lower SI, the Coulomb stress rate that is created by injection results in lower induced seismicity. This combination of effects -- spatially distributed injection in regions of lower SI -- allows for the highest basin-wide injection rates (and largest cumulative injected volume) for a given seismic hazard. 

\section{Discussion}
The combination of physics-based forecasting with optimization management shows promise for future work in mitigating induced seismic hazard at the basin-scale. The optimization framework allows a user to maximize a particular objective (i.e. reduction, safety or economic) while maintaining a specified induced seismic hazard. Our method is also flexible and adaptable to other regions or other types of fluid injection that induce seismicity. The main components are the following:
\begin{enumerate}
    \item \textbf{Physics-based model of pressure and/or stress change.} First, a physics-based model of injection must be built of the region that has good estimates of the relevant reservoir flow parameters. Here, we have built a fully coupled, poroelastic numerical model using the finite-element method calibrated using injection data from reservoir step-rate tests. However, a finite-difference model could also work (e.g. MODFLOW). Any linear system is the key.  Depending on whether the poroelastic stress effects are marginal to the pore pressure effects may influence this decision.
    \item \textbf{Seismogenic Index (SI) Map.} Second, a SI map (see Section \ref{sec:calib}) must be calibrated from the empirical relationship of seismic response to injection. Thus, some degree of prior injection and earthquake history are required for forecasting. Without the SI map, optimization is still possible, but will not be constrained by desired seismic hazard.
    \item \textbf{Response Matrix.} Third, a response matrix of system is built from impulse-responses of the system to a unit injection at each prospective injection site (see Section \ref{resp}). The response matrix allows the optimization to scale injection rates of individual wells to find the combination which both satisfies the constraints and maximizes the objective function.
    \item \textbf{Optimization Framework.} Lastly, an optimization framework of an objective function, constraints and decision variables are input. The model then seeks the optimized solution that will satisfy either a reduction, safety or economic objective and maximize fluid injected.
\end{enumerate}

The adaptability of this method to other regions is possible through the gathering of required basin-specific input data on reservoir flow parameters, injection and seismicity response. In addition, the method is flexible enough to consider any fluid injection that produces a linear poroelastic response. Listed below are some of the potential improvements and limitations of the current framework:
\begin{enumerate}

\item \textbf{Real-time optimization and forecasting:} Once the physics-based model and SI map are initially calibrated the user could develop an optimal injection strategy and continuously update the SI map if seismicity evolves in new areas. The response matrix method allows for quick integration of new constraints without the need to re-run elaborate physical models continuously. Therefore, rapid adjustments in well optimization are possible as the SI adjusts and improves in new areas of the basin.

\item \textbf{Stacked optimization for model uncertainty:} As described in Section \ref{sec:mixedSM}, stacked optimization allows the user to find one set of optimal injection rates that explicitly account for the uncertainty in the physical model. The existing framework contains uncertainty in the seismic hazard due to the Poisson distribution within the SI model. However, stacked optimization allows the user to consider uncertainty within the physical model (i.e. a distribution of flow parameters). Stacked optimization does require more computational power as it requires $N$ (where $N$ is the number of wells) additional model runs for each uncertain distribution to be appended to the response matrix.

\item \textbf{Non-linear programming:} Non-linear programming allows optimization of non-linear objective functions and constraints. Currently, our linear program cannot explicitly optimize injection using seismic hazard ($R$) as a constraint because $R$ is non-linearly related to CFS rate. Therefore, we rely on an iterative approach to optimize injection to a desired seismic hazard (see Section \ref{sec:iteration}). Non-linear programming may be able to address the issue of local-minima in the optimal solution where currently non-unique solutions may be found by a linear program. Our iterative method slowly adjusts the constraint locations one at a time to prevent any over saturation in hazard and injected fluid at any one location in the solution. Non-linear programming may be able to save computational time as compared to the iterative approach.

\item \textbf{Incorporating known fault maps:} A key piece of future work is the integration of known fault maps within the optimization framework. Known faults would serve as additional constraint locations appended to the response matrix and constraint vector, where pressure and/or stress change would be limited. From a practical point of view, known faults in many cases of induced seismicity are not the primary drivers of induced seismic hazard (i.e. Oklahoma), but users may desire to avoid stressing faults when optimizing basin-scale injection. This optimization framework would allow the consideration of both an SI map and fault maps.

\item \textbf{Incorporating risk for policy:} While we looked at the total hazard in the region, it would be possible to constrain hazard spatially depending on seismic risk \cite{schultz2021risk}. For example, agreement might be met with industrial well operations that maximizes the fluid injected while restricting hazard in an area with high risk, like a densely populated area. A scientifically informed policy, for example one that limits the probability of exceeding a M$\geq$5+ earthquake within a high risk zone, could be met while still reaching the economic objective of the well operators.

\end{enumerate}
\section{Conclusions}
    Here, we investigated the relationship between wastewater injection and seismicity in the Raton Basin of Colorado and New Mexico using a physics-based forecasting framework. First, a 3D finite element model of a poroelastic crust is used to estimate time dependent Coulomb stress changes over the more than two decades of Raton Basin injection. The outputs of Coulomb stress rate from our finite element model were combined with a seismogenic index (SI) model to forecast induced seismicity in space and time throughout the basin. Using this hybrid physics-statistical forecasting model we found the following conclusions:
\begin{enumerate}
    \item The recent and ongoing induced seismicity within the Raton Basin is well explained by our physics-based forecasting model. Declining seismicity rates between 2016 - 2022 are forecasted well by the decline in basin-wide injection rate. Despite injection rate declines, modeled Coulomb stress rate is still increasing in several regions of the basin, suggesting that induced seismic hazard is still ongoing. Our model also shows that induced seismicity is driven primarily by the pore pressure component of the poroelastic stresses, with poroelastic stress changes accounting for about 5\% of the driving force.
    \item Using our physics-based forecasting model, we estimated the induced seismic hazard produced by continued Raton Basin injection at May 2022 levels through 2027 (Business As Usual case). Our 5 year forecast estimates the probability to exceed a M$\geq$4+ event is 75\% and M$\geq$5+ event 14\%.
    \item Linear-programming optimization using the response matrix method is implemented successfully using a safety objective framework that reduces seismic hazard for given amount of fluid injection (safety objective) or (b) maximizes fluid injection for a prescribed seismic hazard (economic objective).
    \item Across the different objectives tested, the optimization algorithm tends to spread injection out across the field when compared to the Business As Usual case. In the safety and economic objective cases, we observed the algorithm spreading out higher rate injection wells from one another and to regions lower seismogenic index (SI). We also demonstrate that injection prior to the optimization management period may have differing effects on seismic hazard during the management period. In the reduction and safety cases, we show that prior injection enhanced seismic hazard during the management period, thus decreasing the impact of injection optimization. We conclude that optimization of injection earlier in an induced sequence will allow for better control of seismic hazard during the management period.
\end{enumerate}
\acknowledgments
The authors wish to thank Mark Zoback and Steve Gorelick for both support, advice and feedback on simulation optimization methods during M. Weingarten's postdoctoral research. The authors appreciate Margaret Glasgow for helpful discussions and comments on Raton Basin seismicity. We also appreciate discussion with Robert Guyer and Daniel Trugman at early stages of the work. We acknowledge use of the CSRC high-performance computing cluster and other support from San Diego State University. The wastewater injection data is available from the Colorado Oil and Gas Corporation Commission website (\url{https://ecmc.state.co.us/#/home}). The wastewater injection data is available from the New Mexico Oil Conservation Division Permitting website (\url{https://wwwapps.emnrd.nm.gov/OCD/OCDPermitting/Data/Wells.aspx}).

R.G.H. built the models, performed analysis of the model results, made the figures, and wrote the manuscript. M.W. conceived the experiment and optimization approach, managed the study, provided access to the modelling software, and helped write the manuscript. C.L. assisted with seismogenic index analysis and contributed to the manuscript. Y.F. contributed to the manuscript.


\newcommand{\noopsort}[1]{}
\begin{thebibliography}{}

\bibitem [\protect \citeauthoryear {%
Bachmann%
, Wiemer%
, Goertz-Allmann%
\BCBL {}\ \BBA {} Woessner%
}{%
Bachmann%
\ \protect \BOthers {.}}{%
{\protect \APACyear {2012}}%
}]{%
bachmann2012influence}
\APACinsertmetastar {%
bachmann2012influence}%
\begin{APACrefauthors}%
Bachmann, C\BPBI E.%
, Wiemer, S.%
, Goertz-Allmann, B.%
\BCBL {}\ \BBA {} Woessner, J.%
\end{APACrefauthors}%
\unskip\
\newblock
\APACrefYearMonthDay{2012}{}{}.
\newblock
{\BBOQ}\APACrefatitle {Influence of pore-pressure on the event-size distribution of induced earthquakes} {Influence of pore-pressure on the event-size distribution of induced earthquakes}.{\BBCQ}
\newblock
\APACjournalVolNumPages{Geophysical Research Letters}{39}{9}{}.
\PrintBackRefs{\CurrentBib}

\bibitem [\protect \citeauthoryear {%
Bao%
\ \BBA {} Eaton%
}{%
Bao%
\ \BBA {} Eaton%
}{%
{\protect \APACyear {2016}}%
}]{%
bao2016fault}
\APACinsertmetastar {%
bao2016fault}%
\begin{APACrefauthors}%
Bao, X.%
\BCBT {}\ \BBA {} Eaton, D\BPBI W.%
\end{APACrefauthors}%
\unskip\
\newblock
\APACrefYearMonthDay{2016}{}{}.
\newblock
{\BBOQ}\APACrefatitle {{Fault activation by hydraulic fracturing in western Canada}} {{Fault activation by hydraulic fracturing in western Canada}}.{\BBCQ}
\newblock
\APACjournalVolNumPages{Science}{354}{6318}{1406--1409}.
\PrintBackRefs{\CurrentBib}

\bibitem [\protect \citeauthoryear {%
Barnhart%
, Benz%
, Hayes%
, Rubinstein%
\BCBL {}\ \BBA {} Bergman%
}{%
Barnhart%
\ \protect \BOthers {.}}{%
{\protect \APACyear {2014}}%
}]{%
barnhart2014seismological}
\APACinsertmetastar {%
barnhart2014seismological}%
\begin{APACrefauthors}%
Barnhart, W\BPBI D.%
, Benz, H\BPBI M.%
, Hayes, G\BPBI P.%
, Rubinstein, J\BPBI L.%
\BCBL {}\ \BBA {} Bergman, E.%
\end{APACrefauthors}%
\unskip\
\newblock
\APACrefYearMonthDay{2014}{}{}.
\newblock
{\BBOQ}\APACrefatitle {{Seismological and geodetic constraints on the 2011 Mw5. 3 Trinidad, Colorado earthquake and induced deformation in the Raton Basin}} {{Seismological and geodetic constraints on the 2011 Mw5. 3 Trinidad, Colorado earthquake and induced deformation in the Raton Basin}}.{\BBCQ}
\newblock
\APACjournalVolNumPages{Journal of Geophysical Research: Solid Earth}{119}{10}{7923--7933}.
\PrintBackRefs{\CurrentBib}

\bibitem [\protect \citeauthoryear {%
Belitz%
\ \BBA {} Bredehoeft%
}{%
Belitz%
\ \BBA {} Bredehoeft%
}{%
{\protect \APACyear {1988}}%
}]{%
belitz1988hydrodynamics}
\APACinsertmetastar {%
belitz1988hydrodynamics}%
\begin{APACrefauthors}%
Belitz, K.%
\BCBT {}\ \BBA {} Bredehoeft, J\BPBI D.%
\end{APACrefauthors}%
\unskip\
\newblock
\APACrefYearMonthDay{1988}{}{}.
\newblock
{\BBOQ}\APACrefatitle {{Hydrodynamics of Denver Basin: Explanation of subnormal fluid pressures}} {{Hydrodynamics of Denver Basin: Explanation of subnormal fluid pressures}}.{\BBCQ}
\newblock
\APACjournalVolNumPages{AAPG bulletin}{72}{11}{1334--1359}.
\PrintBackRefs{\CurrentBib}

\bibitem [\protect \citeauthoryear {%
Biot%
}{%
Biot%
}{%
{\protect \APACyear {1941}}%
}]{%
biot_general_1941}
\APACinsertmetastar {%
biot_general_1941}%
\begin{APACrefauthors}%
Biot, M\BPBI A.%
\end{APACrefauthors}%
\unskip\
\newblock
\APACrefYearMonthDay{1941}{}{}.
\newblock
{\BBOQ}\APACrefatitle {General Theory of Three‐Dimensional Consolidation} {General theory of three‐dimensional consolidation}.{\BBCQ}
\newblock
\APACjournalVolNumPages{J. Appl. Phys.}{12}{2}{155--164}.
\newblock
\begin{APACrefDOI} \doi{10.1063/1.1712886} \end{APACrefDOI}
\PrintBackRefs{\CurrentBib}

\bibitem [\protect \citeauthoryear {%
Borja%
}{%
Borja%
}{%
{\protect \APACyear {2006}}%
}]{%
borja2006mechanical}
\APACinsertmetastar {%
borja2006mechanical}%
\begin{APACrefauthors}%
Borja, R\BPBI I.%
\end{APACrefauthors}%
\unskip\
\newblock
\APACrefYearMonthDay{2006}{}{}.
\newblock
{\BBOQ}\APACrefatitle {On the mechanical energy and effective stress in saturated and unsaturated porous continua} {On the mechanical energy and effective stress in saturated and unsaturated porous continua}.{\BBCQ}
\newblock
\APACjournalVolNumPages{International Journal of Solids and Structures}{43}{6}{1764--1786}.
\PrintBackRefs{\CurrentBib}

\bibitem [\protect \citeauthoryear {%
Cacace%
, Hofmann%
\BCBL {}\ \BBA {} Shapiro%
}{%
Cacace%
\ \protect \BOthers {.}}{%
{\protect \APACyear {2021}}%
}]{%
cacace2021projecting}
\APACinsertmetastar {%
cacace2021projecting}%
\begin{APACrefauthors}%
Cacace, M.%
, Hofmann, H.%
\BCBL {}\ \BBA {} Shapiro, S\BPBI A.%
\end{APACrefauthors}%
\unskip\
\newblock
\APACrefYearMonthDay{2021}{}{}.
\newblock
{\BBOQ}\APACrefatitle {Projecting seismicity induced by complex alterations of underground stresses with applications to geothermal systems} {Projecting seismicity induced by complex alterations of underground stresses with applications to geothermal systems}.{\BBCQ}
\newblock
\APACjournalVolNumPages{Scientific Reports}{11}{1}{23560}.
\PrintBackRefs{\CurrentBib}

\bibitem [\protect \citeauthoryear {%
Clark%
, Northrop%
\BCBL {}\ \BBA {} Read%
}{%
Clark%
\ \protect \BOthers {.}}{%
{\protect \APACyear {1966}}%
}]{%
clark1966geology}
\APACinsertmetastar {%
clark1966geology}%
\begin{APACrefauthors}%
Clark, K.%
, Northrop, S.%
\BCBL {}\ \BBA {} Read, C.%
\end{APACrefauthors}%
\unskip\
\newblock
\APACrefYearMonthDay{1966}{}{}.
\newblock
{\BBOQ}\APACrefatitle {{Geology of the Sangre de Cristo Mountains and adjacent areas, between Taos and Raton, New Mexico}} {{Geology of the Sangre de Cristo Mountains and adjacent areas, between Taos and Raton, New Mexico}}.{\BBCQ}
\newblock
\BIn{} \APACrefbtitle {Taos-Raton-Spanish Peaks Country (New Mexico and Colorado): Geological Society 17th Annual Fall Field Conference Guidebook} {Taos-raton-spanish peaks country (new mexico and colorado): Geological society 17th annual fall field conference guidebook}\ (\BPGS\ 56--65).
\PrintBackRefs{\CurrentBib}

\bibitem [\protect \citeauthoryear {%
Cocco%
}{%
Cocco%
}{%
{\protect \APACyear {2002}}%
}]{%
cocco_pore_2002}
\APACinsertmetastar {%
cocco_pore_2002}%
\begin{APACrefauthors}%
Cocco, M.%
\end{APACrefauthors}%
\unskip\
\newblock
\APACrefYearMonthDay{2002}{}{}.
\newblock
{\BBOQ}\APACrefatitle {Pore pressure and poroelasticity effects in Coulomb stress analysis of earthquake interactions} {Pore pressure and poroelasticity effects in coulomb stress analysis of earthquake interactions}.{\BBCQ}
\newblock
\APACjournalVolNumPages{J. Geophys. Res.}{107}{}{2030}.
\PrintBackRefs{\CurrentBib}

\bibitem [\protect \citeauthoryear {%
Dassault~Systemes%
}{%
Dassault~Systemes%
}{%
{\protect \APACyear {2020}}%
}]{%
blabla}
\APACinsertmetastar {%
blabla}%
\begin{APACrefauthors}%
Dassault~Systemes, .%
\end{APACrefauthors}%
\unskip\
\newblock
\APACrefYear{2020}.
\newblock
\APACrefbtitle {ABAQUS (version 2019)} {Abaqus (version 2019)}.
\PrintBackRefs{\CurrentBib}

\bibitem [\protect \citeauthoryear {%
Detournay%
\ \BBA {} Cheng%
}{%
Detournay%
\ \BBA {} Cheng%
}{%
{\protect \APACyear {1993}}%
}]{%
detournay1993fundamentals}
\APACinsertmetastar {%
detournay1993fundamentals}%
\begin{APACrefauthors}%
Detournay, E.%
\BCBT {}\ \BBA {} Cheng, A\BPBI H\BHBI D.%
\end{APACrefauthors}%
\unskip\
\newblock
\APACrefYearMonthDay{1993}{}{}.
\newblock
{\BBOQ}\APACrefatitle {Fundamentals of poroelasticity} {Fundamentals of poroelasticity}.{\BBCQ}
\newblock
\BIn{} \APACrefbtitle {Analysis and design methods} {Analysis and design methods}\ (\BPGS\ 113--171).
\newblock
\APACaddressPublisher{}{Elsevier}.
\PrintBackRefs{\CurrentBib}

\bibitem [\protect \citeauthoryear {%
Duffield%
}{%
Duffield%
}{%
{\protect \APACyear {2007}}%
}]{%
aqtesolv}
\APACinsertmetastar {%
aqtesolv}%
\begin{APACrefauthors}%
Duffield, G.%
\end{APACrefauthors}%
\unskip\
\newblock
\APACrefYear{2007}.
\newblock
\APACrefbtitle {AQTESOLV$^{TM}$ Version 4.5 User's Guide} {Aqtesolv$^{TM}$ version 4.5 user's guide}.
\PrintBackRefs{\CurrentBib}

\bibitem [\protect \citeauthoryear {%
Ellsworth%
}{%
Ellsworth%
}{%
{\protect \APACyear {2013}}%
}]{%
ellsworth2013injection}
\APACinsertmetastar {%
ellsworth2013injection}%
\begin{APACrefauthors}%
Ellsworth, W\BPBI L.%
\end{APACrefauthors}%
\unskip\
\newblock
\APACrefYearMonthDay{2013}{}{}.
\newblock
{\BBOQ}\APACrefatitle {Injection-induced earthquakes} {Injection-induced earthquakes}.{\BBCQ}
\newblock
\APACjournalVolNumPages{Science}{341}{6142}{}.
\PrintBackRefs{\CurrentBib}

\bibitem [\protect \citeauthoryear {%
Fialko%
}{%
Fialko%
}{%
{\protect \APACyear {2004}}%
}]{%
fi04c}
\APACinsertmetastar {%
fi04c}%
\begin{APACrefauthors}%
Fialko, Y.%
\end{APACrefauthors}%
\unskip\
\newblock
\APACrefYearMonthDay{2004}{}{}.
\newblock
{\BBOQ}\APACrefatitle {{Evidence of fluid-filled upper crust from observations of post-seismic deformation due to the 1992 $M_w7.3$ Landers earthquake}} {{Evidence of fluid-filled upper crust from observations of post-seismic deformation due to the 1992 $M_w7.3$ Landers earthquake}}.{\BBCQ}
\newblock
\APACjournalVolNumPages{J. Geophys. Res.}{109}{}{B08401}.
\PrintBackRefs{\CurrentBib}

\bibitem [\protect \citeauthoryear {%
Fialko%
\ \BBA {} Simons%
}{%
Fialko%
\ \BBA {} Simons%
}{%
{\protect \APACyear {2000}}%
}]{%
fi&si00a}
\APACinsertmetastar {%
fi&si00a}%
\begin{APACrefauthors}%
Fialko, Y.%
\BCBT {}\ \BBA {} Simons, M.%
\end{APACrefauthors}%
\unskip\
\newblock
\APACrefYearMonthDay{2000}{}{}.
\newblock
{\BBOQ}\APACrefatitle {Deformation and seismicity in the {Coso} geothermal area, {Inyo County, California}: Observations and modeling using satellite radar interferometry} {Deformation and seismicity in the {Coso} geothermal area, {Inyo County, California}: Observations and modeling using satellite radar interferometry}.{\BBCQ}
\newblock
\APACjournalVolNumPages{J. Geophys. Res.}{105}{}{21781--21793}.
\PrintBackRefs{\CurrentBib}

\bibitem [\protect \citeauthoryear {%
Foulger%
, Wilson%
, Gluyas%
, Julian%
\BCBL {}\ \BBA {} Davies%
}{%
Foulger%
\ \protect \BOthers {.}}{%
{\protect \APACyear {2018}}%
}]{%
foulger2018global}
\APACinsertmetastar {%
foulger2018global}%
\begin{APACrefauthors}%
Foulger, G\BPBI R.%
, Wilson, M\BPBI P.%
, Gluyas, J\BPBI G.%
, Julian, B\BPBI R.%
\BCBL {}\ \BBA {} Davies, R\BPBI J.%
\end{APACrefauthors}%
\unskip\
\newblock
\APACrefYearMonthDay{2018}{}{}.
\newblock
{\BBOQ}\APACrefatitle {Global review of human-induced earthquakes} {Global review of human-induced earthquakes}.{\BBCQ}
\newblock
\APACjournalVolNumPages{Earth-Science Reviews}{178}{}{438--514}.
\PrintBackRefs{\CurrentBib}

\bibitem [\protect \citeauthoryear {%
Ge%
\ \BBA {} Saar%
}{%
Ge%
\ \BBA {} Saar%
}{%
{\protect \APACyear {2022}}%
}]{%
ge2022induced}
\APACinsertmetastar {%
ge2022induced}%
\begin{APACrefauthors}%
Ge, S.%
\BCBT {}\ \BBA {} Saar, M\BPBI O.%
\end{APACrefauthors}%
\unskip\
\newblock
\APACrefYearMonthDay{2022}{}{}.
\newblock
{\BBOQ}\APACrefatitle {Induced seismicity during geoenergy development—A hydromechanical perspective} {Induced seismicity during geoenergy development—a hydromechanical perspective}.{\BBCQ}
\newblock
\APACjournalVolNumPages{Journal of Geophysical Research: Solid Earth}{127}{3}{e2021JB023141}.
\PrintBackRefs{\CurrentBib}

\bibitem [\protect \citeauthoryear {%
Geldon%
}{%
Geldon%
}{%
{\protect \APACyear {1989}}%
}]{%
geldon1989ground}
\APACinsertmetastar {%
geldon1989ground}%
\begin{APACrefauthors}%
Geldon, L.%
\end{APACrefauthors}%
\unskip\
\newblock
\APACrefYearMonthDay{1989}{}{}.
\newblock
{\BBOQ}\APACrefatitle {{Ground-water hydrology of the central Raton Basin, Colorado and New Mexico}} {{Ground-water hydrology of the central Raton Basin, Colorado and New Mexico}}.{\BBCQ}
\newblock

\PrintBackRefs{\CurrentBib}

\bibitem [\protect \citeauthoryear {%
Giardini%
}{%
Giardini%
}{%
{\protect \APACyear {2009}}%
}]{%
giardini2009geothermal}
\APACinsertmetastar {%
giardini2009geothermal}%
\begin{APACrefauthors}%
Giardini, D.%
\end{APACrefauthors}%
\unskip\
\newblock
\APACrefYearMonthDay{2009}{}{}.
\newblock
{\BBOQ}\APACrefatitle {Geothermal quake risks must be faced} {Geothermal quake risks must be faced}.{\BBCQ}
\newblock
\APACjournalVolNumPages{Nature}{462}{7275}{848--849}.
\PrintBackRefs{\CurrentBib}

\bibitem [\protect \citeauthoryear {%
Glasgow%
\ \protect \BOthers {.}}{%
Glasgow%
\ \protect \BOthers {.}}{%
{\protect \APACyear {2021}}%
{\protect \APACexlab {{\protect \BCnt {1}}}}}]{%
Glasgow2021}
\APACinsertmetastar {%
Glasgow2021}%
\begin{APACrefauthors}%
Glasgow, M.%
, Schmandt, B.%
, Wang, R.%
, Zhang, M.%
, Bilek, S.%
\BCBL {}\ \BBA {} Kiser, E.%
\end{APACrefauthors}%
\unskip\
\newblock
\APACrefYearMonthDay{2021{\protect \BCnt {1}}}{{\APACmonth{08}}}{}.
\newblock
\APACrefbtitle {Raton Basin 2016-2020 earthquake catalog.} {Raton basin 2016-2020 earthquake catalog.}
\newblock
\APACaddressPublisher{}{International Seismological Centre}.
\newblock
\begin{APACrefDOI} \doi{10.31905/127xp53r} \end{APACrefDOI}
\PrintBackRefs{\CurrentBib}

\bibitem [\protect \citeauthoryear {%
Glasgow%
\ \protect \BOthers {.}}{%
Glasgow%
\ \protect \BOthers {.}}{%
{\protect \APACyear {2021}}%
{\protect \APACexlab {{\protect \BCnt {2}}}}}]{%
glasgow2021raton}
\APACinsertmetastar {%
glasgow2021raton}%
\begin{APACrefauthors}%
Glasgow, M.%
, Schmandt, B.%
, Wang, R.%
, Zhang, M.%
, Bilek, S\BPBI L.%
\BCBL {}\ \BBA {} Kiser, E.%
\end{APACrefauthors}%
\unskip\
\newblock
\APACrefYearMonthDay{2021{\protect \BCnt {2}}}{}{}.
\newblock
{\BBOQ}\APACrefatitle {{Raton Basin induced seismicity is hosted by networks of short basement faults and mimics tectonic earthquake statistics}} {{Raton Basin induced seismicity is hosted by networks of short basement faults and mimics tectonic earthquake statistics}}.{\BBCQ}
\newblock
\APACjournalVolNumPages{Journal of Geophysical Research: Solid Earth}{126}{11}{e2021JB022839}.
\PrintBackRefs{\CurrentBib}

\bibitem [\protect \citeauthoryear {%
Goertz-Allmann%
, K{\"u}hn%
, Oye%
, Bohloli%
\BCBL {}\ \BBA {} Aker%
}{%
Goertz-Allmann%
\ \protect \BOthers {.}}{%
{\protect \APACyear {2014}}%
}]{%
goertz2014combining}
\APACinsertmetastar {%
goertz2014combining}%
\begin{APACrefauthors}%
Goertz-Allmann, B\BPBI P.%
, K{\"u}hn, D.%
, Oye, V.%
, Bohloli, B.%
\BCBL {}\ \BBA {} Aker, E.%
\end{APACrefauthors}%
\unskip\
\newblock
\APACrefYearMonthDay{2014}{}{}.
\newblock
{\BBOQ}\APACrefatitle {{Combining microseismic and geomechanical observations to interpret storage integrity at the In Salah CCS site}} {{Combining microseismic and geomechanical observations to interpret storage integrity at the In Salah CCS site}}.{\BBCQ}
\newblock
\APACjournalVolNumPages{Geophysical Journal International}{198}{1}{447--461}.
\PrintBackRefs{\CurrentBib}

\bibitem [\protect \citeauthoryear {%
Gorelick%
}{%
Gorelick%
}{%
{\protect \APACyear {1983}}%
}]{%
gorelick1983review}
\APACinsertmetastar {%
gorelick1983review}%
\begin{APACrefauthors}%
Gorelick, S\BPBI M.%
\end{APACrefauthors}%
\unskip\
\newblock
\APACrefYearMonthDay{1983}{}{}.
\newblock
{\BBOQ}\APACrefatitle {A review of distributed parameter groundwater management modeling methods} {A review of distributed parameter groundwater management modeling methods}.{\BBCQ}
\newblock
\APACjournalVolNumPages{Water Resources Research}{19}{2}{305--319}.
\PrintBackRefs{\CurrentBib}

\bibitem [\protect \citeauthoryear {%
Gorelick%
, Freeze%
, Donohue%
, Keely%
\BCBL {}\ \protect \BOthers {.}}{%
Gorelick%
\ \protect \BOthers {.}}{%
{\protect \APACyear {1993}}%
}]{%
gorelick1993groundwater}
\APACinsertmetastar {%
gorelick1993groundwater}%
\begin{APACrefauthors}%
Gorelick, S\BPBI M.%
, Freeze, R\BPBI A.%
, Donohue, D.%
, Keely, J\BPBI F.%
\BCBL {}\ \BOthersPeriod {.}\end{APACrefauthors}%
\unskip\
\newblock
\APACrefYear{1993}.
\newblock
\APACrefbtitle {Groundwater contamination: optimal capture and containment.} {Groundwater contamination: optimal capture and containment.}
\newblock
\APACaddressPublisher{}{Lewis Publishers Inc.}
\PrintBackRefs{\CurrentBib}

\bibitem [\protect \citeauthoryear {%
Gorelick%
\ \BBA {} Remson%
}{%
Gorelick%
\ \BBA {} Remson%
}{%
{\protect \APACyear {1982}}%
}]{%
gorelick1982optimal}
\APACinsertmetastar {%
gorelick1982optimal}%
\begin{APACrefauthors}%
Gorelick, S\BPBI M.%
\BCBT {}\ \BBA {} Remson, I.%
\end{APACrefauthors}%
\unskip\
\newblock
\APACrefYearMonthDay{1982}{}{}.
\newblock
{\BBOQ}\APACrefatitle {Optimal dynamic management of groundwater pollutant sources} {Optimal dynamic management of groundwater pollutant sources}.{\BBCQ}
\newblock
\APACjournalVolNumPages{Water Resources Research}{18}{1}{71--76}.
\PrintBackRefs{\CurrentBib}

\bibitem [\protect \citeauthoryear {%
Gorelick%
\ \BBA {} Zheng%
}{%
Gorelick%
\ \BBA {} Zheng%
}{%
{\protect \APACyear {2015}}%
}]{%
gorelickzhang2015}
\APACinsertmetastar {%
gorelickzhang2015}%
\begin{APACrefauthors}%
Gorelick, S\BPBI M.%
\BCBT {}\ \BBA {} Zheng, C.%
\end{APACrefauthors}%
\unskip\
\newblock
\APACrefYearMonthDay{2015}{}{}.
\newblock
{\BBOQ}\APACrefatitle {Global change and the groundwater management challenge} {Global change and the groundwater management challenge}.{\BBCQ}
\newblock
\APACjournalVolNumPages{Water Resources Research}{51}{5}{3031-3051}.
\PrintBackRefs{\CurrentBib}

\bibitem [\protect \citeauthoryear {%
Grasso%
\ \BBA {} Wittlinger%
}{%
Grasso%
\ \BBA {} Wittlinger%
}{%
{\protect \APACyear {1990}}%
}]{%
grasso1990ten}
\APACinsertmetastar {%
grasso1990ten}%
\begin{APACrefauthors}%
Grasso, J\BHBI R.%
\BCBT {}\ \BBA {} Wittlinger, G.%
\end{APACrefauthors}%
\unskip\
\newblock
\APACrefYearMonthDay{1990}{}{}.
\newblock
{\BBOQ}\APACrefatitle {Ten years of seismic monitoring over a gas field} {Ten years of seismic monitoring over a gas field}.{\BBCQ}
\newblock
\APACjournalVolNumPages{Bulletin of the Seismological Society of America}{80}{2}{450--473}.
\PrintBackRefs{\CurrentBib}

\bibitem [\protect \citeauthoryear {%
Hernandez%
}{%
Hernandez%
}{%
{\protect \APACyear {2020}}%
}]{%
rhernMSthesis}
\APACinsertmetastar {%
rhernMSthesis}%
\begin{APACrefauthors}%
Hernandez, R.%
\end{APACrefauthors}%
\unskip\
\newblock
\APACrefYear{2020}.
\unskip\
\newblock
\APACrefbtitle {{Fluid pressure modeling of faults and simulation-optimization of wastewater injection in the Raton Basin, CO-NM}} {{Fluid pressure modeling of faults and simulation-optimization of wastewater injection in the Raton Basin, CO-NM}}\ \APACtypeAddressSchool {\BUMTh}{}{}.
\unskip\
\newblock
\APACaddressSchool {}{San Diego State University}.
\PrintBackRefs{\CurrentBib}

\bibitem [\protect \citeauthoryear {%
Hernandez%
\ \BBA {} Weingarten%
}{%
Hernandez%
\ \BBA {} Weingarten%
}{%
{\protect \APACyear {2019}}%
}]{%
hernandez2019step}
\APACinsertmetastar {%
hernandez2019step}%
\begin{APACrefauthors}%
Hernandez, R.%
\BCBT {}\ \BBA {} Weingarten, M.%
\end{APACrefauthors}%
\unskip\
\newblock
\APACrefYearMonthDay{2019}{}{}.
\newblock
{\BBOQ}\APACrefatitle {{Step-rate Test Calibration of Primary Injection Reservoir Permeability in a Case of Injection Induced Seismicity, Raton Basin, CO-NM}} {{Step-rate Test Calibration of Primary Injection Reservoir Permeability in a Case of Injection Induced Seismicity, Raton Basin, CO-NM}}.{\BBCQ}
\newblock
\BIn{} \APACrefbtitle {AGU fall meeting abstracts} {Agu fall meeting abstracts}\ (\BVOL\ 2019, \BPGS\ S13E--0493).
\PrintBackRefs{\CurrentBib}

\bibitem [\protect \citeauthoryear {%
Higley%
}{%
Higley%
}{%
{\protect \APACyear {2007}}%
}]{%
higley2007petroleum}
\APACinsertmetastar {%
higley2007petroleum}%
\begin{APACrefauthors}%
Higley, D\BPBI K.%
\end{APACrefauthors}%
\unskip\
\newblock
\APACrefYear{2007}.
\newblock
\APACrefbtitle {{Petroleum systems and assessment of undiscovered oil and gas in the Raton Basin--Sierra Grande Uplift Province, Colorado and New Mexico—USGS Province 41}} {{Petroleum systems and assessment of undiscovered oil and gas in the Raton Basin--Sierra Grande Uplift Province, Colorado and New Mexico—USGS Province 41}}\ (\BNUM\ 69-N).
\newblock
\APACaddressPublisher{}{US Geological Survey}.
\PrintBackRefs{\CurrentBib}

\bibitem [\protect \citeauthoryear {%
R.~Hill%
}{%
R.~Hill%
}{%
{\protect \APACyear {2024}}%
}]{%
HillZenodoRaton}
\APACinsertmetastar {%
HillZenodoRaton}%
\begin{APACrefauthors}%
Hill, R.%
\end{APACrefauthors}%
\unskip\
\newblock
\APACrefYearMonthDay{2024}{}{}.
\newblock
\APACrefbtitle {{Data and Code for Modeling Raton Basin [Data Set]}.} {{Data and Code for Modeling Raton Basin [Data Set]}.}
\newblock
\APACaddressPublisher{}{Zenodo}.
\newblock
\begin{APACrefURL} \url{https://doi.org/10.5281/zenodo.10472485} \end{APACrefURL}
\PrintBackRefs{\CurrentBib}

\bibitem [\protect \citeauthoryear {%
R\BPBI G.~Hill%
, Weingarten%
, Rockwell%
\BCBL {}\ \BBA {} Fialko%
}{%
R\BPBI G.~Hill%
\ \protect \BOthers {.}}{%
{\protect \APACyear {2023}}%
}]{%
hill2023major}
\APACinsertmetastar {%
hill2023major}%
\begin{APACrefauthors}%
Hill, R\BPBI G.%
, Weingarten, M.%
, Rockwell, T\BPBI K.%
\BCBL {}\ \BBA {} Fialko, Y.%
\end{APACrefauthors}%
\unskip\
\newblock
\APACrefYearMonthDay{2023}{}{}.
\newblock
{\BBOQ}\APACrefatitle {{Major southern San Andreas earthquakes modulated by lake-filling events}} {{Major southern San Andreas earthquakes modulated by lake-filling events}}.{\BBCQ}
\newblock
\APACjournalVolNumPages{Nature}{618}{}{761--766}.
\PrintBackRefs{\CurrentBib}

\bibitem [\protect \citeauthoryear {%
Hsu%
\ \BBA {} Yeh%
}{%
Hsu%
\ \BBA {} Yeh%
}{%
{\protect \APACyear {1989}}%
}]{%
hsu1989optimum}
\APACinsertmetastar {%
hsu1989optimum}%
\begin{APACrefauthors}%
Hsu, N\BHBI S.%
\BCBT {}\ \BBA {} Yeh, W\BPBI W\BHBI G.%
\end{APACrefauthors}%
\unskip\
\newblock
\APACrefYearMonthDay{1989}{}{}.
\newblock
{\BBOQ}\APACrefatitle {Optimum experimental design for parameter identification in groundwater hydrology} {Optimum experimental design for parameter identification in groundwater hydrology}.{\BBCQ}
\newblock
\APACjournalVolNumPages{Water Resources Research}{25}{5}{1025--1040}.
\PrintBackRefs{\CurrentBib}

\bibitem [\protect \citeauthoryear {%
L.~Jin%
}{%
L.~Jin%
}{%
{\protect \APACyear {2023}}%
}]{%
jin2023saturated}
\APACinsertmetastar {%
jin2023saturated}%
\begin{APACrefauthors}%
Jin, L.%
\end{APACrefauthors}%
\unskip\
\newblock
\APACrefYearMonthDay{2023}{}{}.
\newblock
{\BBOQ}\APACrefatitle {On A Saturated Poromechanical Framework and Its Relation to Abaqus Soil Mechanics and Biot Poroelasticity Frameworks} {On a saturated poromechanical framework and its relation to abaqus soil mechanics and biot poroelasticity frameworks}.{\BBCQ}
\newblock
\APACjournalVolNumPages{arXiv preprint arXiv:2304.02148}{}{}{}.
\PrintBackRefs{\CurrentBib}

\bibitem [\protect \citeauthoryear {%
L.~Jin%
, Curry%
, Lippoldt%
, Hussenoeder%
\BCBL {}\ \BBA {} Bhargava%
}{%
L.~Jin%
\ \protect \BOthers {.}}{%
{\protect \APACyear {2023}}%
}]{%
jin20233d}
\APACinsertmetastar {%
jin20233d}%
\begin{APACrefauthors}%
Jin, L.%
, Curry, W\BPBI J.%
, Lippoldt, R\BPBI C.%
, Hussenoeder, S\BPBI A.%
\BCBL {}\ \BBA {} Bhargava, P.%
\end{APACrefauthors}%
\unskip\
\newblock
\APACrefYearMonthDay{2023}{}{}.
\newblock
{\BBOQ}\APACrefatitle {3D coupled hydro-mechanical modeling of multi-decadal multi-zone saltwater disposal in layered and faulted poroelastic rocks and implications for seismicity: An example from the Midland Basin} {3d coupled hydro-mechanical modeling of multi-decadal multi-zone saltwater disposal in layered and faulted poroelastic rocks and implications for seismicity: An example from the midland basin}.{\BBCQ}
\newblock
\APACjournalVolNumPages{Tectonophysics}{863}{}{229996}.
\PrintBackRefs{\CurrentBib}

\bibitem [\protect \citeauthoryear {%
Z.~Jin%
, Fialko%
, Zubovich%
\BCBL {}\ \BBA {} Sch{\"o}ne%
}{%
Z.~Jin%
\ \protect \BOthers {.}}{%
{\protect \APACyear {2022}}%
}]{%
jin2022lithospheric}
\APACinsertmetastar {%
jin2022lithospheric}%
\begin{APACrefauthors}%
Jin, Z.%
, Fialko, Y.%
, Zubovich, A.%
\BCBL {}\ \BBA {} Sch{\"o}ne, T.%
\end{APACrefauthors}%
\unskip\
\newblock
\APACrefYearMonthDay{2022}{}{}.
\newblock
{\BBOQ}\APACrefatitle {{Lithospheric deformation due to the 2015 M7.2 Sarez (Pamir) earthquake constrained by 5 years of space geodetic observations}} {{Lithospheric deformation due to the 2015 M7.2 Sarez (Pamir) earthquake constrained by 5 years of space geodetic observations}}.{\BBCQ}
\newblock
\APACjournalVolNumPages{J. Geophys. Res.}{127}{}{e2021JB022461}.
\PrintBackRefs{\CurrentBib}

\bibitem [\protect \citeauthoryear {%
Keranen%
\ \BBA {} Weingarten%
}{%
Keranen%
\ \BBA {} Weingarten%
}{%
{\protect \APACyear {2018}}%
}]{%
keranen2018induced}
\APACinsertmetastar {%
keranen2018induced}%
\begin{APACrefauthors}%
Keranen, K\BPBI M.%
\BCBT {}\ \BBA {} Weingarten, M.%
\end{APACrefauthors}%
\unskip\
\newblock
\APACrefYearMonthDay{2018}{}{}.
\newblock
{\BBOQ}\APACrefatitle {Induced seismicity} {Induced seismicity}.{\BBCQ}
\newblock
\APACjournalVolNumPages{Annual Review of Earth and Planetary Sciences}{46}{}{149--174}.
\PrintBackRefs{\CurrentBib}

\bibitem [\protect \citeauthoryear {%
Keranen%
, Weingarten%
, Abers%
, Bekins%
\BCBL {}\ \BBA {} Ge%
}{%
Keranen%
\ \protect \BOthers {.}}{%
{\protect \APACyear {2014}}%
}]{%
keranen2014sharp}
\APACinsertmetastar {%
keranen2014sharp}%
\begin{APACrefauthors}%
Keranen, K\BPBI M.%
, Weingarten, M.%
, Abers, G\BPBI A.%
, Bekins, B\BPBI A.%
\BCBL {}\ \BBA {} Ge, S.%
\end{APACrefauthors}%
\unskip\
\newblock
\APACrefYearMonthDay{2014}{}{}.
\newblock
{\BBOQ}\APACrefatitle {{Sharp increase in central Oklahoma seismicity since 2008 induced by massive wastewater injection}} {{Sharp increase in central Oklahoma seismicity since 2008 induced by massive wastewater injection}}.{\BBCQ}
\newblock
\APACjournalVolNumPages{Science}{345}{6195}{448--451}.
\PrintBackRefs{\CurrentBib}

\bibitem [\protect \citeauthoryear {%
King%
, Stein%
\BCBL {}\ \BBA {} Lin%
}{%
King%
\ \protect \BOthers {.}}{%
{\protect \APACyear {1994}}%
}]{%
king94}
\APACinsertmetastar {%
king94}%
\begin{APACrefauthors}%
King, G\BPBI C\BPBI P.%
, Stein, R\BPBI C.%
\BCBL {}\ \BBA {} Lin, J.%
\end{APACrefauthors}%
\unskip\
\newblock
\APACrefYearMonthDay{1994}{}{}.
\newblock
{\BBOQ}\APACrefatitle {Static Stress Change and the Triggering of Earthquakes} {Static stress change and the triggering of earthquakes}.{\BBCQ}
\newblock
\APACjournalVolNumPages{Bull. Seism. Soc. Am.}{84}{}{935--953}.
\PrintBackRefs{\CurrentBib}

\bibitem [\protect \citeauthoryear {%
LaBonte%
, Brown%
\BCBL {}\ \BBA {} Fialko%
}{%
LaBonte%
\ \protect \BOthers {.}}{%
{\protect \APACyear {2009}}%
}]{%
labonte+09a}
\APACinsertmetastar {%
labonte+09a}%
\begin{APACrefauthors}%
LaBonte, A.%
, Brown, K.%
\BCBL {}\ \BBA {} Fialko, Y.%
\end{APACrefauthors}%
\unskip\
\newblock
\APACrefYearMonthDay{2009}{}{}.
\newblock
{\BBOQ}\APACrefatitle {{Hydrogeologic detection and finite-element modeling of a slow-slip event in the Costa Rica prism toe}} {{Hydrogeologic detection and finite-element modeling of a slow-slip event in the Costa Rica prism toe}}.{\BBCQ}
\newblock
\APACjournalVolNumPages{J. Geophys. Res.}{114}{}{B00A02}.
\PrintBackRefs{\CurrentBib}

\bibitem [\protect \citeauthoryear {%
Langenbruch%
, Dinske%
\BCBL {}\ \BBA {} Shapiro%
}{%
Langenbruch%
\ \protect \BOthers {.}}{%
{\protect \APACyear {2011}}%
}]{%
langenbruch2011inter}
\APACinsertmetastar {%
langenbruch2011inter}%
\begin{APACrefauthors}%
Langenbruch, C.%
, Dinske, C.%
\BCBL {}\ \BBA {} Shapiro, S.%
\end{APACrefauthors}%
\unskip\
\newblock
\APACrefYearMonthDay{2011}{}{}.
\newblock
{\BBOQ}\APACrefatitle {Inter event times of fluid induced earthquakes suggest their Poisson nature} {Inter event times of fluid induced earthquakes suggest their poisson nature}.{\BBCQ}
\newblock
\APACjournalVolNumPages{Geophysical Research Letters}{38}{21}{}.
\PrintBackRefs{\CurrentBib}

\bibitem [\protect \citeauthoryear {%
Langenbruch%
, Weingarten%
\BCBL {}\ \BBA {} Zoback%
}{%
Langenbruch%
\ \protect \BOthers {.}}{%
{\protect \APACyear {2018}}%
}]{%
langenbruch2018physics}
\APACinsertmetastar {%
langenbruch2018physics}%
\begin{APACrefauthors}%
Langenbruch, C.%
, Weingarten, M.%
\BCBL {}\ \BBA {} Zoback, M\BPBI D.%
\end{APACrefauthors}%
\unskip\
\newblock
\APACrefYearMonthDay{2018}{}{}.
\newblock
{\BBOQ}\APACrefatitle {{Physics-based forecasting of man-made earthquake hazards in Oklahoma and Kansas}} {{Physics-based forecasting of man-made earthquake hazards in Oklahoma and Kansas}}.{\BBCQ}
\newblock
\APACjournalVolNumPages{Nature communications}{9}{1}{1--10}.
\PrintBackRefs{\CurrentBib}

\bibitem [\protect \citeauthoryear {%
Langenbruch%
\ \BBA {} Zoback%
}{%
Langenbruch%
\ \BBA {} Zoback%
}{%
{\protect \APACyear {2016}}%
}]{%
langenbruch2016will}
\APACinsertmetastar {%
langenbruch2016will}%
\begin{APACrefauthors}%
Langenbruch, C.%
\BCBT {}\ \BBA {} Zoback, M\BPBI D.%
\end{APACrefauthors}%
\unskip\
\newblock
\APACrefYearMonthDay{2016}{}{}.
\newblock
{\BBOQ}\APACrefatitle {{How will induced seismicity in Oklahoma respond to decreased saltwater injection rates?}} {{How will induced seismicity in Oklahoma respond to decreased saltwater injection rates?}}{\BBCQ}
\newblock
\APACjournalVolNumPages{Science advances}{2}{11}{e1601542}.
\PrintBackRefs{\CurrentBib}

\bibitem [\protect \citeauthoryear {%
Macartney%
\ \BBA {} O’Farrell%
}{%
Macartney%
\ \BBA {} O’Farrell%
}{%
{\protect \APACyear {2010}}%
}]{%
macartney2010raton}
\APACinsertmetastar {%
macartney2010raton}%
\begin{APACrefauthors}%
Macartney, H.%
\BCBT {}\ \BBA {} O’Farrell, C.%
\end{APACrefauthors}%
\unskip\
\newblock
\APACrefYearMonthDay{2010}{}{}.
\newblock
{\BBOQ}\APACrefatitle {{A Raton Basin geothermal prospect}} {{A Raton Basin geothermal prospect}}.{\BBCQ}
\newblock
\APACjournalVolNumPages{AAPG, Durango, CO}{}{}{}.
\PrintBackRefs{\CurrentBib}

\bibitem [\protect \citeauthoryear {%
Majer%
\ \BBA {} Peterson%
}{%
Majer%
\ \BBA {} Peterson%
}{%
{\protect \APACyear {2007}}%
}]{%
majer2007impact}
\APACinsertmetastar {%
majer2007impact}%
\begin{APACrefauthors}%
Majer, E\BPBI L.%
\BCBT {}\ \BBA {} Peterson, J\BPBI E.%
\end{APACrefauthors}%
\unskip\
\newblock
\APACrefYearMonthDay{2007}{}{}.
\newblock
{\BBOQ}\APACrefatitle {{The impact of injection on seismicity at The Geysers, California Geothermal Field}} {{The impact of injection on seismicity at The Geysers, California Geothermal Field}}.{\BBCQ}
\newblock
\APACjournalVolNumPages{International Journal of Rock Mechanics and Mining Sciences}{44}{8}{1079--1090}.
\PrintBackRefs{\CurrentBib}

\bibitem [\protect \citeauthoryear {%
McGarr%
}{%
McGarr%
}{%
{\protect \APACyear {2014}}%
}]{%
mcgarr2014maximum}
\APACinsertmetastar {%
mcgarr2014maximum}%
\begin{APACrefauthors}%
McGarr, A.%
\end{APACrefauthors}%
\unskip\
\newblock
\APACrefYearMonthDay{2014}{}{}.
\newblock
{\BBOQ}\APACrefatitle {Maximum magnitude earthquakes induced by fluid injection} {Maximum magnitude earthquakes induced by fluid injection}.{\BBCQ}
\newblock
\APACjournalVolNumPages{Journal of Geophysical Research: solid earth}{119}{2}{1008--1019}.
\PrintBackRefs{\CurrentBib}

\bibitem [\protect \citeauthoryear {%
Mignan%
, Landtwing%
, K{\"a}stli%
, Mena%
\BCBL {}\ \BBA {} Wiemer%
}{%
Mignan%
\ \protect \BOthers {.}}{%
{\protect \APACyear {2015}}%
}]{%
mignan2015induced}
\APACinsertmetastar {%
mignan2015induced}%
\begin{APACrefauthors}%
Mignan, A.%
, Landtwing, D.%
, K{\"a}stli, P.%
, Mena, B.%
\BCBL {}\ \BBA {} Wiemer, S.%
\end{APACrefauthors}%
\unskip\
\newblock
\APACrefYearMonthDay{2015}{}{}.
\newblock
{\BBOQ}\APACrefatitle {{Induced seismicity risk analysis of the 2006 Basel, Switzerland, Enhanced Geothermal System project: Influence of uncertainties on risk mitigation}} {{Induced seismicity risk analysis of the 2006 Basel, Switzerland, Enhanced Geothermal System project: Influence of uncertainties on risk mitigation}}.{\BBCQ}
\newblock
\APACjournalVolNumPages{Geothermics}{53}{}{133--146}.
\PrintBackRefs{\CurrentBib}

\bibitem [\protect \citeauthoryear {%
Nakai%
, Sheehan%
\BCBL {}\ \BBA {} Bilek%
}{%
Nakai%
, Sheehan%
\BCBL {}\ \BBA {} Bilek%
}{%
{\protect \APACyear {2017}}%
}]{%
nakai2017seismicity}
\APACinsertmetastar {%
nakai2017seismicity}%
\begin{APACrefauthors}%
Nakai, J.%
, Sheehan, A.%
\BCBL {}\ \BBA {} Bilek, S.%
\end{APACrefauthors}%
\unskip\
\newblock
\APACrefYearMonthDay{2017}{}{}.
\newblock
{\BBOQ}\APACrefatitle {{Seismicity of the rocky mountains and Rio Grande Rift from the EarthScope Transportable Array and CREST temporary seismic networks, 2008--2010}} {{Seismicity of the rocky mountains and Rio Grande Rift from the EarthScope Transportable Array and CREST temporary seismic networks, 2008--2010}}.{\BBCQ}
\newblock
\APACjournalVolNumPages{Journal of Geophysical Research: Solid Earth}{122}{3}{2173--2192}.
\PrintBackRefs{\CurrentBib}

\bibitem [\protect \citeauthoryear {%
Nakai%
, Weingarten%
, Sheehan%
, Bilek%
\BCBL {}\ \BBA {} Ge%
}{%
Nakai%
, Weingarten%
\BCBL {}\ \protect \BOthers {.}}{%
{\protect \APACyear {2017}}%
}]{%
nakai2017possible}
\APACinsertmetastar {%
nakai2017possible}%
\begin{APACrefauthors}%
Nakai, J.%
, Weingarten, M.%
, Sheehan, A.%
, Bilek, S.%
\BCBL {}\ \BBA {} Ge, S.%
\end{APACrefauthors}%
\unskip\
\newblock
\APACrefYearMonthDay{2017}{}{}.
\newblock
{\BBOQ}\APACrefatitle {{A possible causative mechanism of Raton Basin, New Mexico and Colorado earthquakes using recent seismicity patterns and pore pressure modeling}} {{A possible causative mechanism of Raton Basin, New Mexico and Colorado earthquakes using recent seismicity patterns and pore pressure modeling}}.{\BBCQ}
\newblock
\APACjournalVolNumPages{Journal of Geophysical Research: Solid Earth}{122}{10}{8051--8065}.
\PrintBackRefs{\CurrentBib}

\bibitem [\protect \citeauthoryear {%
Nelson%
, Gianoutsos%
\BCBL {}\ \BBA {} Anna%
}{%
Nelson%
\ \protect \BOthers {.}}{%
{\protect \APACyear {2013}}%
}]{%
nelson2013outcrop}
\APACinsertmetastar {%
nelson2013outcrop}%
\begin{APACrefauthors}%
Nelson, P\BPBI H.%
, Gianoutsos, N\BPBI J.%
\BCBL {}\ \BBA {} Anna, L\BPBI O.%
\end{APACrefauthors}%
\unskip\
\newblock
\APACrefYearMonthDay{2013}{}{}.
\newblock
{\BBOQ}\APACrefatitle {{Outcrop control of basin-scale underpressure in the Raton Basin, Colorado and New Mexico}} {{Outcrop control of basin-scale underpressure in the Raton Basin, Colorado and New Mexico}}.{\BBCQ}
\newblock

\PrintBackRefs{\CurrentBib}

\bibitem [\protect \citeauthoryear {%
Pearse%
\ \BBA {} Fialko%
}{%
Pearse%
\ \BBA {} Fialko%
}{%
{\protect \APACyear {2010}}%
}]{%
pearse&fi10a}
\APACinsertmetastar {%
pearse&fi10a}%
\begin{APACrefauthors}%
Pearse, J.%
\BCBT {}\ \BBA {} Fialko, Y.%
\end{APACrefauthors}%
\unskip\
\newblock
\APACrefYearMonthDay{2010}{}{}.
\newblock
{\BBOQ}\APACrefatitle {{Mechanics of active magmatic intraplating in the Rio Grande Rift near Socorro, New Mexico}} {{Mechanics of active magmatic intraplating in the Rio Grande Rift near Socorro, New Mexico}}.{\BBCQ}
\newblock
\APACjournalVolNumPages{J. Geophys. Res.}{115}{}{B07413}.
\PrintBackRefs{\CurrentBib}

\bibitem [\protect \citeauthoryear {%
Qin%
, Chen%
, Ma%
\BCBL {}\ \BBA {} Chen%
}{%
Qin%
\ \protect \BOthers {.}}{%
{\protect \APACyear {2022}}%
}]{%
qin2022forecasting}
\APACinsertmetastar {%
qin2022forecasting}%
\begin{APACrefauthors}%
Qin, Y.%
, Chen, T.%
, Ma, X.%
\BCBL {}\ \BBA {} Chen, X.%
\end{APACrefauthors}%
\unskip\
\newblock
\APACrefYearMonthDay{2022}{}{}.
\newblock
{\BBOQ}\APACrefatitle {{Forecasting induced seismicity in Oklahoma using machine learning methods}} {{Forecasting induced seismicity in Oklahoma using machine learning methods}}.{\BBCQ}
\newblock
\APACjournalVolNumPages{Scientific Reports}{12}{1}{9319}.
\PrintBackRefs{\CurrentBib}

\bibitem [\protect \citeauthoryear {%
Rice%
\ \BBA {} Cleary%
}{%
Rice%
\ \BBA {} Cleary%
}{%
{\protect \APACyear {1976}}%
}]{%
rice_basic_1976}
\APACinsertmetastar {%
rice_basic_1976}%
\begin{APACrefauthors}%
Rice, J\BPBI R.%
\BCBT {}\ \BBA {} Cleary, M\BPBI P.%
\end{APACrefauthors}%
\unskip\
\newblock
\APACrefYearMonthDay{1976}{}{}.
\newblock
{\BBOQ}\APACrefatitle {Some basic stress diffusion solutions for fluid-saturated elastic porous media with compressible constituents} {Some basic stress diffusion solutions for fluid-saturated elastic porous media with compressible constituents}.{\BBCQ}
\newblock
\APACjournalVolNumPages{Rev. Geophys.}{14}{2}{227}.
\newblock
\begin{APACrefDOI} \doi{10.1029/RG014i002p00227} \end{APACrefDOI}
\PrintBackRefs{\CurrentBib}

\bibitem [\protect \citeauthoryear {%
Rubinstein%
, Ellsworth%
, McGarr%
\BCBL {}\ \BBA {} Benz%
}{%
Rubinstein%
\ \protect \BOthers {.}}{%
{\protect \APACyear {2014}}%
}]{%
rubinstein20142001}
\APACinsertmetastar {%
rubinstein20142001}%
\begin{APACrefauthors}%
Rubinstein, J\BPBI L.%
, Ellsworth, W\BPBI L.%
, McGarr, A.%
\BCBL {}\ \BBA {} Benz, H\BPBI M.%
\end{APACrefauthors}%
\unskip\
\newblock
\APACrefYearMonthDay{2014}{}{}.
\newblock
{\BBOQ}\APACrefatitle {{The 2001--present induced earthquake sequence in the Raton Basin of northern New Mexico and southern Colorado}} {{The 2001--present induced earthquake sequence in the Raton Basin of northern New Mexico and southern Colorado}}.{\BBCQ}
\newblock
\APACjournalVolNumPages{Bulletin of the Seismological Society of America}{104}{5}{2162--2181}.
\PrintBackRefs{\CurrentBib}

\bibitem [\protect \citeauthoryear {%
Rudnicki%
}{%
Rudnicki%
}{%
{\protect \APACyear {1986}}%
}]{%
RUDNICKI1986383}
\APACinsertmetastar {%
RUDNICKI1986383}%
\begin{APACrefauthors}%
Rudnicki, J\BPBI W.%
\end{APACrefauthors}%
\unskip\
\newblock
\APACrefYearMonthDay{1986}{}{}.
\newblock
{\BBOQ}\APACrefatitle {Fluid mass sources and point forces in linear elastic diffusive solids} {Fluid mass sources and point forces in linear elastic diffusive solids}.{\BBCQ}
\newblock
\APACjournalVolNumPages{Mechanics of Materials}{5}{4}{383-393}.
\PrintBackRefs{\CurrentBib}

\bibitem [\protect \citeauthoryear {%
Rutqvist%
, Rinaldi%
, Cappa%
\BCBL {}\ \BBA {} Moridis%
}{%
Rutqvist%
\ \protect \BOthers {.}}{%
{\protect \APACyear {2015}}%
}]{%
rutqvist2015modeling}
\APACinsertmetastar {%
rutqvist2015modeling}%
\begin{APACrefauthors}%
Rutqvist, J.%
, Rinaldi, A\BPBI P.%
, Cappa, F.%
\BCBL {}\ \BBA {} Moridis, G\BPBI J.%
\end{APACrefauthors}%
\unskip\
\newblock
\APACrefYearMonthDay{2015}{}{}.
\newblock
{\BBOQ}\APACrefatitle {Modeling of fault activation and seismicity by injection directly into a fault zone associated with hydraulic fracturing of shale-gas reservoirs} {Modeling of fault activation and seismicity by injection directly into a fault zone associated with hydraulic fracturing of shale-gas reservoirs}.{\BBCQ}
\newblock
\APACjournalVolNumPages{Journal of Petroleum Science and Engineering}{127}{}{377--386}.
\PrintBackRefs{\CurrentBib}

\bibitem [\protect \citeauthoryear {%
Schultz%
, Beroza%
\BCBL {}\ \BBA {} Ellsworth%
}{%
Schultz%
\ \protect \BOthers {.}}{%
{\protect \APACyear {2021}}%
}]{%
schultz2021risk}
\APACinsertmetastar {%
schultz2021risk}%
\begin{APACrefauthors}%
Schultz, R.%
, Beroza, G\BPBI C.%
\BCBL {}\ \BBA {} Ellsworth, W\BPBI L.%
\end{APACrefauthors}%
\unskip\
\newblock
\APACrefYearMonthDay{2021}{}{}.
\newblock
{\BBOQ}\APACrefatitle {A risk-based approach for managing hydraulic fracturing--induced seismicity} {A risk-based approach for managing hydraulic fracturing--induced seismicity}.{\BBCQ}
\newblock
\APACjournalVolNumPages{Science}{372}{6541}{504--507}.
\PrintBackRefs{\CurrentBib}

\bibitem [\protect \citeauthoryear {%
Segall%
}{%
Segall%
}{%
{\protect \APACyear {2010}}%
}]{%
segall2010earthquake}
\APACinsertmetastar {%
segall2010earthquake}%
\begin{APACrefauthors}%
Segall, P.%
\end{APACrefauthors}%
\unskip\
\newblock
\APACrefYear{2010}.
\newblock
\APACrefbtitle {{Earthquake and Volcano Deformation}} {{Earthquake and Volcano Deformation}}.
\newblock
\APACaddressPublisher{}{Princeton University Press}.
\PrintBackRefs{\CurrentBib}

\bibitem [\protect \citeauthoryear {%
Segall%
\ \BBA {} Lu%
}{%
Segall%
\ \BBA {} Lu%
}{%
{\protect \APACyear {2015}}%
}]{%
segall2015injection}
\APACinsertmetastar {%
segall2015injection}%
\begin{APACrefauthors}%
Segall, P.%
\BCBT {}\ \BBA {} Lu, S.%
\end{APACrefauthors}%
\unskip\
\newblock
\APACrefYearMonthDay{2015}{}{}.
\newblock
{\BBOQ}\APACrefatitle {Injection-induced seismicity: Poroelastic and earthquake nucleation effects} {Injection-induced seismicity: Poroelastic and earthquake nucleation effects}.{\BBCQ}
\newblock
\APACjournalVolNumPages{Journal of Geophysical Research: Solid Earth}{120}{7}{5082--5103}.
\PrintBackRefs{\CurrentBib}

\bibitem [\protect \citeauthoryear {%
Shapiro%
, Dinske%
, Langenbruch%
\BCBL {}\ \BBA {} Wenzel%
}{%
Shapiro%
\ \protect \BOthers {.}}{%
{\protect \APACyear {2010}}%
}]{%
shapiro2010seismogenic}
\APACinsertmetastar {%
shapiro2010seismogenic}%
\begin{APACrefauthors}%
Shapiro, S\BPBI A.%
, Dinske, C.%
, Langenbruch, C.%
\BCBL {}\ \BBA {} Wenzel, F.%
\end{APACrefauthors}%
\unskip\
\newblock
\APACrefYearMonthDay{2010}{}{}.
\newblock
{\BBOQ}\APACrefatitle {Seismogenic index and magnitude probability of earthquakes induced during reservoir fluid stimulations} {Seismogenic index and magnitude probability of earthquakes induced during reservoir fluid stimulations}.{\BBCQ}
\newblock
\APACjournalVolNumPages{The Leading Edge}{29}{3}{304--309}.
\PrintBackRefs{\CurrentBib}

\bibitem [\protect \citeauthoryear {%
Shirzaei%
, Ellsworth%
, Tiampo%
, Gonz{\'a}lez%
\BCBL {}\ \BBA {} Manga%
}{%
Shirzaei%
\ \protect \BOthers {.}}{%
{\protect \APACyear {2016}}%
}]{%
shirzaei2016surface}
\APACinsertmetastar {%
shirzaei2016surface}%
\begin{APACrefauthors}%
Shirzaei, M.%
, Ellsworth, W\BPBI L.%
, Tiampo, K\BPBI F.%
, Gonz{\'a}lez, P\BPBI J.%
\BCBL {}\ \BBA {} Manga, M.%
\end{APACrefauthors}%
\unskip\
\newblock
\APACrefYearMonthDay{2016}{}{}.
\newblock
{\BBOQ}\APACrefatitle {{Surface uplift and time-dependent seismic hazard due to fluid injection in eastern Texas}} {{Surface uplift and time-dependent seismic hazard due to fluid injection in eastern Texas}}.{\BBCQ}
\newblock
\APACjournalVolNumPages{Science}{353}{6306}{1416--1419}.
\PrintBackRefs{\CurrentBib}

\bibitem [\protect \citeauthoryear {%
Shmonov%
, Vitiovtova%
, Zharikov%
\BCBL {}\ \BBA {} Grafchikov%
}{%
Shmonov%
\ \protect \BOthers {.}}{%
{\protect \APACyear {2003}}%
}]{%
shmonov2003permeability}
\APACinsertmetastar {%
shmonov2003permeability}%
\begin{APACrefauthors}%
Shmonov, V.%
, Vitiovtova, V.%
, Zharikov, A.%
\BCBL {}\ \BBA {} Grafchikov, A.%
\end{APACrefauthors}%
\unskip\
\newblock
\APACrefYearMonthDay{2003}{}{}.
\newblock
{\BBOQ}\APACrefatitle {Permeability of the continental crust: implications of experimental data} {Permeability of the continental crust: implications of experimental data}.{\BBCQ}
\newblock
\APACjournalVolNumPages{Journal of Geochemical Exploration}{78}{}{697--699}.
\PrintBackRefs{\CurrentBib}

\bibitem [\protect \citeauthoryear {%
Snee%
\ \BBA {} Zoback%
}{%
Snee%
\ \BBA {} Zoback%
}{%
{\protect \APACyear {2022}}%
}]{%
snee2022state}
\APACinsertmetastar {%
snee2022state}%
\begin{APACrefauthors}%
Snee, J\BHBI E\BPBI L.%
\BCBT {}\ \BBA {} Zoback, M\BPBI D.%
\end{APACrefauthors}%
\unskip\
\newblock
\APACrefYearMonthDay{2022}{}{}.
\newblock
{\BBOQ}\APACrefatitle {{State of stress in areas of active unconventional oil and gas development in North America}} {{State of stress in areas of active unconventional oil and gas development in North America}}.{\BBCQ}
\newblock
\APACjournalVolNumPages{AAPG Bulletin}{106}{2}{355--385}.
\PrintBackRefs{\CurrentBib}

\bibitem [\protect \citeauthoryear {%
Stokes%
\ \protect \BOthers {.}}{%
Stokes%
\ \protect \BOthers {.}}{%
{\protect \APACyear {2023}}%
}]{%
stokes2023pore}
\APACinsertmetastar {%
stokes2023pore}%
\begin{APACrefauthors}%
Stokes, S\BPBI M.%
, Ge, S.%
, Brown, M\BPBI R.%
, Menezes, E\BPBI A.%
, Sheehan, A\BPBI F.%
\BCBL {}\ \BBA {} Tiampo, K\BPBI F.%
\end{APACrefauthors}%
\unskip\
\newblock
\APACrefYearMonthDay{2023}{}{}.
\newblock
{\BBOQ}\APACrefatitle {Pore Pressure Diffusion and Onset of Induced Seismicity} {Pore pressure diffusion and onset of induced seismicity}.{\BBCQ}
\newblock
\APACjournalVolNumPages{Journal of Geophysical Research: Solid Earth}{128}{3}{e2022JB026012}.
\PrintBackRefs{\CurrentBib}

\bibitem [\protect \citeauthoryear {%
Tariq%
}{%
Tariq%
}{%
{\protect \APACyear {1987}}%
}]{%
tariq1987evaluation}
\APACinsertmetastar {%
tariq1987evaluation}%
\begin{APACrefauthors}%
Tariq, S\BPBI M.%
\end{APACrefauthors}%
\unskip\
\newblock
\APACrefYearMonthDay{1987}{}{}.
\newblock
{\BBOQ}\APACrefatitle {Evaluation of flow characteristics of perforations including nonlinear effects with the finite-element method} {Evaluation of flow characteristics of perforations including nonlinear effects with the finite-element method}.{\BBCQ}
\newblock
\APACjournalVolNumPages{SPE Production Engineering}{2}{02}{104--112}.
\PrintBackRefs{\CurrentBib}

\bibitem [\protect \citeauthoryear {%
Terzaghi%
, Peck%
\BCBL {}\ \BBA {} Mesri%
}{%
Terzaghi%
\ \protect \BOthers {.}}{%
{\protect \APACyear {1996}}%
}]{%
terzaghi1996soil}
\APACinsertmetastar {%
terzaghi1996soil}%
\begin{APACrefauthors}%
Terzaghi, K.%
, Peck, R\BPBI B.%
\BCBL {}\ \BBA {} Mesri, G.%
\end{APACrefauthors}%
\unskip\
\newblock
\APACrefYear{1996}.
\newblock
\APACrefbtitle {Soil mechanics in engineering practice} {Soil mechanics in engineering practice}.
\newblock
\APACaddressPublisher{}{John wiley \& sons}.
\PrintBackRefs{\CurrentBib}

\bibitem [\protect \citeauthoryear {%
Toda%
, Stein%
\BCBL {}\ \BBA {} Sagiya%
}{%
Toda%
\ \protect \BOthers {.}}{%
{\protect \APACyear {2002}}%
}]{%
toda2002evidence}
\APACinsertmetastar {%
toda2002evidence}%
\begin{APACrefauthors}%
Toda, S.%
, Stein, R\BPBI S.%
\BCBL {}\ \BBA {} Sagiya, T.%
\end{APACrefauthors}%
\unskip\
\newblock
\APACrefYearMonthDay{2002}{}{}.
\newblock
{\BBOQ}\APACrefatitle {{Evidence from the AD 2000 Izu islands earthquake swarm that stressing rate governs seismicity}} {{Evidence from the AD 2000 Izu islands earthquake swarm that stressing rate governs seismicity}}.{\BBCQ}
\newblock
\APACjournalVolNumPages{Nature}{419}{6902}{58--61}.
\PrintBackRefs{\CurrentBib}

\bibitem [\protect \citeauthoryear {%
Townend%
\ \BBA {} Zoback%
}{%
Townend%
\ \BBA {} Zoback%
}{%
{\protect \APACyear {2000}}%
}]{%
townend2000faulting}
\APACinsertmetastar {%
townend2000faulting}%
\begin{APACrefauthors}%
Townend, J.%
\BCBT {}\ \BBA {} Zoback, M\BPBI D.%
\end{APACrefauthors}%
\unskip\
\newblock
\APACrefYearMonthDay{2000}{}{}.
\newblock
{\BBOQ}\APACrefatitle {How faulting keeps the crust strong} {How faulting keeps the crust strong}.{\BBCQ}
\newblock
\APACjournalVolNumPages{Geology}{28}{5}{399--402}.
\PrintBackRefs{\CurrentBib}

\bibitem [\protect \citeauthoryear {%
Van~der Elst%
, Page%
, Weiser%
, Goebel%
\BCBL {}\ \BBA {} Hosseini%
}{%
Van~der Elst%
\ \protect \BOthers {.}}{%
{\protect \APACyear {2016}}%
}]{%
van2016induced}
\APACinsertmetastar {%
van2016induced}%
\begin{APACrefauthors}%
Van~der Elst, N\BPBI J.%
, Page, M\BPBI T.%
, Weiser, D\BPBI A.%
, Goebel, T\BPBI H.%
\BCBL {}\ \BBA {} Hosseini, S\BPBI M.%
\end{APACrefauthors}%
\unskip\
\newblock
\APACrefYearMonthDay{2016}{}{}.
\newblock
{\BBOQ}\APACrefatitle {Induced earthquake magnitudes are as large as (statistically) expected} {Induced earthquake magnitudes are as large as (statistically) expected}.{\BBCQ}
\newblock
\APACjournalVolNumPages{Journal of Geophysical Research: Solid Earth}{121}{6}{4575--4590}.
\PrintBackRefs{\CurrentBib}

\bibitem [\protect \citeauthoryear {%
van Thienen-Visser%
\ \BBA {} Breunese%
}{%
van Thienen-Visser%
\ \BBA {} Breunese%
}{%
{\protect \APACyear {2015}}%
}]{%
van2015induced}
\APACinsertmetastar {%
van2015induced}%
\begin{APACrefauthors}%
van Thienen-Visser, K.%
\BCBT {}\ \BBA {} Breunese, J.%
\end{APACrefauthors}%
\unskip\
\newblock
\APACrefYearMonthDay{2015}{}{}.
\newblock
{\BBOQ}\APACrefatitle {{Induced seismicity of the Groningen gas field: History and recent developments}} {{Induced seismicity of the Groningen gas field: History and recent developments}}.{\BBCQ}
\newblock
\APACjournalVolNumPages{The Leading Edge}{34}{6}{664--671}.
\PrintBackRefs{\CurrentBib}

\bibitem [\protect \citeauthoryear {%
Walsh~III%
\ \BBA {} Zoback%
}{%
Walsh~III%
\ \BBA {} Zoback%
}{%
{\protect \APACyear {2015}}%
}]{%
walsh2015oklahoma}
\APACinsertmetastar {%
walsh2015oklahoma}%
\begin{APACrefauthors}%
Walsh~III, F\BPBI R.%
\BCBT {}\ \BBA {} Zoback, M\BPBI D.%
\end{APACrefauthors}%
\unskip\
\newblock
\APACrefYearMonthDay{2015}{}{}.
\newblock
{\BBOQ}\APACrefatitle {{Oklahoma’s recent earthquakes and saltwater disposal}} {{Oklahoma’s recent earthquakes and saltwater disposal}}.{\BBCQ}
\newblock
\APACjournalVolNumPages{Science advances}{1}{5}{e1500195}.
\PrintBackRefs{\CurrentBib}

\bibitem [\protect \citeauthoryear {%
H.~Wang%
}{%
H.~Wang%
}{%
{\protect \APACyear {2000}}%
}]{%
wang00a}
\APACinsertmetastar {%
wang00a}%
\begin{APACrefauthors}%
Wang, H.%
\end{APACrefauthors}%
\unskip\
\newblock
\APACrefYear{2000}.
\newblock
\APACrefbtitle {Theory of Linear Poroelasticity: With Applications to Geomechanics and Hydrogeology} {Theory of linear poroelasticity: With applications to geomechanics and hydrogeology}.
\newblock
\APACaddressPublisher{Princeton, New Jersey}{287 pp., Princeton Univ. Press}.
\PrintBackRefs{\CurrentBib}

\bibitem [\protect \citeauthoryear {%
R.~Wang%
\ \protect \BOthers {.}}{%
R.~Wang%
\ \protect \BOthers {.}}{%
{\protect \APACyear {2020}}%
}]{%
wang2020injection}
\APACinsertmetastar {%
wang2020injection}%
\begin{APACrefauthors}%
Wang, R.%
, Schmandt, B.%
, Zhang, M.%
, Glasgow, M.%
, Kiser, E.%
, Rysanek, S.%
\BCBL {}\ \BBA {} Stairs, R.%
\end{APACrefauthors}%
\unskip\
\newblock
\APACrefYearMonthDay{2020}{}{}.
\newblock
{\BBOQ}\APACrefatitle {{Injection-induced earthquakes on complex fault zones of the Raton Basin illuminated by machine-learning phase picker and dense nodal array}} {{Injection-induced earthquakes on complex fault zones of the Raton Basin illuminated by machine-learning phase picker and dense nodal array}}.{\BBCQ}
\newblock
\APACjournalVolNumPages{Geophysical Research Letters}{47}{14}{e2020GL088168}.
\PrintBackRefs{\CurrentBib}

\bibitem [\protect \citeauthoryear {%
W.~Wang%
\ \protect \BOthers {.}}{%
W.~Wang%
\ \protect \BOthers {.}}{%
{\protect \APACyear {2022}}%
}]{%
wang2022tidal}
\APACinsertmetastar {%
wang2022tidal}%
\begin{APACrefauthors}%
Wang, W.%
, Shearer, P\BPBI M.%
, Vidale, J\BPBI E.%
, Xu, X.%
, Trugman, D\BPBI T.%
\BCBL {}\ \BBA {} Fialko, Y.%
\end{APACrefauthors}%
\unskip\
\newblock
\APACrefYearMonthDay{2022}{}{}.
\newblock
{\BBOQ}\APACrefatitle {Tidal modulation of seismicity at the Coso geothermal field} {Tidal modulation of seismicity at the coso geothermal field}.{\BBCQ}
\newblock
\APACjournalVolNumPages{Earth and Planetary Science Letters}{579}{}{117335}.
\PrintBackRefs{\CurrentBib}

\bibitem [\protect \citeauthoryear {%
Weingarten%
, Ge%
, Godt%
, Bekins%
\BCBL {}\ \BBA {} Rubinstein%
}{%
Weingarten%
\ \protect \BOthers {.}}{%
{\protect \APACyear {2015}}%
}]{%
weingarten2015high}
\APACinsertmetastar {%
weingarten2015high}%
\begin{APACrefauthors}%
Weingarten, M.%
, Ge, S.%
, Godt, J\BPBI W.%
, Bekins, B\BPBI A.%
\BCBL {}\ \BBA {} Rubinstein, J\BPBI L.%
\end{APACrefauthors}%
\unskip\
\newblock
\APACrefYearMonthDay{2015}{}{}.
\newblock
{\BBOQ}\APACrefatitle {High-rate injection is associated with the increase in US mid-continent seismicity} {High-rate injection is associated with the increase in us mid-continent seismicity}.{\BBCQ}
\newblock
\APACjournalVolNumPages{Science}{348}{6241}{1336--1340}.
\PrintBackRefs{\CurrentBib}

\bibitem [\protect \citeauthoryear {%
White%
\ \BBA {} Foxall%
}{%
White%
\ \BBA {} Foxall%
}{%
{\protect \APACyear {2016}}%
}]{%
white2016assessing}
\APACinsertmetastar {%
white2016assessing}%
\begin{APACrefauthors}%
White, J\BPBI A.%
\BCBT {}\ \BBA {} Foxall, W.%
\end{APACrefauthors}%
\unskip\
\newblock
\APACrefYearMonthDay{2016}{}{}.
\newblock
{\BBOQ}\APACrefatitle {Assessing induced seismicity risk at CO2 storage projects: Recent progress and remaining challenges} {Assessing induced seismicity risk at co2 storage projects: Recent progress and remaining challenges}.{\BBCQ}
\newblock
\APACjournalVolNumPages{International Journal of Greenhouse Gas Control}{49}{}{413--424}.
\PrintBackRefs{\CurrentBib}

\bibitem [\protect \citeauthoryear {%
Wu%
}{%
Wu%
}{%
{\protect \APACyear {1976}}%
}]{%
wu1976soil}
\APACinsertmetastar {%
wu1976soil}%
\begin{APACrefauthors}%
Wu, T\BPBI H.%
\end{APACrefauthors}%
\unskip\
\newblock
\APACrefYearMonthDay{1976}{}{}.
\newblock
{\BBOQ}\APACrefatitle {Soil mechanics} {Soil mechanics}.{\BBCQ}
\newblock
\APACjournalVolNumPages{Publication of: Allyn and Bacon, Incorporated}{}{}{}.
\PrintBackRefs{\CurrentBib}

\bibitem [\protect \citeauthoryear {%
Zbinden%
, Rinaldi%
, Urpi%
\BCBL {}\ \BBA {} Wiemer%
}{%
Zbinden%
\ \protect \BOthers {.}}{%
{\protect \APACyear {2017}}%
}]{%
zbinden2017physics}
\APACinsertmetastar {%
zbinden2017physics}%
\begin{APACrefauthors}%
Zbinden, D.%
, Rinaldi, A\BPBI P.%
, Urpi, L.%
\BCBL {}\ \BBA {} Wiemer, S.%
\end{APACrefauthors}%
\unskip\
\newblock
\APACrefYearMonthDay{2017}{}{}.
\newblock
{\BBOQ}\APACrefatitle {On the physics-based processes behind production-induced seismicity in natural gas fields} {On the physics-based processes behind production-induced seismicity in natural gas fields}.{\BBCQ}
\newblock
\APACjournalVolNumPages{Journal of Geophysical Research: Solid Earth}{122}{5}{3792--3812}.
\PrintBackRefs{\CurrentBib}

\bibitem [\protect \citeauthoryear {%
Zhai%
, Shirzaei%
, Manga%
\BCBL {}\ \BBA {} Chen%
}{%
Zhai%
\ \protect \BOthers {.}}{%
{\protect \APACyear {2019}}%
}]{%
zhai2019pore}
\APACinsertmetastar {%
zhai2019pore}%
\begin{APACrefauthors}%
Zhai, G.%
, Shirzaei, M.%
, Manga, M.%
\BCBL {}\ \BBA {} Chen, X.%
\end{APACrefauthors}%
\unskip\
\newblock
\APACrefYearMonthDay{2019}{}{}.
\newblock
{\BBOQ}\APACrefatitle {{Pore-pressure diffusion, enhanced by poroelastic stresses, controls induced seismicity in Oklahoma}} {{Pore-pressure diffusion, enhanced by poroelastic stresses, controls induced seismicity in Oklahoma}}.{\BBCQ}
\newblock
\APACjournalVolNumPages{Proceedings of the National Academy of Sciences}{116}{33}{16228--16233}.
\PrintBackRefs{\CurrentBib}

\end{thebibliography}
\newcommand{\noopsort}[1]{}

\section{Open Research}
Data of Abaqus files, post-processing scripts, SI model scripts, optimization methodology scripts, and figure generation scripts are available online at Hill, R. \citeyear{HillZenodoRaton} (\url{https://doi.org/10.5281/zenodo.10472485}).
\section{Supplementary}
\subsection{Data}
Wastewater injection well data for Las Animas County, Colorado was retrieved from Colorado Oil and Gas Corporation Commission Website (\url{https://ecmc.state.co.us/#/home}), (Accessed: 20223-10-10). Wastewater injection well data for Colfax County, New Mexico was retrieved from New Mexico Oil Conservation Division Permitting Website (\url{https://wwwapps.emnrd.nm.gov/OCD/OCDPermitting/Data/Wells.aspx}) (Accessed: Accessed: 20223-10-10). In this study we convert injection well data from bbl/month to m$^3$/day across 29 wells from Nomber 1994 to May 2022 (See Supplementary Data).

Multiple seismic studies have taken place in the Raton Basin. We leverage these combined data sets to form a comprehensive catalog of earthquakes up to July-2020. Earthquakes from 1963–2013 are given from Rubinstein et al., 2014, which include recorded earthquakes by the USGS temporary seismic networks from 2001-2011 \cite{rubinstein20142001}. Earthquakes from 2008-2010 were recorded by the EarthScope Transportable Array \cite{nakai2017seismicity}. Past 2013, we rely on cataloged earthquakes from USGS National Earthquake Information Center (NEIC). Furthermore, from July 2016 to July 2020 Earthquakes are provided from a combined broadband seismometer and geohpone node study available from the International Seismological Centre \cite{glasgow2021raton,Glasgow2021}.
\subsection{Step Rate Tests}
\begin{figure}[h]
    \centering
    \includegraphics[scale=.30]{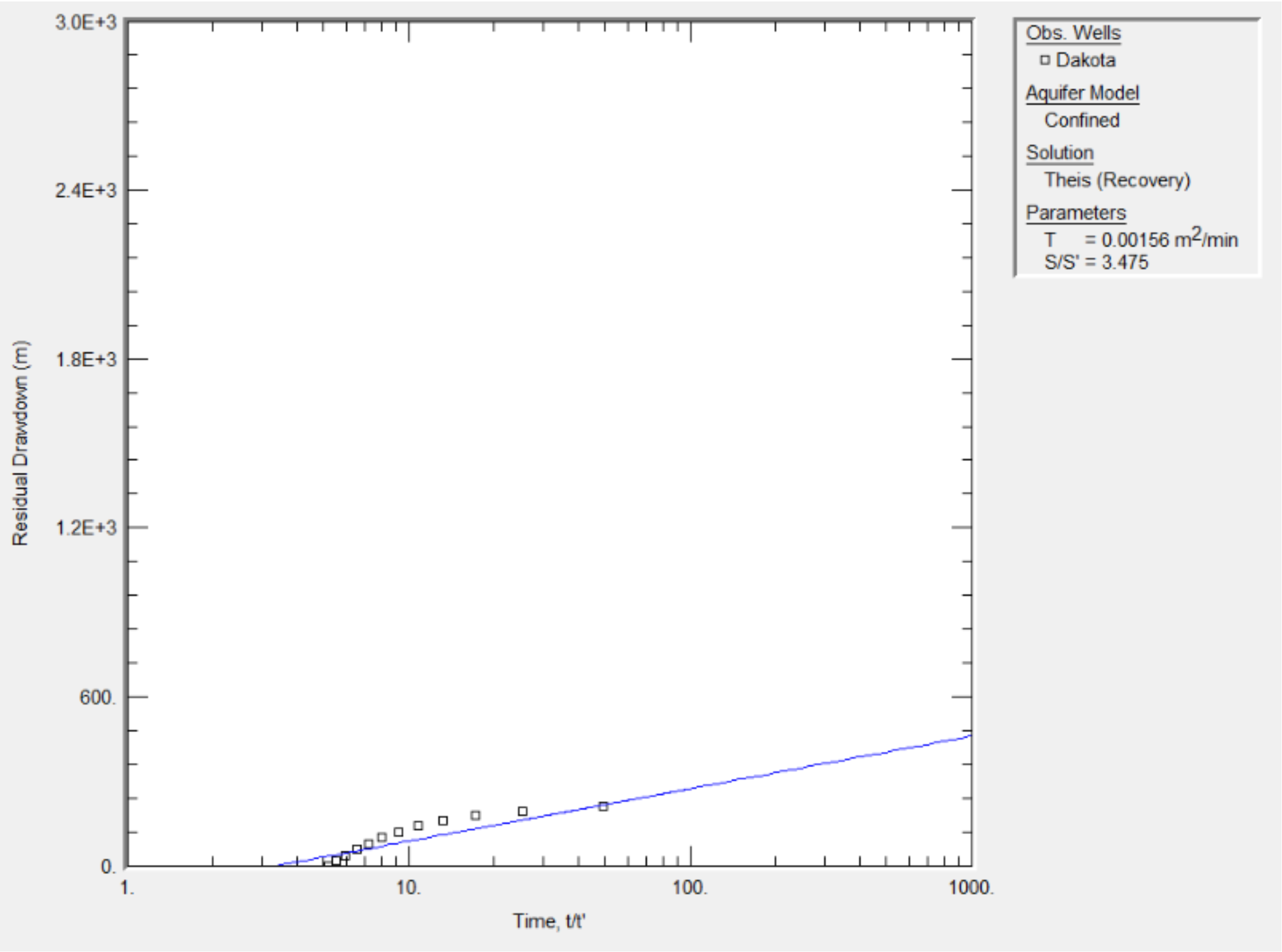}
    \caption{\textbf{Dakota Formation (Recovery).} Dakota formation residual drawdown over $log(t/t')$ calculated fit from AQTESOLV. Transmissivites were converted to permeability (Table 1).}
    \label{fig:dakRecover}
\end{figure}

\begin{figure}[h]
    \centering
    \includegraphics[scale=.30]{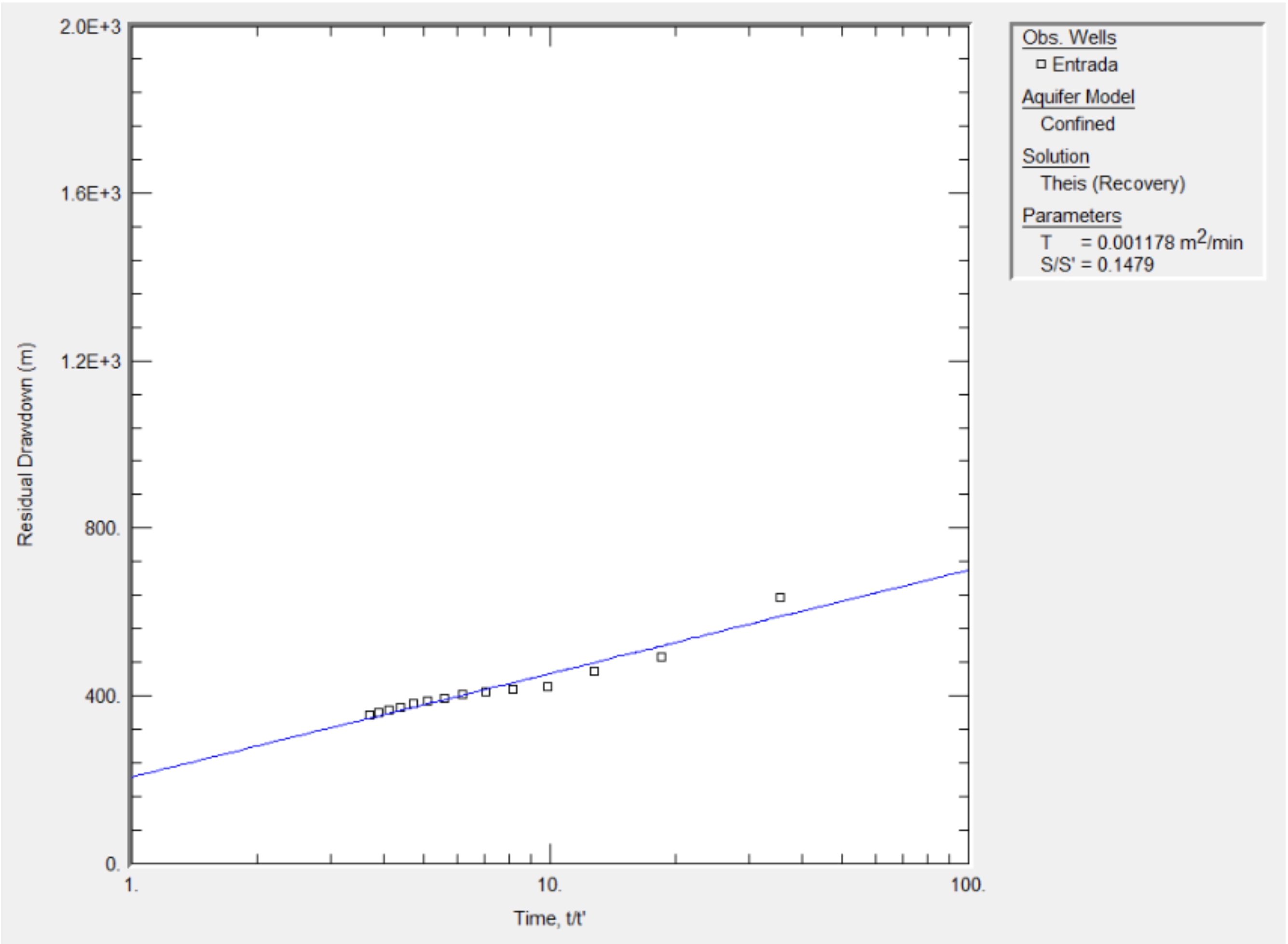}
    \caption{\textbf{Entrada Formation (Recovery).} Entrada formation residual drawdown over $log(t/t')$ calculated fit from AQTESOLV. Transmissivites were converted to permeability (Table 1).}
    \label{fig:entRecover}
\end{figure}

Prior work calibrated reservoir permeability in the main injection reservoirs, the Dakota Formation and the Entrada Formation, from injection-recovery step rate tests \cite{hernandez2019step}. A step rate test determines how pressures within a formation change as a result of small-scale injection. The pressure changes can be converted to input parameters for AQTESOLV which utilizes a Theis step-drawdown test to approximate hydraulic properties \cite{aqtesolv}. SM Table 1 provides the permeability values obtained from AQTESOLV for different cases: case 1 considered the lowest values of psi from each step, case 2 considered the highest values of psi, case 3 used an incremental increase per minute, and case 4 was simply the recovery data. Case 4 provided the lowest mean residual for both reservoirs (SM Figure \ref{fig:dakRecover}-\ref{fig:entRecover}) and was chosen as the preferred permeability for the model. We include plots from within AQTESOLV of the data and transmissivity solution The calculated permeability are within those reported by previous studies \cite{belitz1988hydrodynamics,nakai2017possible}.
\subsection{Simplified Optimization Example (no SI map required)}\label{sec:simple}
There is strong evidence to suggest that stressing rate and accumulated stress, the latter which is related to the total injected volume ($\int_V\Delta \tau(P)\sim \Delta V$) \cite{van2016induced}, are key factors that influence the occurrence of induced seismicity \cite{mcgarr2014maximum,weingarten2015high,toda2002evidence,qin2022forecasting}. As an example, the following management model uses the prior total Coulomb stress $\tau$ and Coulomb stress rate $\dot{\tau}$ at locations in the Raton Basin that were associated with injection induced M$\geq$ 4+ events. We make the assumption that all former M$\geq$ 4+ events occurred at the mean seismogenic depth where model results are output. This management model can be thought of as a retroactive example since we exclusively let the previous stress conditions of past large earthquakes inform the management model solution. Therefore, this method does not require an SI map to forecast the hazard, although the solution to the injection rates $q$ can be used to forward solve the hazard if desired. The following steps describe the methodology, generalized for application to other studies:
\begin{enumerate}
    \item Resolve the stress and pore pressure spatiotemporal evolution from the numerical domain based on the full well injection history.
    \item Record the $\tau$ and $\dot{\tau}$ at each M$\geq$4+ earthquake location in the numerical domain during the time step it occurred. These will provide the constraints for the $x_{\tau}$ and $x_{\dot{\tau}}$ respectively.
    \item Generate response matrix $R_{\tau}$ and $R_{\dot{\tau}}$ (See Appendix) for both $\tau$ and $\dot{\tau}$ then stack them vertically; This requires running Q individual models based on Q wells for the length of the management period desired.
    \item Solve the linear program management model: $\begin{bmatrix} 
    R_{\tau} \\
    R_{\dot{\tau}}
    \end{bmatrix}q\leq \begin{bmatrix} 
    x_{\tau} \\
    x_{\dot{\tau}}
    \end{bmatrix}$
\end{enumerate}

\begin{figure}[h]
    \centering
    \includegraphics[scale=.30]{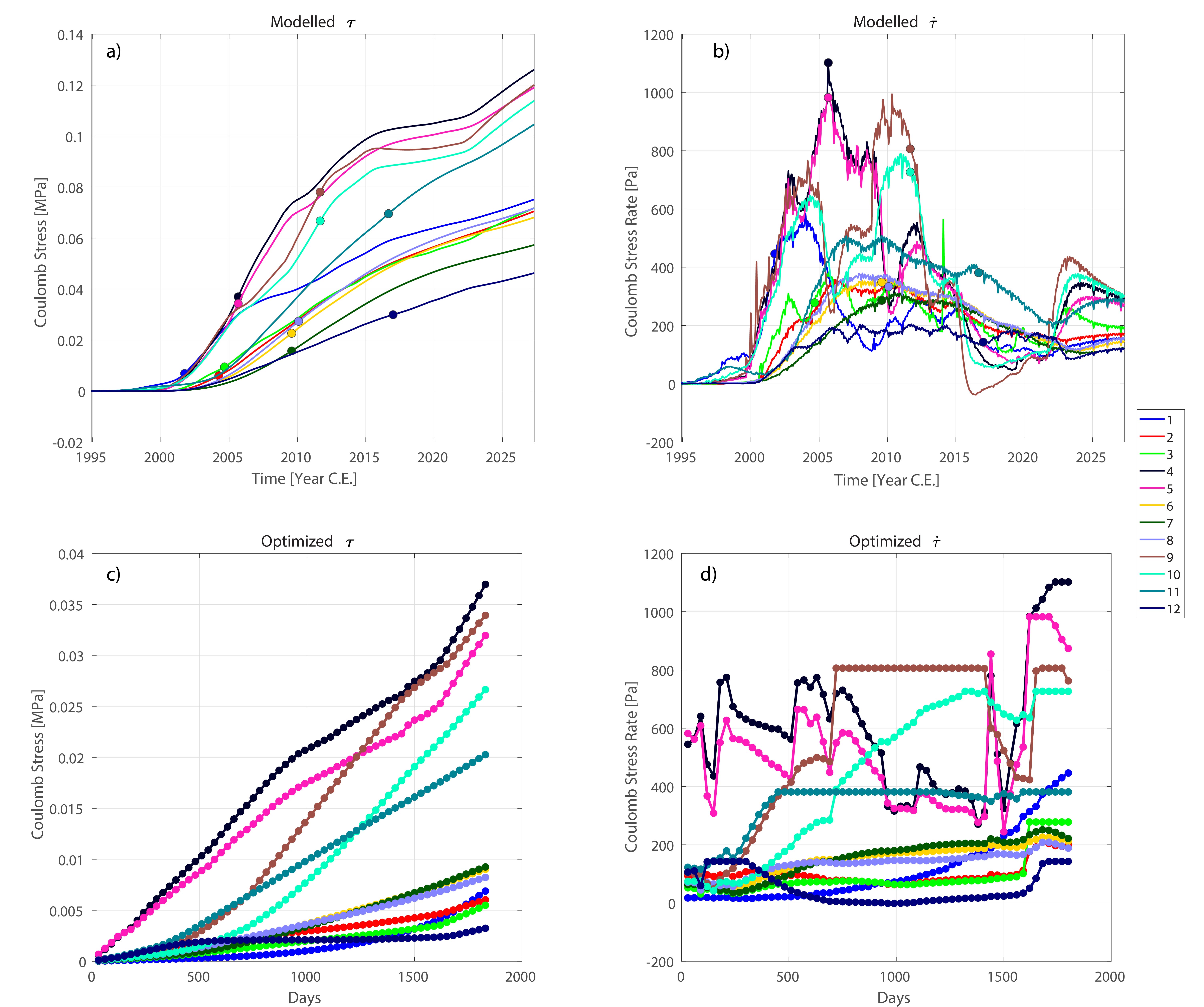}
    \caption{\textbf{Simple Optimization Example (12 points)} Panel a) is the modelled Coulomb stress at the 12 model output points with dots at the time of previously recorded M$\geq$4+ earthquakes at that location. Panel b) is the modelled Coulomb stress rate at the 12 model output points with dots at the time of the M$\geq$4+ earthquake at that location. Notice how the Coulomb stress rate at the model output points 4,5,8,6,7,10 coincide when rates were peaking indicating good, and entirely independent, agreement between Coulomb stress rate and timing of seismicity. Panel c) is the optimized Coulomb stress which is considerably lower than the modelled stress. Panel d) is the optimized Coulomb stress rate. Notice how some locations clearly reach the maximum allowed rate for some time steps. Individual model output locations compared to the overall and rate constraints through time are provided in the Supplementary (SM Figure \ref{fig:sm12}).}
    \label{fig:12p1234}
\end{figure}

SM Figure \ref{fig:12p1234} describes the derived constraints at each of the earthquake locations and the resulting optimization of the $\tau$ and $\dot{\tau}$ at each of the locations during the management period. Note that the total Coulomb stress and Coulomb stressing rate thresholds are never exceeded. The cumulative injection rate is also reduced.
Another important feature of the optimization is the shape of the $\tau$ and $\dot{\tau}$ at each of the locations during the management period. Notice that $\tau$ steadily increases and that $\dot{\tau}$ increases near the end. The optimization only considers the 5 year management period, and therefore does not consider what ramping the injection rates and subsequent $\tau$ and $\dot{\tau}$ near the end of the management period would do for the months following the management period. We present the solution this way to introduce the response matrix method and reveal the inherent flaws in the optimization since this exact solution would not be ideal for practical use. However, there are a variety of solutions that makes use of mixed-integer programming to control the behavior of the injection wells to avoid this type of solution which we elaborate on in the main text and incorporate for prospective case `Reduction'.

\subsection{Mixed-Integer Programming and Additional Constraints}\label{sec:mixedSM}
Monotonic decreasing/increasing is an injection scenario by which the injection for all the wells is only ever decreasing/increasing and never increasing/decreasing. The construction of the mixed-integer $R^*$ matrix for a monotonically decreasing scenario is simple. If we consider $q_{jk}$ to represent the injection rate for well $j$ at managment period $k$, then for all $k$ the constraint $q_{j,k+1}\leq q_{j,k}$ must be satisfied for monotonically decreasing rates. To ensure that this constraint is met $x^*$ must equal a column vector of zeros with length $m$, and the integer matrix $R^*$ would contain $-1$s across the diagonal and $1$s offset from the diagonal by the number of wells. Similarly, for the monotonic increasing scenario the constraint that $q_{j,k}\leq q_{j,k+1}$ must be satisfied. To achieve this the integer matrix $R^*$ would contain $1$s along the diagonal and $-1$s offset from the diagonal by the number of wells, but with the important inclusion that the diagonals associated with the last time step at all well locations is 0 because otherwise $q_{j,k}\leq 0$ which would result in zero injection rates for all time. We include the monotonically decreasing constraint for Prospectice case `Reduction'.

Running average constrains the injection rates to to be equivalent to an average over $t$ management periods such that the constraint $\frac{q_{j,k+1}+q_{j,k+2}+\cdots q_{j,k+t}}{t} \leq q_{j,k}$ is satisfied. The running average is useful if smoothing of injection rates through time is desired. The mixed integer construction still results in a column vector of zeros with length $m$ for $x^*$. The integer matrix $R^*$ therefore contains diagonal integer values equivalent to $-t$ and $t$ $1$s offset from the diagonal by the number of wells times t.

Exclusion of certain wells is another constraint that is necessary for typical injection management practices. The construction of the integer matrix $R^*$ is similar to the monotonic scenario. In order to satisfy the constraint for specific wells such that $q_{j,k} \leq 0 $ wells at specified management periods in the $R^*$ matrix are represented with 1s since Eq. (\ref{eq:linprog}) limits the injection rate $q$ as nonnegative. The combination of monotonic, running average, and exclusion of wells allows for a wide variety of variable injection scenarios that are all possible to optimize for.

Furthermore, uncertainty in the simulation model is also possible to incorporate into the management solution. While not included in this study, the concept is similar to the previous management model controls. For example, in our model of the Raton Basin, if there was significant uncertainty in the fault permeability structure we could recreate an entirely new response matrix based on an altered simulation model where the fault zone permeability in the model was changed. This would require 29 (each well) different unit-source solutions ie. model runs. The newly formed response matrix is appended with the primary response matrix and also the constraint vector is appended. The linear program will find an optimal solution again, but with the inclusion that the uncertainty in permeability is accounted for. Uncertainties in any of the material parameters is accountable for different model realizations which are `stackable' \emph{ad infinitum}.  It is important to note that solving the linear program in this way means that the solution finds the optimal injection solution \emph{to} the uncertainty instead of \emph{with} uncertainty. The only 'free' uncertainty that does not require additional simulation model realizations is that of the fault geometry. Additional $\tau$ response matrices are calculable for different receiver fault geometries and concatenated in the same way as any other uncertainties.
\subsection{Iterative Method}\label{sec:iteration}
\begin{figure}[h]
    \centering
    \includegraphics[scale=.40]{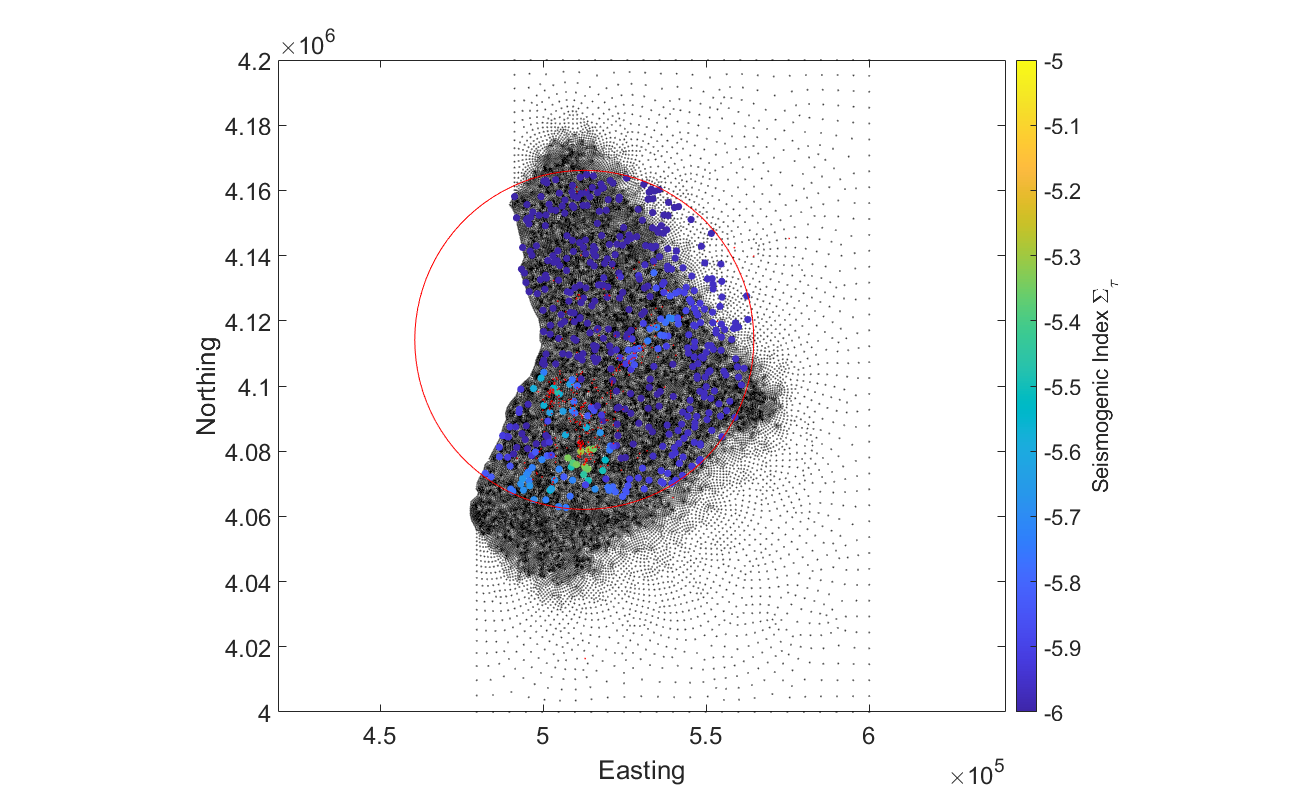}
    \caption{Uniform random distribution of points (500) used for management model 2 examples. The red dots represent the earthquakes in the basin with M$\geq$2.5. The red circle represents the subset of the model points used such that all seismicity is within it ensuring that the random points chosen for the initialization of the optimization are not irrelevant.}
    \label{fig:urandpoints}
\end{figure}

The iteration technique is designed to slowly adjust the rate constraints at the subset of model output locations such that the forward solution of the constraints and subsequent seismicity rate and seismic hazard across the entire basin arrives at the desired threshold. The technique is not exhaustive or optimized, but was found to work adequately for our efforts.
\begin{enumerate}
    \item Given the forward solution of rate constraints $x$ from the optimized injection rates $q$ resolve the total seismicity rate and subsequent hazard across the basin. If within the tolerance of the desired threshold finish the iteration. If not within the tolerance of the desired threshold continue to the next step.
    \item Find the locations $l$ used in the optimization (ie. the 500 subset of points used in optimization (SM Figure \ref{fig:urandpoints})) that for all time during management period (ie. 5 years) reached their constraints, even one time step. 
    \item For the specific locations $l$, increase their constraints (for all time) by a small amount. That is to say use a multiplier that increases the constraint. The amount is based on how far way from the desired solution the current total probability is. If close, then the scaling is low, but if far the scaling can be larger if desired by the user. Otherwise if the probability is too high reduce all constraints by an adjustable percentage.
    \item Solve the optimization again with the adjusted constraints which will produce a new $q$ array.
    \item Forward solve a solution for the rate constraints $x$ given the new $q$.
    \item If you are incorporating previous remnant stress fields, add those stress rates to $x$ now. This is for Prospective Case \#2.
    \item Return to step 1.
\end{enumerate}
\subsection{Abaqus Soil Mechanics}
Prior work has shown that the soil mechanics framework of Abaqus is a limiting-case solution of a fully coupled poromechanical framework- a generalization of Biot poroelasticity \cite{biot_general_1941,jin2023saturated,jin20233d}. Here, we summarize the underpinnings of the Abaqus framework and how, in a limiting-case, it is equivalent to that of a linear poroelastic framework. We further include numerical sensitivity test compared to well known analytical solutions to confirm the robustness of the software for our study region.

Abaqus actually provides a nonlinear unsaturated soil mechanics framework for a multi-phase material (solid grains, wetting fluid, and air). Futhermore, Abaqus considers fluid entrapment and thermal expansion of the fluid and solid grains. Abaqus adopts an effective stress principle defined by (equation 2.8.1-1 Abaqus Theory Manual)
\begin{equation}\label{stress}
         \boldsymbol{\Bar{\sigma}} = \boldsymbol{\sigma}  + (\chi u_w + (1-\chi )u_a)\boldsymbol{I}
\end{equation}
Where $\boldsymbol{\Bar{\sigma}}$ is the effective stress, $\boldsymbol{\sigma}$ is the total stress, $u_w$ is the average pressure stress of the wetting liqud, $u_a$ is the the average pressure stress in the other liquid (air), and $\chi$ is a factor that depends on saturation and on the surface tension of the liquid/solid system \cite{wu1976soil}.

The force equilibrium in Abaqus is then defined by (equation 2.8.2-1 Abaqus Theory Manual):
\begin{equation}\label{forceEq}
    \int_V \boldsymbol{\sigma}:\delta \epsilon \,dV = \int_S \boldsymbol{t} \cdot \delta \boldsymbol{v} \,dS + \int_V \boldsymbol{f} \cdot \delta \boldsymbol{v} \,dV + \int_V (n_w+n_t)\rho_w \boldsymbol{g}\cdot \delta \boldsymbol{v} \,dV
\end{equation}
Where $\delta \boldsymbol{v}$ is a virtual velocity field, $\delta \epsilon = sym(\partial \delta \boldsymbol{v} / \partial \boldsymbol{x})$ is the virtual rate of deformation, $\boldsymbol{t}$ are the surface tractions per unit area, $\boldsymbol{f}$ are all the body forces besides the wetting fluid, $s$ is the saturation of wetting fluid and $n$ the porosity with $sn=n_w$, $n_w+n_t$ is the total volume of the wetting liquid (free $+$ trapped) per unit of current volume, $\rho_w$ is the density of the wetting fluid, and $\boldsymbol{g}$ is gravitational acceleration. This formulation in weak form is expressed by writing the principle of virtual work for the volume under consideration in its current configuration at time t.

A second conservation law for the fluid mass is required and Abaqus defines the continuity statement for the wetting liquid phase in a porous medium (in similar weak form as the force equilibrium) by (Section 2.8.4 Abaqus Theory Manual):
\begin{equation}\label{fluid}
    \int_V \delta u_w \frac{1}{J}\frac{d}{dt}(J \rho_w(n_w+n_t)) \,dV + \int_V \delta u_w \frac{\partial}{\partial \boldsymbol{x}} \cdot (\rho_w n_w \boldsymbol{v}_w) \,dV = 0
\end{equation}
where $\delta u_w$ is an arbitrary, continuous, variational field, and $J = |{\frac{dV}{dV^0}}|$ is the ratio of the medium's volume in the current configuration to its volume in the reference configuration, and $\boldsymbol{v}_w$ is the average velocity of the wetting liquid.

Now, several constitutive relationships are defined by Abaqus to describe the mechanical behavior of the porous medium. First, the liquid response is defined by (2.8.3-1):
\begin{equation}
    \frac{\rho_w}{\rho^0_w} \approx 1 + \frac{u_w}{K_w} - \epsilon^{th}_w
\end{equation}
where $\rho^0_w$ is the density of wetting liquid in the reference configuration, $K_w$ is the liquid's bulk modulus and  $\epsilon^{th}_w$ is the volumetric expansion of the liquid caused by temperature change. Similarly, the grain's response is defined by (2.8.3-2):
\begin{equation}
    \frac{\rho_g}{\rho^0_g} \approx 1 + \frac{1}{K_g}(su_w + \frac{\Bar{p}}{1-n-n_t})- \epsilon^{th}_g
\end{equation}
where $\rho_g$ is the density of the grains, $K_g$ is the bulk modulus of the solid matter, and $\epsilon^{th}_g$ is the volumetric thermal strain.

The effective strain is also introduced for solid grains defined by (2.8.3-6):
\begin{equation}
    \Bar{\epsilon} = \epsilon + (\frac{1}{3}(\frac{s u_w}{K_g}-\epsilon_g^{th})-\frac{1}{3} \ln(1+Jn_t-n^0_t))\boldsymbol{I}-\epsilon^{ms}(s)
\end{equation}
Where $\epsilon^{ms}(s)$ is a saturation driven moisture swelling strain that represents the volumetric swelling of the solid skeleton in partially saturated flow conditions. Note, in Abaqus, it is this effective strain that modifies the effective stress:
\begin{equation}\label{hookEffective}
    \Bar{\boldsymbol{\sigma}} = \mathbb{C}^e : \Bar{\epsilon}
\end{equation}

The final constitutive relationship is that of the pore fluid flow that is governed by Forchheimer's law or Darcy's law (linear version of Forchheimer's law). Abaqus defines Forchheimer's law by:
\begin{equation}
n_w\boldsymbol{v}_w(1+\beta\sqrt{\boldsymbol{v}_w \cdot \boldsymbol{v}_w}) = -\boldsymbol{\hat{k}} \cdot \frac{\partial \phi}{\partial \boldsymbol{x}}
\end{equation}
where $\boldsymbol{\hat{k}}$ is the `hydraulic conductivity' of the medium, $\phi = z + \frac{u_w}{g \rho_w}$ is the piezometric head, and $\beta$ is a velocity coefficient \cite{tariq1987evaluation}. We see that, as the fluid velocity tends to zero (most geomechanical problems), Forchheimer's law approaches Darcy's law and is equivalent when $\beta=0$.

Now, when we consider no fluid entrapment, isothermal conditions, full saturation, and that the solid skeleton is linearly elastic the Abaqus sets of equations reduce to a familar and equivalent form of typical isothermal, single-phase fluid, fully saturated fully coupled linear poroelasticity. Our limiting-case assumptions reduce the effective stress formulation of equation \ref{stress} to:
\begin{equation}\label{stressReduce}
    \boldsymbol{\Bar{\sigma}} = \boldsymbol{\sigma} + u_w \boldsymbol{I}
\end{equation}
and reduces the effective strain formulation of equation to:
\begin{equation}\label{strainReduce}
    \Bar{\epsilon} = \epsilon + \frac{1}{3}\frac{u_w}{K_g}\boldsymbol{I}
\end{equation}
Taking the strong form of equation \ref{forceEq} and then substituting equation \ref{stressReduce} then \ref{hookEffective} and \ref{strainReduce} yields:
\begin{align}
\nabla \cdot \boldsymbol{\sigma} + (\rho_w n_w + \rho_g (1-n_w))\boldsymbol{g}=0 \\
\nabla \cdot \Bar{\boldsymbol{\sigma}} - \nabla u_w + (\rho_w n_w + \rho_g (1-n_w))\boldsymbol{g}=0 \\
\nabla \cdot \bigr(\mathbb{C}^e:\epsilon + \frac{1}{3} \frac{u_w}{K_g} \mathbb{C}^e:\boldsymbol{I}\bigr) - \nabla u_w + (\rho_w n_w + \rho_g (1-n_w))\boldsymbol{g}=0 \\
\nabla \cdot (\mathbb{C}^e:\epsilon ) + \nabla \cdot \bigr( \frac{K}{K_g}u_w \boldsymbol{I}\bigr) - \nabla u_w + (\rho_w n_w + \rho_g (1-n_w))\boldsymbol{g}=0 \\ 
\nabla \cdot (\mathbb{C}^e:\epsilon ) - \bigr( 1 - \frac{K}{K_g}\bigr)\nabla u_w + (\rho_w n_w + \rho_g (1-n_w))\boldsymbol{g}=0\label{eq:15}
\end{align}
Which in the limit of full saturation and quasi-static particle motion becomes equivalent to a rigorously derived balance of linear momentum for an unsaturated fluid-solid mixture given by \cite<equation 3.14, >{borja2006mechanical}:
\begin{equation} \label{eq:borja}
    \nabla \cdot \boldsymbol{\sigma} + \rho_{w,g} \boldsymbol{g} = 0
\end{equation}
Where $\rho_{w,g}$ is the total mass density of the mixture comprised of the wetting density and solid phase density ie. $\rho_{w,g} = \rho_w n_w + \rho_g(1-n_w)$. The equivalency is clear in the case of effective stress defined as $\boldsymbol{\sigma '} = \boldsymbol{\sigma} + \bigr(1 - \frac{K}{K_g} \bigr)u_w \boldsymbol{I}$ and $\boldsymbol{\sigma '} = \mathbb{C}^e:\epsilon$ which when subsituted into equation \ref{eq:borja} equals equation \ref{eq:15}.

Lastly, Abaqus resolves these equations using the coupled pore fluid diffusion and stress analysis (i.e. the `*SOILS' keyword). The weak forms of the fully coupled governing equations (\ref{stress} and \ref{fluid}) define the state of the porous medium. The force equilibrium and mass conservation equations are approximated with a finite set of equations by introducing interpolation functions and use implicit integration at the end of each time increment. Newton's method is often used for their solution. Since both equations are interpolated simultaneously the solver is fully coupled as well.

The robustness of Abaqus' poromechnics is verified in a variety of ways. One trivial benchmark is the one-dimensional Terzaghi consolidation problem \cite{terzaghi1996soil}. The solution and sample code is provided in the Abaqus Benchmark Manual (see 1.14.1 in Abaqus Benchmark Manual). 
Since our work involves multiple injectors we further validate Abaqus to a well known analytical solution of a point source injection \cite{RUDNICKI1986383}.

Rudnicki (1986) solved the homogeneous diffusion equation based on Biot (1941) and Rice and Cleary (1976) formulation for the case of continuous fluid injection at one point into an infinite linear poroelastic isothermal fully saturated medium. The solutions for pore pressure $P$ and stress $\sigma_{ij}$ are defined by:
\begin{align*}
P(\mathbf{x},t) = \frac{\Phi}{\rho_f c} \frac{1}{4\pi r}[\frac{(\lambda_u - \lambda)(\lambda+2\mu)}{\alpha^2(\lambda_u+2\mu)}]erfc(\frac{1}{2}\xi) \\
\sigma_{ij}(\mathbf{x},t) = \frac{\Phi}{\rho_f c}\frac{(\lambda_u-\lambda)\mu}{4 \pi r \alpha (\lambda_u+2\mu)}\bigg\{ \delta_{ij} [erfc(\frac{1}{2}\xi)-\frac{2}{\xi^2}g(\xi)]+(\frac{x_ix_j}{r^2})[erfc(\frac{1}{2}\xi)+\frac{6}{\xi^2}g(\xi)] \bigg\} \\
\end{align*}
where 
\begin{align*}
    g(\xi) = \frac{1}{2\sqrt{\pi}}\int_0^{\xi}exp(-\frac{1}{4}s^2)ds = erf(\frac{1}{2}\xi)-\frac{\xi}{\sqrt{\pi}}exp(-\frac{1}{4}\xi^2)
\end{align*}
and where $\Phi/\rho_f$ is the injected fluid volume flow rate, r = $\sqrt{(x_k x_k})$, $\xi = \frac{r}{\sqrt{ct}}$,
with diffusivity $c=K(\lambda_u-\lambda)(\lambda+2\mu)/\zeta^2(\lambda_u+2\mu)$, and $\alpha = 1-(\frac{K}{K_g})$ the Biot-Willis coefficient. Where $\lambda,\lambda_u$ is the Lam\'e and undrained Lam\'e parameter respectively, $\mu$ is the shear modulus, $K$ is the hydraulic conductivity, $\rho_f$ is the fluid density.

\begin{figure}[h]
    \centering
    \includegraphics[scale=.13]{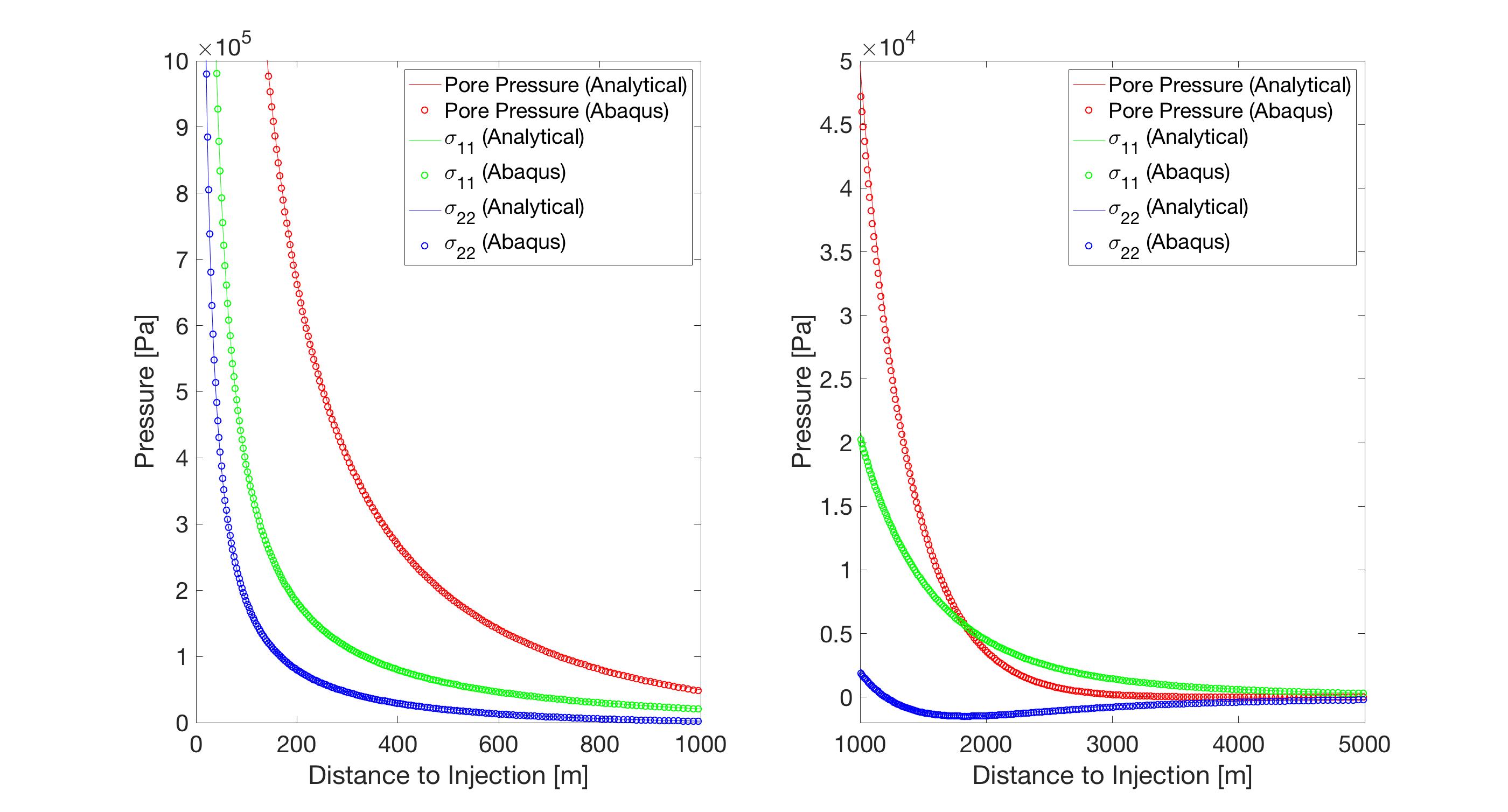}
\caption{\textbf{Analytical Validation} Pore pressure, radial stress $\sigma_{11}$, and perpendicular tangential stress $\sigma_{22}$ as a function of distance from injection ($r=0$) after 34 days of constant injection. The Abaqus material parameters used are $E=26.4$ GPa, $\nu=0.3$, $K=1\cdot10^{-7}$ m/s, void ratio $e=0.1$, wetting fluid weight is 10 kN/m$^3$, $K_g=40$ GPa, and $K_f=2.2$ GPa. }
    \label{fig:analytical}
\end{figure}

We compare these solutions to a simple toy Abaqus model for a variety of different material parameters. We find negligible differences between the analytical and numerical solutions. We provide an example of the benchmark in SM Figure \ref{fig:analytical}.
\section{SM Figures and Tables}
\begin{table}[h]
\begin{tabular}{l|l|l}
\multicolumn{1}{l|}{\textbf{}}  & \multicolumn{2}{c}{Permeability $m^2$}                                \\ \hline
\multicolumn{1}{l|}{Test Case} & \multicolumn{1}{c}{Dakota} & \multicolumn{1}{c}{Entrada} \\ \hline
Low Displacement                & 6.825 $\cdot$ 10$^{-14}$                         & 5.892 $\cdot$ 10$^{-14}$                          \\
High Displacement               & 6.415 $\cdot$ 10$^{-14}$                         & 6.164 $\cdot$ 10$^{-14}$                          \\
Increasing Displacement         & 6.607 $\cdot$ 10$^{-14}$                         & 5.836 $\cdot$ 10$^{-14}$                          \\
Recovery                        & 6.667 $\cdot$ 10$^{-14}$                         & 8.924 $\cdot$ 10$^{-14}$     \\ \hline                    
\end{tabular}
\begin{flushleft}

Table SM 1: \textbf{Permeability Calibration.} Calculated permeabilities in each step rate case test for the Dakota and Entrada formations obtained from AQTESOLV \cite{rhernMSthesis}.

\end{flushleft}
\end{table}

\begin{figure}[h]
    \centering
    \includegraphics[scale=.60]{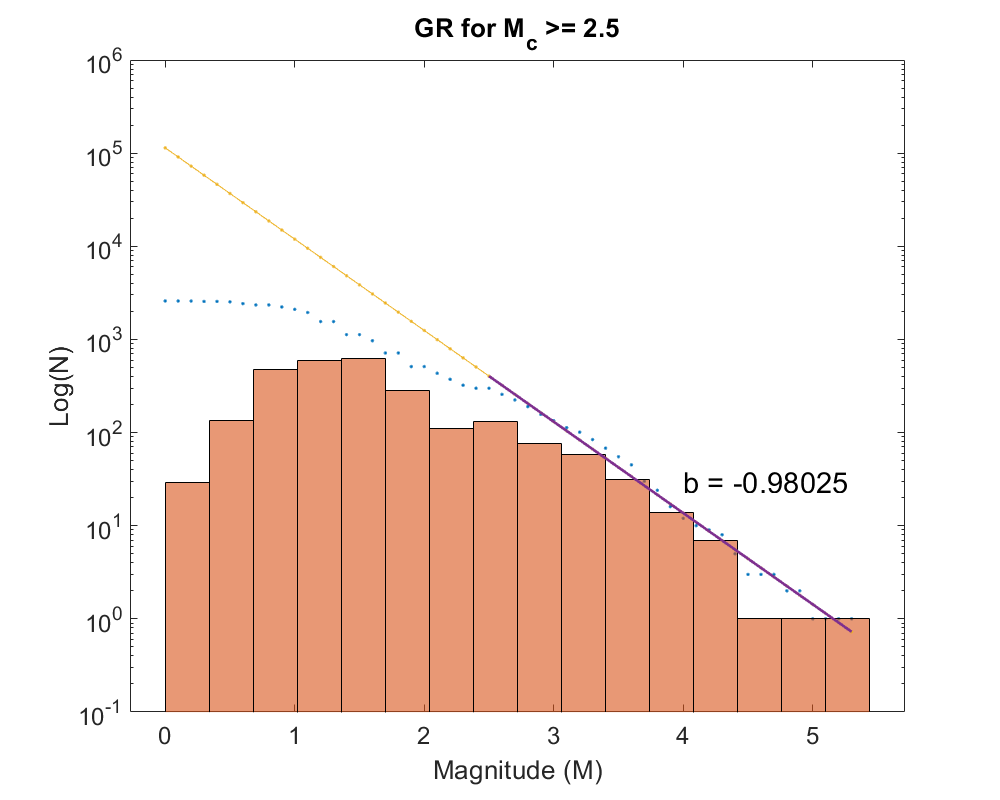}
    \caption{GR law of earthquake catalog prior to higher resolution data from Glagow et al., 2021 \cite{glasgow2021raton} (see Data). A magnitude cut-off of $M_c$=2.5 is chosen from visual inspection where the frequency of events experience 'roll-off` from b-value estimate.}
    \label{fig:mcplot}
\end{figure}

\begin{figure}[h]
    \centering
    \includegraphics[scale=.25]{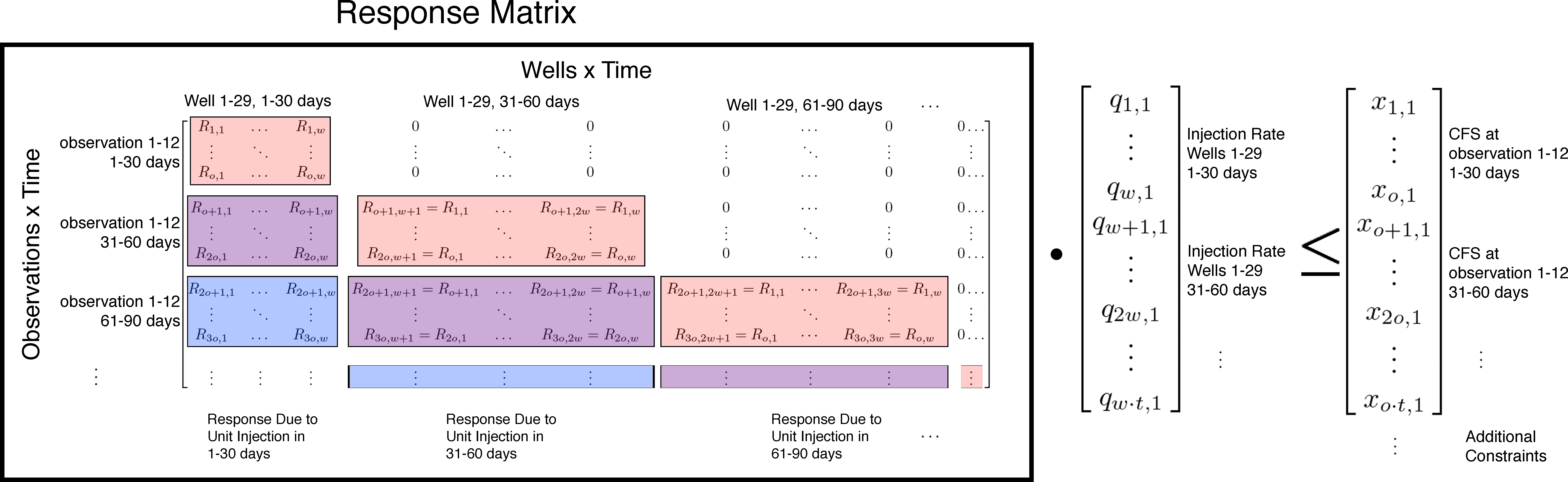}
    \caption{\textbf{Response Matrix.} As an example, we denote the Coulomb stress response matrix as $R_{mn}$, where m$=$732 is the number of rows that equals the number of model output locations (12) times the number of time steps (61), and where n$=$1769 is the number of columns that equals the number of wells (29) times the number of time steps (61). Steps to form the response matrix for the Coulomb stress \textit{rate} are provided in the Appendix. If we denote $q$ as the injection rates at each of the 29 wells for all time steps (61), we can multiply $Rq$ to produce the resulting Coulomb stress at each of the observed locations for each time step. This is the foundation for the management model and linear program optimization. An example of using 12 model output locations is presented in the Supplementary Methods \ref{sec:simple}).}
    \label{fig:response}
\end{figure}

\begin{figure}[h]
    \centering
    \includegraphics[scale=.23]{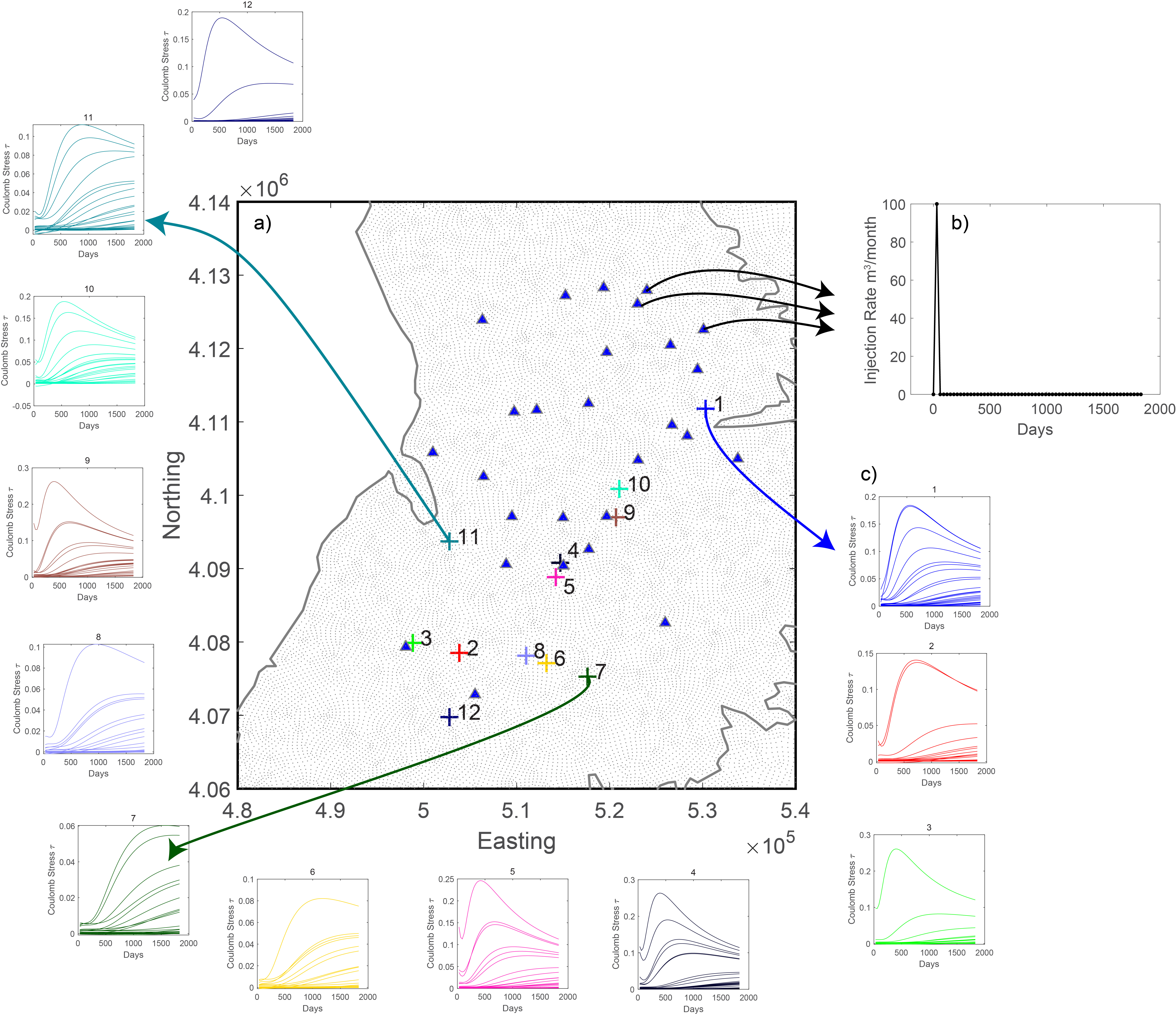}
    \caption{\textbf{Visualization of Response Matrix Generation.} Panel a) shows a zoomed in portion of the model with corresponding well locations (blue triangles) and 12 model output locations where prior M$\geq$4+ events occured in the basin. Panel b) is the unit impulse injection rate. We create 29 separate models that follow this injection profile for each well. The impulse response is an injection of 100 m$^3$ immediately followed by zero injection rate with no injection at the other well locations. Note that the unit impulse response shares both the number of time steps and total time length of the management model. We then record the response at the entire basin (model points in grey). For this example we choose 12 points associated with prior earthquakes. Panel c) shows the 29 responses that each unit response has on each given location. These response values are combined in the response matrix (Figure \ref{fig:response}).}
    \label{fig:respGen}
\end{figure}

\pagebreak
\appendix
\section{Coulomb Stress Rate Response Matrix}
The \textit{rate} response matrix is represented as differences in the original Response Matrix between adjacent time intervals, analogous to a derivative. It is helpful to define components of the original Response Matrix as $R_{n,w,t}$ corresponding to the response at $n$ model output points by $w$ wells during the time interval $t$. Similarly, the injection rate for all wells $w$ during the time interval $t$ is given by $q_{w,t}$ One of these components is equivalent to the colored blocks in (\ref{fig:response}).

It is informative to expand on the derivation of the \textit{rate} response matrix by working out how each time step portion is generated, and its relation to the rate constraint. First, the the initial time step is simply:
\begin{align*}
R_{n,w,1}q_{w,1} \leq \dot{x}_{n,1}
\end{align*}
Then, for the second time step, rate constraint $x_{n,2}$ must satisfy the difference between the response generated in step 2 from the response generated prior. In other words, the difference between the second `row' of the Response Matrix (the response at t=1) and the response at t=2:
\begin{align*}
(R_{n,w,2}q_{w,1} + R_{n,w,1}q_{w,2}) - (R_{n,w,1}q_{w,1}) \leq \dot{x}_{n,2}
\end{align*}
Which, we then factor out the independent injection rates at specific time steps from:
\begin{align*}
(R_{n,w,2}-R_{n,w,1})q_{w,1} + (R_{n,w,1})q_{w,2} \leq \dot{x}_{n,2}
\end{align*}
Repeating the two steps above for the next time step, a pattern begins to emerge:
\begin{align*}
(R_{n,w,3}q_{w,1}+R_{n,w,2}q_{w,2}+R_{n,w,1}q_{w,3})-(R_{n,w,2}q_{w,1}+R_{n,w,1}q_{w,2}) &\leq \dot{x}_{n,3} \\
(R_{n,w,3}-R_{n,w,2})q_{w,1} + (R_{n,w,2}-R_{n,w,1})q_{w,2} + (R_{n,w,1})q_{w,3} &\leq \dot{x}_{n,3} 
\end{align*}
so that in general the rows for each time step $t$ of the \textit{rate} response matrix $\dot{R}_{n,w,t}$ are appended by:
\begin{align*}
\dot{R}_{n,w,t} = \sum_{N=0}^{t-1} (R_{n,w,t-N}-R_{n,w,t-N-1})
\end{align*}
Therefore, the coefficients for each $q_{w,t}$ factor can be combined in a \textit{rate} response matrix which only requires the individual $R_{n,w,t}$ components from the original response matrix to generate. Once generated, if desired, you can choose to optimize the injection rate from the rate constraints exclusively or combined with other constraints.

\newpage

\end{document}